\documentclass[usenatbib]{mn2e}
\usepackage{amsmath,amssymb,amsfonts}    
\usepackage{graphics}                    
\usepackage{mathrsfs}
\usepackage{topcapt}
\usepackage{nicefrac}
\usepackage[usenames,dvipsnames]{xcolor}
\usepackage{graphicx}
\usepackage{footnote}
\usepackage{rotating}
\usepackage{slashbox}
\usepackage{hyperref}                 
\usepackage{natbib}

\DeclareFontFamily{U}{euc}{}
\DeclareFontShape{U}{euc}{m}{n}{<-6>eurm5<6-8>eurm7<8->eurm10}{}%
\DeclareSymbolFont{AMSc}{U}{euc}{m}{n} 
\DeclareMathSymbol{\upmu}{\mathord}{AMSc}{"16}

\def\gtrsim{\mathrel{\hbox{\rlap{\hbox{\lower4pt\hbox{$\sim$}}}\hbox{$>$}}}}
\def\lessim{\mathrel{\hbox{\rlap{\hbox{\lower4pt\hbox{$\sim$}}}\hbox{$<$}}}}
\newcommand{\liso}{L_{iso}}
\newcommand{\eiso}{E_{iso}}
\newcommand{\epkz}{E_{p,z}}
\newcommand{\durz}{T_{90,z}}
\newcommand{\pbol}{P_{bol}}
\newcommand{\sbol}{S_{bol}}
\newcommand{\epk}{E_{p}}
\newcommand{\dur}{T_{90}}
\newcommand{\pph}{P_{50-300}}
\newcommand{\diff}{d}

\title[Short vs. Long Gamma-Ray Bursts]{Short vs. Long Gamma-Ray Bursts: A Comprehensive Study of Energetics and Prompt Gamma-Ray Correlations}
\author[Shahmoradi and Nemiroff]{Amir Shahmoradi$~^{1}$\thanks{E-mail: amir@physics.utexas.edu (AS); nemiroff@mtu.edu (RJN)}, Robert J. Nemiroff$~^{2}$\footnotemark[1] \\
$^{1}$Department of Physics \& Institute for Fusion Studies, The University of Texas at Austin, TX 78712, USA \\
$^{2}$Department of Physics, Michigan Technological University, Houghton, MI 49931, USA}

\begin{document}

\date{\date{Accepted ... Received ... ; in original form ...}}

\pagerange{\pageref{firstpage}--\pageref{lastpage}} \pubyear{2014}

\maketitle

\label{firstpage}

\begin{abstract}
    We present the results of a comprehensive study of the luminosity function, energetics, prompt gamma-ray correlations, and classification methodology of short--hard and long--soft GRBs (SGRBs \& LGRBs), based on observational data in the largest catalog of GRBs available to this date: BATSE catalog of 2702 GRBs. We find that: ~1. The least-biased classification method of GRBs into short and long, solely based on prompt--emission properties, appears to be the ratio of the observed spectral peak energy to the observed duration ($R=\epk/\dur$) with the dividing line at $R\simeq50[keV~s^{-1}]$.  ~2. Once data is carefully corrected for the effects of the detection threshold of gamma-ray instruments, the population distribution of SGRBs and LGRBs can be individually well described as multivariate log-normal distribution in the $4$--dimensional space of the isotropic peak gamma-ray luminosity, total isotropic gamma-ray emission, the intrinsic spectral peak energy, and the intrinsic duration.  ~3. Relatively large fractions of SGRBs and LGRBs with moderate-to-low spectral peak energies have been missed by BATSE detectors. ~4. Relatively strong and highly significant intrinsic hardness--brightness and duration--brightness correlations likely exist in both populations of SGRBs and LGRBs, once data is corrected for selection effects. The strengths of these correlations are very similar in both populations, implying similar mechanisms at work in both GRB classes, leading to the emergence of these prompt gamma-ray correlations.
\end{abstract}


\section{Introduction}

    The field of Gamma-Ray Bursts (GRBs) has witnessed rapid growth over the past decades, in particular, following the launches of NASA Compton Gamma-Ray Observatory \citep{meegan_spatial_1992}, Swift \citep{gehrels_swift_2004} \& Fermi \citep{michelson_fermi_2010} missions. Early hints to the existence of distinct populations of gamma-ray transients \citep[e.g.,][]{mazets_catalog_1981, norris_frequency_1984}, and at least two classes of short-hard (Type-I) \& long-soft (Type-II) GRBs \citep{dezalay_short_1992} have now been extensively corroborated and confirmed by the prompt-emission data from independent gamma-ray detector missions \citep[e.g.,][]{kouveliotou_identification_1993, gehrels_gamma-ray_2009, zhang_revisiting_2012} or follow-up observations of the afterglows or host galaxies \citep[e.g.,][]{zhang_discerning_2009, berger_environments_2011, berger_short-duration_2014}. Although the possibility of more than two classes of GRBs with distinct progenitors has been extensively discussed and considered \citep[e.g.,][]{horvath_third_1998, mukherjee_three_1998, hakkila_tools_2001, balastegui_reclassification_2001, hakkila_subgroups_2004, horvath_new_2006, gehrels_new_2006, chattopadhyay_statistical_2007, horvath_classification_2008, virgili_low-luminosity_2009, gao_new_2010, horvath_classification_2012, levan_new_2014, kobori_investigation_2014}, it has remained a matter of debate and speculation to this date \citep[e.g.,][]{hakkila_fluence_2000, hakkila_gamma-ray_2000, hakkila_properties_2000, hakkila_how_2003, hakkila_dual_2004, shahmoradi_multivariate_2013, zhang_how_2014, levan_new_2014}.

    Beginning with the influential work of \citet{kouveliotou_identification_1993}, GRBs have been traditionally classified into two populations of Short and Long GRBs (SGRBs \& LGRBs respectively) based on a sharp cutoff on the bimodal distribution of the observed duration ($\dur$) of prompt gamma-ray emission, generally set to $\dur\sim2-3[s]$. Nevertheless, the detector \& energy dependence of the observed GRB duration \citep[e.g.,][]{fenimore_gamma-ray_1995, nemiroff_pulse_2000, qin_comprehensive_2013} has prompted many studies in search of alternative less-biased methods of GRB classification, typically based on a combination of the prompt gamma-ray emission and$/$or afterglow$/$host properties \citep[e.g.,][]{gehrels_gamma-ray_2009, zhang_discerning_2009, goldstein_new_2010, lu_new_2010, shahmoradi_possible_2011, zhang_correlation_2012, zhang_revisiting_2012, shahmoradi_multivariate_2013, lu_`amplitude_2014} or using the prompt-emission spectral correlations in conjunction with the traditional method of classification \citep[e.g.,][]{qin_statistical_2013}.

    Ideally, a phenomenological classification method of GRBs should be based on their intrinsic (i.e., rest-frame) properties, free from potential biases due to data analysis, detector specifications, observational selection effects and sample incompleteness. Such method is currently far from reach due to detector-induced heterogeneity in available GRB catalogs \citep[e.g.,][]{qin_comprehensive_2013} and complex selection effects in the detection, analysis, and redshift measurement processes \citep[e.g.,][c.f., Shahmoradi \& Nemiroff 2011 for a comprehensive review of relevant literature]{hakkila_fluence_2000, hakkila_how_2003, band_testing_2005, nakar_outliers_2005, nava_peak_2008, butler_complete_2007, butler_generalized_2009, butler_cosmic_2010, shahmoradi_how_2009, shahmoradi_possible_2011, shahmoradi_multivariate_2013, shahmoradi_gamma-ray_2013, coward_swift_2013}.

    Historically, the intrinsic properties of short GRBs have also been less studied compared to long GRBs, potentially due to lack of redshift information for the majority of SGRBs \citep[e.g.,][]{coward_swift_2012}. This has led to an additional obstacle towards a quantitative GRB taxonomy. While the population properties of long GRBs and their progenitors have been extensively researched and fairly well constrained \citep[e.g.,][c.f., Shahmoradi 2013 for a comprehensive review of the literature]{dainotti_towards_2011, shahmoradi_multivariate_2013, shahmoradi_gamma-ray_2013, lien_probing_2013, dainotti_slope_2013, lien_probing_2014, howell_constraining_2014, pescalli_luminosity_2015, littlejohns_investigating_2014, dainotti_selection_2014}, the cosmic rate and the intrinsic prompt-emission properties of short GRBs are far less understood and investigated \citep[e.g.,][]{guetta_luminosity_2005, guetta_batse-_2006, nakar_local_2006, salvaterra_short_2008, chapman_short_2009, virgili_are_2011, czerny_observational_2011, wanderman_rate_2014}.

    Motivated by the existing gap and uncertainties in the current knowledge of the intrinsic population properties of short GRBs and the lack of an efficient, quantitative, bias-free classification method for gamma-ray bursts into long and short subgroups, here we present a methodology and model to constrain the energetics, luminosity function and the joint distributions and correlations of the prompt gamma-ray emission of SGRBs. Despite lacking a complete knowledge of the true cosmic rate and redshift distribution, here we argue and show that short GRBs exhibit very similar prompt-emission correlations and population distribution to those of long GRBs as presented by \citet{shahmoradi_multivariate_2013}, both qualitatively and, under plausible cosmic rate assumptions, also quantitatively. This result, along with other recent works on the time-resolved and time-integrated spectral properties of SGRBs \citep[e.g.,][]{ghirlanda_spectral_2011, shahmoradi_possible_2011, zhang_correlation_2012, tsutsui_possible_2013, calderone_there_2014} points towards the possibility of a unified mechanism responsible for the prompt gamma-ray emission of the two GRB classes, independently of the diverse progenitor candidates for the two GRB populations.

    The presented work also paves the way towards a detector-independent minimally-biased phenomenological classification method for GRBs solely based on the observed prompt gamma-ray data of individual events. Towards this, we focus our attention on the largest sample of uniformly-detected gamma-ray bursts to this date: the BATSE catalog of 2702 GRBs \citep{paciesas_fourth_1999, goldstein_batse_2013}.  In the following sections, we present an example of data mining on BATSE data that showcases the tremendous amount of useful, yet unexplored information buried in this seemingly archaic GRB catalog.

     We devote Sec. \ref{sec:samsel} of this manuscript to presenting an elaborate method of classification for the {\it observed} sample of Gamma-Ray Bursts into two subclasses of short-hard and long-soft GRBs. In Sec. \ref{sec:MC}, we elaborate on the construction of a GRB world model that is capable of describing BATSE SGRB data. This is followed by a discussion of the procedure for fitting the model to BATSE data in Sec. \ref{sec:MF}. The predictions of the model together with univariate and multivariate goodness-of-fit tests are presented in Sec. \ref{sec:results}, followed by a discussion of the similarities and differences in the population distributions of SGRBs \& LGRBs in Sec. \ref{sec:discussion}. The main findings of the presented GRB world model are summarized in Sec. \ref{sec:CR}.

\section{GRB World Model}
    \label{sec:GWM}

    \subsection{Sample Selection}
    \label{sec:samsel}

        \begin{figure*}
            \centering
            \includegraphics[scale=0.83]{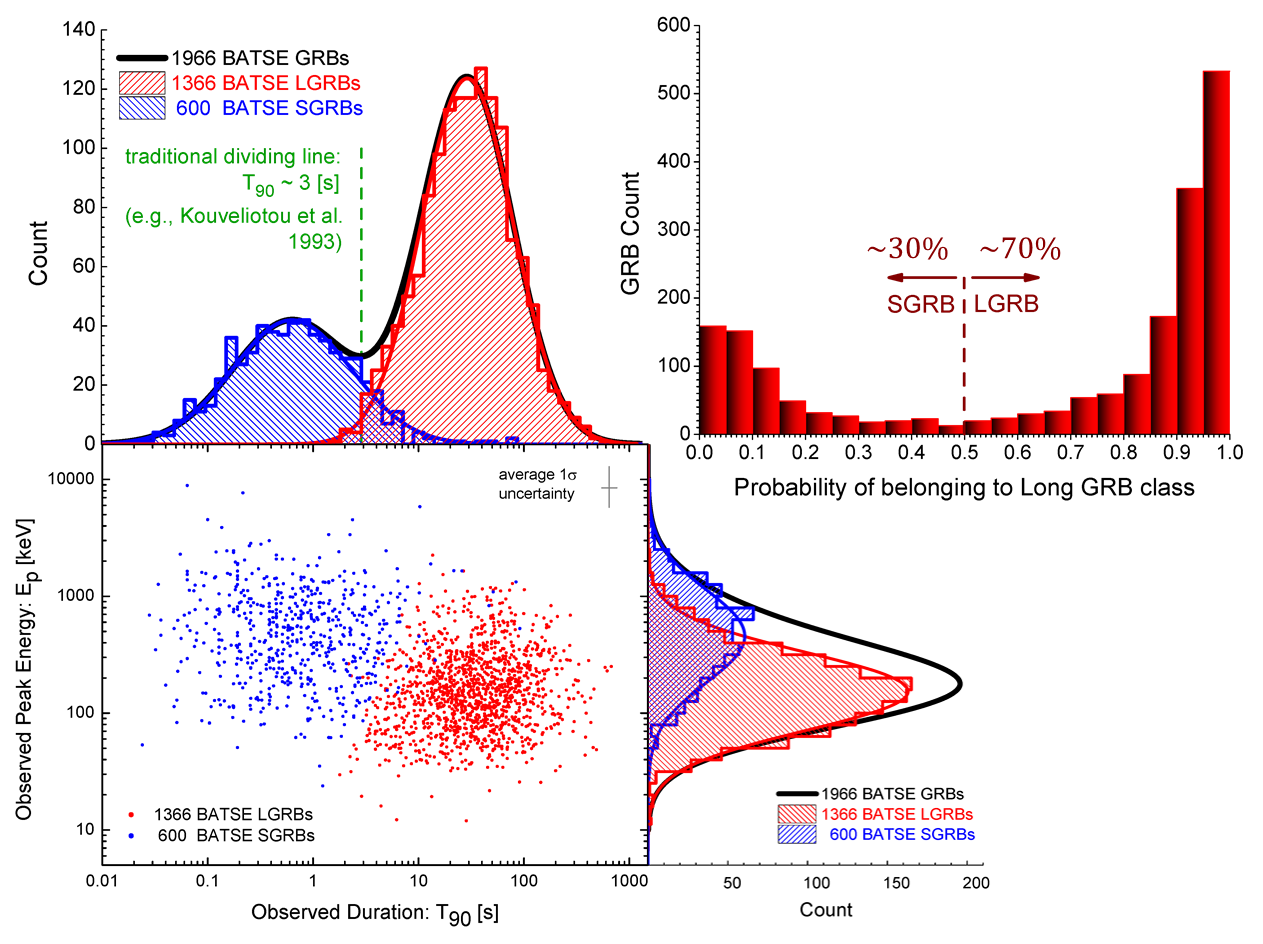}
            \caption{Classification of $1966$ BATSE GRBs with measured peak flux ($\pbol$), fluence ($\sbol$), and observed duration ($\dur$) taken from the current BATSE catalog and measured spectral peak energies ($\epk$) taken from \citet{shahmoradi_hardness_2010}. The segregation methodology is based on fuzzy C-means clustering algorithm using the two GRB variables $\epk$ \& $\dur$ that are least affected by the triggering threshold of the BATSE Large Area Detectors (c.f., Sec. \ref{sec:GWM}). \label{fig:classification}}
        \end{figure*}

        Depending on the triggering criteria, some gamma-ray detectors facilitate the detection of one class of bursts over the others \citep[e.g.,][]{paciesas_fourth_1999, hakkila_batse_2003, band_comparison_2003, band_postlaunch_2006, band_exists_2008, lien_probing_2014}. For example, the specific detector sensitivity of the Burst Alert Telescope (BAT) onboard Swift satellite results in better detections of LGRBs over SGRBs \citep[e.g.,][]{band_comparison_2003, band_postlaunch_2006}. Compared to BAT however, BATSE Large Area Detectors (LAD) had an increased {\it relative} sensitivity to short GRBs \citep[e.g.,][]{hakkila_batse_2003, band_comparison_2003}. Several studies have already offered new methods for a global detector-independent phenomenological classification of GRBs based on empirical relations that are believed to hold for only a specific class of GRBs or based on the prompt or afterglow emission data \citep[e.g.,][]{zhang_discerning_2009, lu_new_2010, qin_statistical_2013}. These methods however, either suffer from sample incompleteness or require information, such as redshift, that might not be available for the majority of GRBs \citep[e.g.,][]{coward_swift_2012, coward_swift_2013}.

        Here in this work, we ensure a minimally-biased analysis of short-duration class of GRBs by following the classification approach of \citet{shahmoradi_multivariate_2013}, which to the extent of our knowledge provides the least biased classification of BATSE GRBs, solely based on prompt emission data\footnote{The authors have already investigated multiple classification methodologies, on a variety of GRB characteristics, that are documented in \citet{shahmoradi_possible_2011} \& \citet{shahmoradi_multivariate_2013}.}. The word `bias' here refers to the systematic contamination of data and analysis that might be introduced when using the traditional definition of GRB classes, based on a sharp cutoff on the duration variable $\dur$ \citep[][]{kouveliotou_identification_1993}, as it is generally used by many GRB researchers \citep[e.g.,][]{guetta_luminosity_2005, butler_cosmic_2010, campisi_redshift_2010, wanderman_luminosity_2010}.

        The classification method used here is based on fuzzy C-means clustering algorithm of \citet{dunn_fuzzy_1973, bezdek_pattern_1981}:  Each BATSE GRB is assigned a probability (i.e., class coefficient) of belonging to LGRB (versus SGRB) population, the value of which depends on the set of GRB variables used in classification. This can be any combinations of the peak flux ($\pph~[photon/s/cm^2]$) in BATSE detection energy range $[50-300~keV]$, in three different time scales: $64$, $256$ \& $1024~[ms]$; bolometric fluence ($\sbol~[erg/cm^2]$); the observer-frame spectral peak energy ($\epk~[keV]$), for which we use estimates by \citet{shahmoradi_hardness_2010}, and the observed duration ($\dur$~[s]).\footnote{All BATSE GRB data are taken from the current BATSE Catalog and the spectral peak energies ($\epk$) are taken from \citet{shahmoradi_hardness_2010}, publicly available at: \url{https://sites.google.com/site/amshportal/research/aca/in-the-news/lgrb-world-model}.}   Then, GRBs with LGRB class coefficient $>0.5$ are flagged as long-duration bursts. Following \citet{shahmoradi_multivariate_2013}, we make use of only two GRB variables $\epk$ \& $\dur$ for classification of the sample into long and short GRBs.  Unlike $\epk$ \& $\dur$ which are weakly coupled to (i.e., correlated with) the variable peak flux ($\pbol$), the population distribution of the two other GRB prompt variables ($\sbol$ \& $\pbol$) are strongly affected by the detector threshold and not suitable for classifications based on fuzzy C-means algorithms. This is mainly due to the sensitivity of C-means clustering method to different subgroup sizes, orientations, and asymmetries \citep[c.f.,][Sec. 2.1 \& Appendix A therein for details]{shahmoradi_multivariate_2013}. In addition, the two GRB populations are most distinctively separated in the plane of $\dur-\epk$. This classification led to the initial selection of $1366$ events as LGRBs and $600$ events as SGRBs in our sample of $1966$ BATSE GRBs with complete prompt emission data as illustrated in Figure \ref{fig:classification}, also in section 2.1 of \citet{shahmoradi_multivariate_2013}.

        To ensure minimal contamination of the two GRB classes, the lightcurves of $291$ bursts among $1966$ BATSE GRBs with SGRB class coefficients in the range of $0.3-0.7$ were also visually inspected in the four main energy channels of BATSE Large Area Detectors. This led to reclassification of some events as Soft Gamma Repeaters (SGRs) or from one class to another, thus reducing the size of the original SGRB sample to $565$ (Table \ref{tab:data}). \footnote{It is notable that the same simulation protocol for the original $600$ BATSE events flagged as SGRBs did not result in any qualitative change in the conclusions of this work, although the strength and significance of the measured parameters of the model are affected by the presence potential non-SGRB events in data.} \citet{shahmoradi_multivariate_2013} finds that the inclusion of the uncertainties on the two GRB variables $\dur$ \& $\epk$ has marginal effects on the derived samples of the two GRB classes discussed above. Also, a classification based on $T_{50}$ instead of $\dur$ resulted in about the same sample sizes for the two GRB classes with a minimal difference of $\sim0.7\%$ (c.f., section 2.1, Appendix A and Figure (1) in \citet{shahmoradi_multivariate_2013} for further comparison and details of different classification methods).

\begin{table*}
\begin{center}
\caption{BATSE catalog GRB trigger numbers classified as SGRBs. \label{tab:data}}
\begin{tabular}{|p{0.85cm}|p{0.85cm}|p{0.85cm}|p{0.85cm}|p{0.85cm}|p{0.85cm}|p{0.85cm}|p{0.85cm}|p{0.85cm}|p{0.85cm}|p{0.85cm}|p{0.85cm}|p{0.85cm}|p{0.85cm}|}
{\small Trigger} & {\small Trigger} & {\small Trigger} & {\small Trigger} & {\small Trigger} & {\small Trigger} & {\small Trigger} & {\small Trigger} & {\small Trigger} & {\small Trigger} & {\small Trigger} & {\small Trigger} & {\small Trigger} & {\small Trigger} \\
\hline
\hline
108   &  138   &  185   &  207   &  218   &  229   &  254   &  289   &  297   &  373   &  432   &  474   &  480   &  486   \\
491   &  508   &  512   &  537   &  547   &  551   &  555   &  568   &  575   &  603   &  677   &  729   &  734   &  788   \\
799   &  809   &  830   &  834   &  836   &  845   &  856   &  867   &  878   &  906   &  909   &  929   &  936   &  942   \\
974   &  1051  &  1073  &  1076  &  1088  &  1096  &  1097  &  1102  &  1112  &  1128  &  1129  &  1154  &  1211  &  1223  \\
1289  &  1308  &  1346  &  1359  &  1404  &  1435  &  1443  &  1453  &  1461  &  1481  &  1518  &  1546  &  1553  &  1566  \\
1588  &  1634  &  1635  &  1636  &  1637  &  1659  &  1662  &  1665  &  1679  &  1680  &  1683  &  1694  &  1719  &  1736  \\
1741  &  1747  &  1760  &  1851  &  1953  &  1968  &  2003  &  2037  &  2040  &  2041  &  2043  &  2044  &  2049  &  2056  \\
2068  &  2099  &  2103  &  2115  &  2117  &  2125  &  2126  &  2132  &  2142  &  2145  &  2146  &  2155  &  2159  &  2161  \\
2163  &  2167  &  2201  &  2205  &  2206  &  2217  &  2220  &  2265  &  2268  &  2273  &  2283  &  2288  &  2312  &  2326  \\
2327  &  2330  &  2332  &  2352  &  2353  &  2357  &  2358  &  2360  &  2365  &  2368  &  2372  &  2377  &  2382  &  2384  \\
2395  &  2401  &  2424  &  2434  &  2448  &  2449  &  2454  &  2485  &  2487  &  2502  &  2504  &  2512  &  2513  &  2523  \\
2529  &  2536  &  2564  &  2583  &  2585  &  2597  &  2599  &  2614  &  2615  &  2623  &  2632  &  2633  &  2649  &  2679  \\
2680  &  2690  &  2693  &  2701  &  2715  &  2728  &  2748  &  2755  &  2757  &  2760  &  2776  &  2788  &  2795  &  2799  \\
2800  &  2801  &  2810  &  2814  &  2821  &  2823  &  2828  &  2834  &  2844  &  2846  &  2849  &  2851  &  2860  &  2861  \\
2873  &  2879  &  2892  &  2894  &  2910  &  2918  &  2933  &  2952  &  2964  &  2966  &  2973  &  2975  &  2977  &  2978  \\
2987  &  2988  &  2995  &  3016  &  3027  &  3037  &  3038  &  3039  &  3043  &  3051  &  3066  &  3073  &  3078  &  3087  \\
3094  &  3113  &  3114  &  3118  &  3121  &  3137  &  3144  &  3146  &  3152  &  3155  &  3160  &  3164  &  3173  &  3215  \\
3218  &  3266  &  3278  &  3280  &  3282  &  3286  &  3293  &  3294  &  3297  &  3308  &  3323  &  3333  &  3335  &  3338  \\
3340  &  3342  &  3349  &  3359  &  3374  &  3379  &  3384  &  3437  &  3441  &  3476  &  3477  &  3487  &  3494  &  3502  \\
3510  &  3530  &  3545  &  3606  &  3611  &  3640  &  3642  &  3665  &  3668  &  3722  &  3728  &  3735  &  3737  &  3742  \\
3751  &  3770  &  3774  &  3782  &  3791  &  3799  &  3810  &  3866  &  3867  &  3868  &  3888  &  3889  &  3894  &  3895  \\
3902  &  3904  &  3910  &  3919  &  3921  &  3936  &  3939  &  3940  &  4327  &  4660  &  4744  &  4776  &  4807  &  4871  \\
4955  &  5079  &  5206  &  5212  &  5277  &  5339  &  5439  &  5448  &  5453  &  5456  &  5458  &  5459  &  5461  &  5467  \\
5469  &  5471  &  5485  &  5488  &  5491  &  5498  &  5499  &  5500  &  5501  &  5527  &  5528  &  5529  &  5533  &  5536  \\
5537  &  5546  &  5547  &  5556  &  5560  &  5562  &  5564  &  5576  &  5592  &  5599  &  5607  &  5619  &  5620  &  5633  \\
5638  &  5647  &  5650  &  5664  &  5724  &  5730  &  5733  &  5740  &  5770  &  5992  &  6091  &  6096  &  6105  &  6117  \\
6120  &  6123  &  6135  &  6136  &  6145  &  6153  &  6166  &  6178  &  6180  &  6182  &  6204  &  6205  &  6215  &  6216  \\
6219  &  6230  &  6237  &  6238  &  6251  &  6263  &  6265  &  6275  &  6281  &  6284  &  6292  &  6299  &  6301  &  6307  \\
6314  &  6331  &  6338  &  6341  &  6342  &  6343  &  6347  &  6354  &  6361  &  6368  &  6372  &  6376  &  6385  &  6386  \\
6398  &  6401  &  6411  &  6412  &  6427  &  6436  &  6439  &  6443  &  6445  &  6447  &  6452  &  6462  &  6469  &  6486  \\
6488  &  6497  &  6535  &  6540  &  6542  &  6543  &  6547  &  6562  &  6569  &  6571  &  6573  &  6579  &  6580  &  6586  \\
6591  &  6606  &  6634  &  6635  &  6638  &  6641  &  6643  &  6645  &  6659  &  6662  &  6671  &  6679  &  6682  &  6689  \\
6693  &  6697  &  6700  &  6710  &  6715  &  6718  &  6753  &  6757  &  6786  &  6787  &  6788  &  6800  &  6824  &  6866  \\
6867  &  6870  &  6904  &  6916  &  6931  &  7009  &  7060  &  7063  &  7078  &  7102  &  7106  &  7133  &  7142  &  7148  \\
7159  &  7173  &  7187  &  7227  &  7240  &  7281  &  7283  &  7287  &  7290  &  7292  &  7294  &  7297  &  7305  &  7329  \\
7344  &  7353  &  7359  &  7361  &  7366  &  7367  &  7375  &  7378  &  7427  &  7430  &  7440  &  7447  &  7449  &  7453  \\
7455  &  7456  &  7472  &  7495  &  7496  &  7508  &  7514  &  7526  &  7547  &  7554  &  7559  &  7581  &  7584  &  7595  \\
7599  &  7601  &  7602  &  7626  &  7663  &  7671  &  7706  &  7710  &  7734  &  7745  &  7753  &  7754  &  7775  &  7784  \\
7789  &  7793  &  7800  &  7805  &  7827  &  7830  &  7901  &  7912  &  7922  &  7939  &  7943  &  7952  &  7970  &  7979  \\
7980  &  7988  &  7995  &  7999  &  8018  &  8027  &  8035  &  8041  &  8047  &  8072  &  8076  &  8077  &  8079  &  8082  \\
8085  &  8089  &  8097  &  8104  &  8120  &  ---   &  ---   &  ---   &  ---   &  ---   &  ---   &  ---   &  ---   &  ---   \\
\hline
\hline
\end{tabular}
\end{center}
\end{table*}

    \subsection{Model Construction}
    \label{sec:MC}

        Our primary goal in this work is to derive a multivariate statistical model that, subject to BATSE detection threshold, is capable of reproducing the observational data of $565$ BATSE SGRBs. Examples of multivariate treatment of GRB luminosity function and energetics are rare in studies of Gamma-Ray Bursts.  Conversely, many authors have focused primarily on the univariate distribution of the spectral parameters, most importantly on the luminosity function. This is particularly true for the short population of GRBs where the prompt, afterglow and redshift information of individual events are scarcely available. In \citet{shahmoradi_multivariate_2013} we argued that an accurate modelling of the luminosity function of LGRBs requires at least two GRB observable incorporated in the model: the bolometric peak flux ($\pbol$) and the observed peak energy ($\epk$). The parameter $\epk$ is required, since most gamma-ray detectors are photon counters, a quantity that depends on not only $\pbol$ but also $\epk$ of the burst. This leads to the requirement of using a bivariate distribution as the minimum acceptable model for long GRBs, for the purpose of constraining the luminosity function.

        For the class of short GRBs, the duration distribution (e.g., $\dur$) of the population spans a wide range from milliseconds to tens of seconds. The wide duration distribution is particularly important in modeling BATSE Large Area Detectors, since short GRBs could be potentially triggered on any of three triggering timescales: $64$ms, $256$ms \& $1024$ms. Therefore, a SGRB world model should minimally incorporate the joint {\it trivariate} distribution of $\pbol$, $\epk$ and an appropriate definition of the observed duration (e.g., $\dur$). The duration variable is required in order to correctly account for the detection threshold of BATSE LADs. In addition, the definition of the observer-frame parameter {\it bolometric peak flux } ($\pbol$) and the corresponding rest-frame parameter $\liso$ merits special attention in the study of SGRBs. Here we use the $64ms$ definition of peak flux, $P_{\text{bol},64}$, for SGRBs taken from BATSE catalog data, in contrast to the two other common definitions: $256ms$ \& $1024ms$. Unlike the case for LGRBs \citep[e.g.,][]{shahmoradi_multivariate_2013}, we show in the addendum to this article (Appendix \ref{App:BDT} \& \ref{App:SB}) that $P_{bol,64}$ is the least biased measure of peak flux for the majority of BATSE SGRBs, also the best definition for an appropriate modelling of the BATSE detection threshold in case of short GRBs.

        Hereafter in the text and figures, the two parameters $\pbol$ and $\liso$ implicitly refer to a $64ms$ definition of peak flux and luminosity wherever used for SGRBs and to the commonly used $1024ms$ definitions wherever used for LGRBs.

        Following the arguments of \citet{shahmoradi_multivariate_2013} for LGRBs, we propose the multivariate log-normal distribution as the simplest natural candidate model capable of describing BATSE SGRB data. The motivation behind this choice of model comes from the available observational data that closely resembles a joint multivariate log-normal distribution for four most widely studied temporal and spectral parameters of both GRB classes in the observer-frame: $\pbol$, $\sbol$ (bolometric fluence), $\epk$, $\dur$: Since most SGRBs are expected to originate from low redshifts $z\lesssim3$, the convolution of these observer-frame parameters with the redshift distribution results in negligible variation in the shape of the rest-frame joint distribution of the same SGRB parameters. Therefore, the redshift-convoluted 4-Dimensional (4D) rest-frame distribution can be well approximated as a linear translation of the observer-frame parameters to the rest-frame parameter space, keeping the shape of the distribution almost intact \citep[e.g.,][]{balazs_difference_2003}. This implies that the joint distribution of the intrinsic SGRB variables: the isotropic peak luminosity ($\liso$), the total isotropic emission ($\eiso$), the rest-frame time-integrated spectral peak energy ($\epkz$) and the rest-frame duration ($\durz$) might be indeed well described as a multivariate log-normal distribution.

        We model the process of SGRB observation as a non-homogeneous Poisson process whose mean rate parameter -- the cosmic SGRB differential rate, $\mathcal{R}_{cosmic}$ -- is the product of the differential comoving SGRB rate density $\dot\zeta(z)$ with a $p=4$ dimensional log-normal Probability Density Function (pdf), $\mathcal{LN}$, of four SGRB variables: $\liso$, $\eiso$, $\epkz$ and $\durz$, with location vector $\vec \mu$ and the scale (i.e., covariance) matrix $\Sigma$,

        \begin{eqnarray}
            \label{eq:cosmicrate}
            \mathcal{R}_{cosmic} &=& \frac{\diff N}{\diff\liso~\diff\eiso~\diff\epkz~\diff\durz~\diff z} \\
                           &\propto& \mathcal{LN}\bigg(\liso,\eiso,\epkz,\durz\big|\vec \mu,\boldsymbol\Sigma\bigg) \nonumber  \\
                           &\times & \frac{\dot\zeta(z)\nicefrac{\diff V}{\diff z}}{(1+z)}, \nonumber
        \end{eqnarray}

        where $\dot\zeta(z)$ is the comoving SGRB rate density, and the factor $(1+z)$ in the denominator accounts for cosmological time dilation. The comoving volume element per unit redshift, $\nicefrac{\diff V}{\diff z}$, is given by,

        \begin{equation}
            \label{eq:dvdz}
            \frac{\diff V}{\diff z} = \frac{C}{H_0}\frac{4\pi {D_L}^2(z)}{(1+z)^2\bigg[\Omega_M(1+z)^3+\Omega_\Lambda\bigg]^{1/2}},
        \end{equation}

        with $D_L$ standing for the luminosity distance,

        \begin{equation}
            \label{eq:lumdis}
            D_L(z)=\frac{C}{H_0}(1+z)\int^{z}_{0}dz'\bigg[(1+z')^{3}\Omega_{M}+\Omega_{\Lambda}\bigg]^{-1/2},
        \end{equation}

        assuming a flat $\Lambda$CDM cosmology, with parameters set to $h=0.70$, $\Omega_M=0.27$ and $\Omega_\Lambda=0.73$ \citep{jarosik_seven-year_2011} for consistency with the work of \citet{shahmoradi_multivariate_2013}. The parameters $C$ \& $H_{0}=100h~[Km/s/MPc]$ stand for the speed of light and the Hubble constant respectively.

        The 4-dimensional log-normal distribution of Equation (\ref{eq:cosmicrate}), $\mathcal{LN}$, has an intimate connection to the multivariate Gaussian distribution in the logarithmic space of SGRB observable parameters \citep[c.f., Appendix D in ][]{shahmoradi_multivariate_2013}.

        Finally, in order to obtain the observed rate ($\mathcal{R}_{obs}$) of SGRBs detected by BATSE LADs, the cosmic SGRB rate, $\mathcal{R}_{cosmic}$, in Equation (\ref{eq:cosmicrate}) must be convolved with an accurate model of BATSE trigger efficiency for short GRBs, $\eta$,

        \begin{equation}
            \label{eq:obsrate}
            \mathcal{R}_{obs}=\eta(\liso,\epkz,\durz,z)\times\mathcal{R}_{cosmic}
        \end{equation}

        In reality, the variable $\eta$ is a highly complex function of observational conditions and prompt emission characteristics, almost unique to each individual GRB. Nevertheless, we show in Appendix \ref{App:BDT} that it can be approximated as a generic function of the burst's redshift ($z$), isotropic luminosity ($\liso$), the rest-frame spectral peak energy ($\epkz$) and the rest-frame duration ($\durz$).

        \subsection{The SGRB Rate Density}
        \label{sec:SGRBrate}

        The largest source of uncertainty in population studies of short GRBs originates from the lack of an accurate knowledge of their cosmic rate. Only a small fraction of heterogeneously-detected SGRBs have measured redshifts to this date \citep[e.g.,][]{coward_swift_2012} and redshift completeness often limits studies to the brightest events \citep[e.g.,][]{davanzo_complete_2014}. It is therefore, perceivable that the current observed redshift distribution of SGRBs is likely strongly biased and not representative of the entire population of SGRBs \citep[e.g.,][]{nakar_short-hard_2007, coward_swift_2013}. An alternative approach to empirical determination of the rate of SGRBs is through population synthesis simulations \citep[e.g.,][]{belczynski_study_2006}, based on the assumption of `compact binary mergers' as the progenitor of the majority of SGRBs \citep[e.g.,][]{narayan_gamma-ray_1992, eichler_nucleosynthesis_1989, paczynski_gamma-ray_1986}. In this scenario, the cosmic rate of SGRBs follows the Star Formation Rate (SFR) convolved with a distribution of the {\it delay time} between the formation of a binary system and its coalescence due to gravitational radiation.

        There is currently no consensus on the statistical moments and shape of the delay time distribution, solely based on observations of individual events and their host galaxies. The median delays vary widely in the range of $\sim0.1-7$ billion years depending on the assumptions involved in estimation methods or in the dominant binary formation channels considered \citep[e.g.,][]{belczynski_study_2006, hopman_redshift_2006, bogomazov_evolution_2007, berger_new_2007, zheng_deducing_2007, berger_environments_2011, hao_progenitor_2013, guelbenzu_another_2014}. Recent results from population synthesis simulations however, favor very short delay times of a few hundred million years with a long negligible tail towards several billion years \citep[e.g.,][]{oshaughnessy_short_2008, belczynski_origin_2010}.

        The computational expenses and limitations imposed on this work strongly limit the number of possible scenarios that could be considered for the cosmic rate of short GRBs. Thus, in order to approximate the comoving rate density $\dot\zeta(z)$ of SGRBs, we adopt the Star Formation Rate (SFR) of \citet{hopkins_normalization_2006} in the form of a piecewise power-law function,

        \begin{equation}
            \mathcal{SFR}(z)  \propto  \begin{cases}
                                            (1+z)^{\gamma_0} & z<z_0 \\
                                            (1+z)^{\gamma_1} & z_0<z<z_1 \\
                                            (1+z)^{\gamma_2} & z>z_1, \\
                                       \end{cases}
            \label{eq:zeta}
        \end{equation}

         with parameters $(z_0,z_1,\gamma_0,\gamma_1,\gamma_2)$ set to the best-fit values $(0.993,3.8,3.3,0.055,-4.46)$ of an updated SFR fit by \citet{li_star_2008}. The SFR is then convolved with a log-normal model of the delay time distribution \citep[e.g.,][]{nakar_short-hard_2007},

		 \begin{equation}
            \mathcal{LN}(\tau|\mu,\sigma) \propto \frac{1}{\tau\sigma}e^{-\frac{(\ln\tau-\mu)^2}{2\sigma^2}}
            \label{eq:lognormal}
        \end{equation}
		
        with parameters $[\mu,\sigma] = [\log(0.1),1.12]$ in units of billion years (Gyrs) estimated from the population synthesis simulation results of \citep[e.g.,][]{belczynski_origin_2010}, such that the comoving rate density of SGRBs is calculated as,

        \begin{equation}
            \dot\zeta(z) \propto \int_z^{\infty} \mathcal{SFR}\big(z^\prime\big)\mathcal{LN}\big(t(z)-t(z^\prime)\big)\frac{\diff t}{\diff z^\prime} \diff z^\prime,
            \label{eq:ratedensity}
        \end{equation}

        with the universe's age $t(z)$ at redshift $z$ given by,

        \begin{equation}
            t(z) = \frac{1}{H_0}\int_{z}^{\infty}\frac{dz'}{(1+z')\sqrt{(1+z')^{3}\Omega_{M}+\Omega_{\Lambda}}},
        \end{equation}

        We also fit data with alternative cosmic rates of SGRBs, assuming that SGRBs follow SFR of \citet{hopkins_normalization_2006} or the convolution of SFR with a long merger delay time of log-normal form with parameters $[\mu,\sigma] = [\log(4.0Gyrs),0.3]$ as suggested by \citet{nakar_short-hard_2007}. For all redshift scenarios, we find that the resulting best-fit parameters are qualitatively the same, although some parameters may exhibit quantitative differences at $>1\sigma$ significance level.

    \subsection{Model Fitting}
    \label{sec:MF}

        Now, with a statistical model at hand for the observed rate of short GRBs (i.e., Equation \ref{eq:obsrate}), we proceed to obtain the best fit parameters of the model to BATSE short GRBs data. In principle, any model fitting must take into account the observational uncertainties and any prior knowledge from independent sources, which can be achieved via Bayesian multilevel methodology \citep[e.g.,][]{hobson_bayesian_2010}. This is done by first constructing the likelihood function, taking into account the uncertainties in observational data \citep[e.g.,][]{loredo_accounting_2004}:  Under the assumption of symmetric Gaussian uncertainties, as it is the case with BATSE catalog data, the full Poisson likelihood of data $\boldsymbol{\mathcal{O}}$ given the parameters $\{\vec\mu,\boldsymbol\Sigma\}$ of the SGRB world model in Eqns. \ref{eq:cosmicrate} \& \ref{eq:obsrate} can be written as,

         \begin{eqnarray}
            \label{eq:likelihood}
            && ~{\mathcal L}\big(\boldsymbol{\mathcal{O}} ~|~ \vec\mu,{\boldsymbol\Sigma} \big) = \mathcal{A}^N \nonumber \\
            && ~\times \exp\bigg(-\mathcal{A}\int_{\vec O space}\mathcal{R}_{obs}\big(\vec O ~|~ \vec\mu,{\boldsymbol\Sigma},\eta \big)~\diff\vec O\bigg) \nonumber \\
            && ~\times \prod_{i=1}^{565} \int_{\vec O space} \mathcal{R}_{cosmic}\big(\vec O ~|~ \vec\mu,{\boldsymbol \Sigma}\big) \nonumber \\
            && ~\times \mathcal{L}_i\big(\vec O ~|~ \hat O(\vec \mu_{i},\boldsymbol\Sigma_i)\big)~\diff\vec O,
        \end{eqnarray}

        in which $\mathcal{A}$ is a factor that properly normalizes the cosmic rate of SGRBs ($\mathcal{R}_{cosmic}$) and the vector $\hat O(\vec \mu_i,\boldsymbol\Sigma_i)$, standing for the $i$th SGRB Observation in BATSE catalog, has the likelihood $\mathcal{L}_i$ of having the true parameters $\vec O\equiv\big[\liso,\eiso,\epkz,\durz\big]$ in the rest-frame 4-dimensional observation space ($\vec O space$) that can be described as a Gaussian probability density function with parameters $\{\vec\mu_i,\boldsymbol\Sigma_i\}$ obtained from BATSE catalog such that,

        \begin{equation}
            \mathcal{L}_i\big(\vec O \big)~\sim~\mathcal{N}(\vec O|\vec\mu_i,\boldsymbol\Sigma_i),
        \end{equation}

        In this sense, the term $\mathcal{R}_{cosmic}$ in Equation \ref{eq:likelihood} acts as a Bayesian prior for $\mathcal{L}_i$. In the absence of this prior knowledge however, as it is the case with BATSE short GRBs, the Empirical Bayes approach provides an alternative solution, in which an ad hoc estimate of the model parameters $\{\vec\mu,{\boldsymbol\Sigma}\}$ based on the observed data -- excluding uncertainties -- serves as the prior for the same data -- including uncertainties -- at the second level of analysis \citep[e.g.,][]{hobson_bayesian_2010}. The joint posterior of the unknown parameters of the model can be then written as,

        \begin{equation}
            \label{eq:posterior}
            \mathcal{P}\big(\vec\mu,\boldsymbol\Sigma|\boldsymbol{\mathcal O}\big) = \mathcal{P}\big(\vec\mu,\boldsymbol{\Sigma}\big) \times {\mathcal L}\big(\boldsymbol{\mathcal{O}} | \vec\mu,{\boldsymbol\Sigma} \big),
        \end{equation}

        As for the choice of {\it hyperprior}, $\mathcal{P}\big(\vec\mu,\boldsymbol{\Sigma}\big)$, we adopt the noninformative uniform prior for the mean vector $\vec\mu$. A variety of noninformative or weakly-informative priors for the covariance matrix $\boldsymbol\Sigma$ have been already proposed and considered in the literature,  with Inverse Wishart familty of distributions among the most popular choices (c.f., \citet{john_barnard_modeling_2000} and references therein). Here, to avoid problems and complications associated with Inverse Wishart priors, we adopt a separation strategy \citep[e.g.,][]{browne_mcmc_2006} by decomposing the covariance matrix $\boldsymbol\Sigma$ into a correlation matrix and a set of standard deviations. We then use uniform priors on the log-transformation of all standard deviations, also on the Fisher-transformation \citep{fisher_frequency_1915} of all correlation coefficients.

        Due to the complex truncation imposed on SGRB data and the world model by BATSE detection threshold (c.f., Appendix \ref{App:BDT}), maximization of the likelihood function of Eqn. \ref{eq:likelihood} is analytically intractable. Calculation of the normalization factor $\mathcal A$ by itself requires a multivariate integral over the 4-dimensional space of SGRB variables at any given redshift. In addition, due to lack of redshift ($z$) information for BATSE SGRBs, the probability for observation of each SGRB given the model parameters must be marginalized over all possible redshifts. These numerical integrations make sampling from the posterior distribution of Eqn. \ref{eq:posterior} an extremely difficult task. Given the potential presence of unknown systematic biases in BATSE catalog data as discussed in Appendix \ref{App:SB}, \citep[also by][Appendix C]{koshut_systematic_1996, hakkila_fluence_2000, hakkila_how_2003, shahmoradi_multivariate_2013} and the high level of uncertainty in the redshift distribution of short GRBs, we take a bold but reasonable and highly simplifying step and drop data uncertainties in the calculation of the likelihood function (Eqn. \ref{eq:likelihood}) in order to bring the problem into the realm of current computational technologies. The joint posterior distribution of the model parameters is then obtained by iterative sampling using a variant of Markov Chain Monte Carlo (MCMC) techniques known as Adaptive Metropolis-Hastings \citep[e.g.,][]{haario_adaptive_2001}. To further the efficiency of MCMC sampling, we implement all algorithms in Fortran \citep{backus_history_1978, metcalf_modern_2011} and approximate the numerical integration in the definition of the luminosity distance of Eqn. \ref{eq:lumdis} by the analytical expressions of \citet{wickramasinghe_analytical_2010}. This integration is encountered on the order of billion times during MCMC sampling from the posterior distribution (c.f., Appendix C in \citet{shahmoradi_multivariate_2013} for further details of the MCMC sampling method).\footnote{The entire simulation codes and algorithms will be available for download at \\ https://bitbucket.org/AmirShahmoradi/grbworldmodel.}

    \section{Results \& Goodness-of-Fit Tests}
    \label{sec:results}

\begin{table} 
    \begin{center}
    \caption{Mean best-fit parameters of SGRB World Model, compared to LGRB world model of \citet{shahmoradi_multivariate_2013}. \label{tab:BFP}}
    \begin{tabular}{c c c}
    \hline
    \hline
    Parameter     & SGRBs World Model  & LGRBs World Model     \\
    \hline
    \multicolumn{3}{c}{Redshift Parameters (Equation \ref{eq:zeta})} \\
    \hline
    $z_0$                   & $0.993$         & $0.993$        \\
    $z_1$                   & $3.8$           & $3.8$          \\
    $\gamma_0$              & $3.3$           & $3.3$          \\
    $\gamma_1$              & $0.0549$        & $0.0549$       \\
    $\gamma_2$              & $-4.46$         & $-4.46$        \\
    \hline
    \multicolumn{3}{c}{Log-normal Merger Delay (Equation \ref{eq:lognormal})} \\
    \hline
    $\mu_{delay}$           & 0.1             & --             \\
    $\sigma_{delay}$        & 1.12            & --             \\
    \hline
    \multicolumn{3}{c}{Location Parameters} \\
    \hline
    $\log(\liso)$           & $51.88\pm0.16$  & $51.54\pm0.18$ \\
    $\log(\eiso)$           & $50.93\pm0.19$  & $51.98\pm0.18$ \\
    $\log(\epkz)$           & $2.98 \pm0.05$  & $2.48\pm0.05$  \\
    $\log(\durz)$           & $-0.74\pm0.08$  & $1.12\pm0.03$  \\
    \hline
    \multicolumn{3}{c}{Scale Parameters} \\
    \hline
    $\log(\sigma_{\liso})$  & $-0.36\pm0.06$  & $-0.25\pm0.06$ \\
    $\log(\sigma_{\eiso})$  & $-0.10\pm0.04$  & $-0.08\pm0.03$ \\
    $\log(\sigma_{\epkz})$  & $-0.39\pm0.02$  & $-0.44\pm0.02$ \\
    $\log(\sigma_{\durz})$  & $-0.24\pm0.02$  & $-0.37\pm0.01$ \\
    \hline
    \multicolumn{3}{c}{Correlation Coefficients} \\
    \hline
    $\rho_{\liso-\eiso}$    & $0.91\pm0.03$  & $0.94\pm0.01$   \\
    $\rho_{\liso-\epkz}$    & $0.51\pm0.10$  & $0.45\pm0.07$   \\
    $\rho_{\liso-\durz}$    & $0.50\pm0.09$  & $0.48\pm0.09$   \\
    $\rho_{\eiso-\epkz}$    & $0.60\pm0.06$  & $0.58\pm0.04$   \\
    $\rho_{\eiso-\durz}$    & $0.63\pm0.05$  & $0.60\pm0.05$   \\
    $\rho_{\epkz-\durz}$    & $0.12\pm0.06$  & $0.31\pm0.04$   \\
    \hline
    \multicolumn{3}{c}{BATSE Detection Efficiency (Eqn. \ref{eq:eta})} \\
    \hline
    $\mu_{thresh}$          & $-0.25\pm0.03$ & $-0.45\pm0.02$  \\
    $\log(\sigma_{thresh})$ & $-0.86\pm0.05$ & $-0.90\pm0.05$  \\
    \hline
    \hline
    \end{tabular}
    \end{center}
    {Note.--- The full Markov Chain sampling of the above parameters from the 16-dimensional parameter space of the likelihood function are available for download at \url{https://sites.google.com/site/amshportal/research/aca/in-the-news/lgrb-world-model} for the LGRB world model and at \url{https://bitbucket.org/AmirShahmoradi/grbworldmodel} for SGRBs world model.}
\end{table}

        \begin{figure*}
            \centering
            \begin{tabular}{cc}
            \includegraphics[scale=0.31]{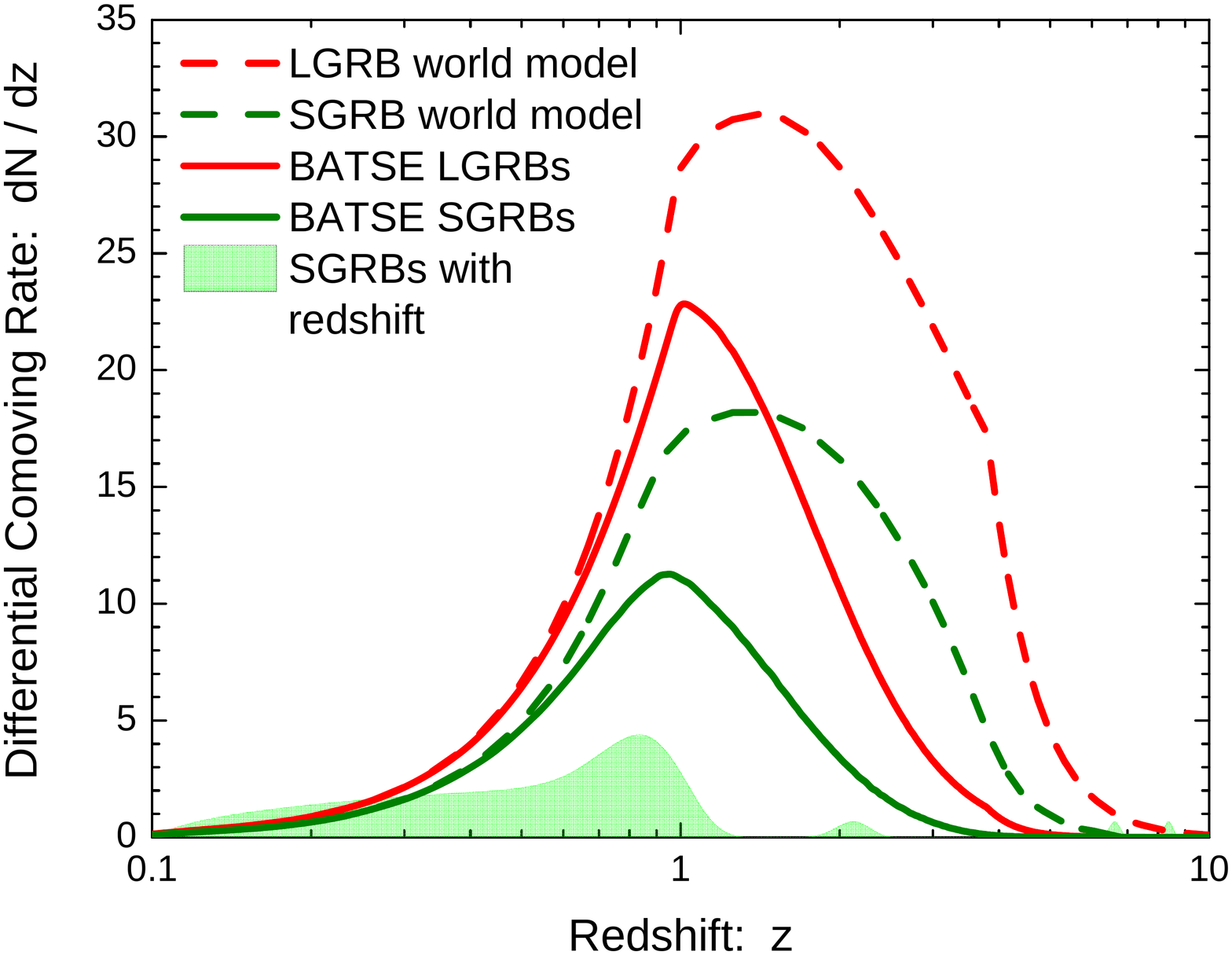} & \includegraphics[scale=0.31]{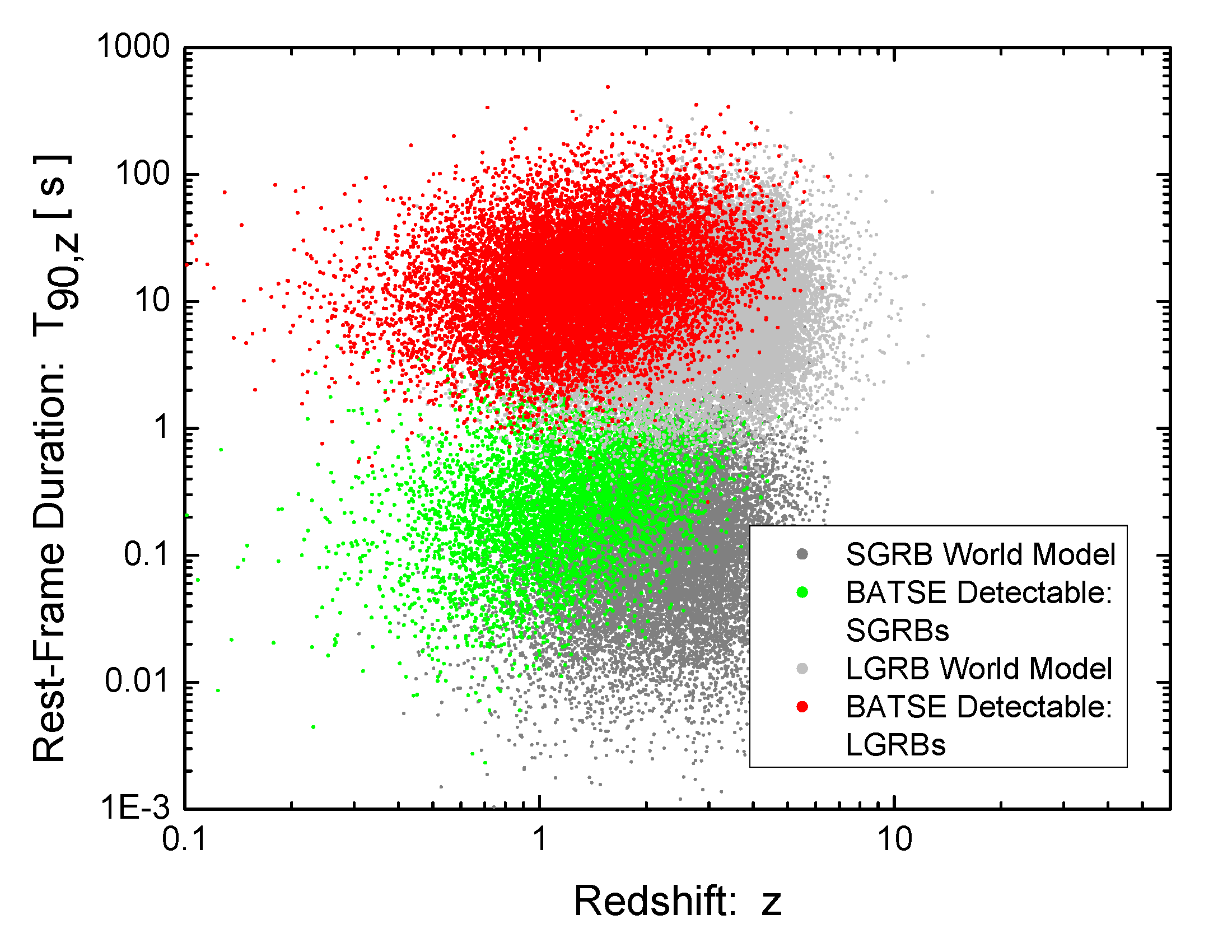} \\
            \end{tabular}
            \caption{{\bf Left:} The predicted redshift distribution of BATSE \textcolor{OliveGreen}{SGRBs} and \textcolor{red}{LGRBs}. The green-colored histogram represents the sample of SGRBs with known redshift taken from \citet{tsutsui_possible_2013}. Assuming an average merger delay time of $\sim0.1\operatorname{Gyrs}$ \citep[e.g.,][]{belczynski_origin_2010}, about $60\%$ of BATSE SGRBs originate from redshifts $z\gtrsim1.0$, highlighting the potential existence of strong selection effects in redshift measurements of SGRBs. {\bf Right:} The intrinsic duration distribution of LGRBs and SGRBs vs. redshift.  Corroborating the independent findings of \citet{littlejohns_investigating_2014} based on Swift data, the presented GRB world model predicts that the shortest duration LGRBs \& SGRBs at high redshifts generally have the lowest likelihood of detection by gamma-ray instruments such as BATSE LAD, Fermi GBM \& Swift BAT detectors. \label{fig:redshift}}
        \end{figure*}

        The resulting {\it mean} and $1\sigma$ standard deviations of the model parameters are tabulated in Table \ref{tab:BFP}. It is notable that the parameters of the model exhibit strong covariance as illustrated in the upper triangle of the correlation matrix of Table \ref{tab:cormat}. For comparison, the lower triangle of the table represents the correlation matrix of the same parameters for the LGRB world model based on $1366$ BATSE LGRBs (Table $3$ in \citet{shahmoradi_multivariate_2013}). All location parameters of the model appear to strongly correlate with each other, so do the scale parameters. The correlations among the four variables of the world models also weaken with increasing the location parameters. Therefore, given BATSE data, a higher cosmic rates of LGRBs and SGRBs at high redshifts generally implies weaker correlations among the prompt gamma-ray emission variables, in particular, the brightness-hardness type of relations. A comparison of the predicted redshift distribution of BATSE SGRBs with LGRBs is illustrated in the {\it left plot} of Figure \ref{fig:redshift}.

\begin{figure*}
            \centering
            \includegraphics[scale=0.097]{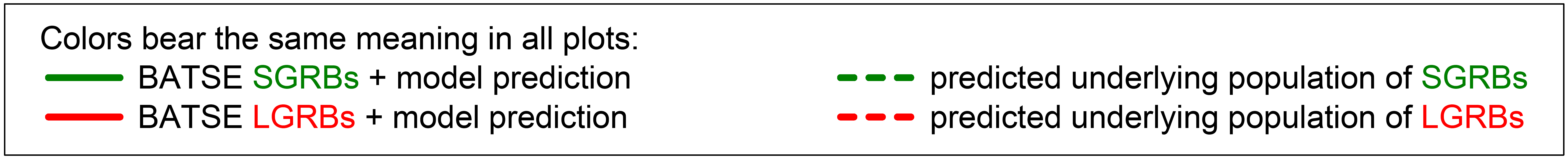}
            \begin{tabular}{cc}
                \includegraphics[scale=0.31]{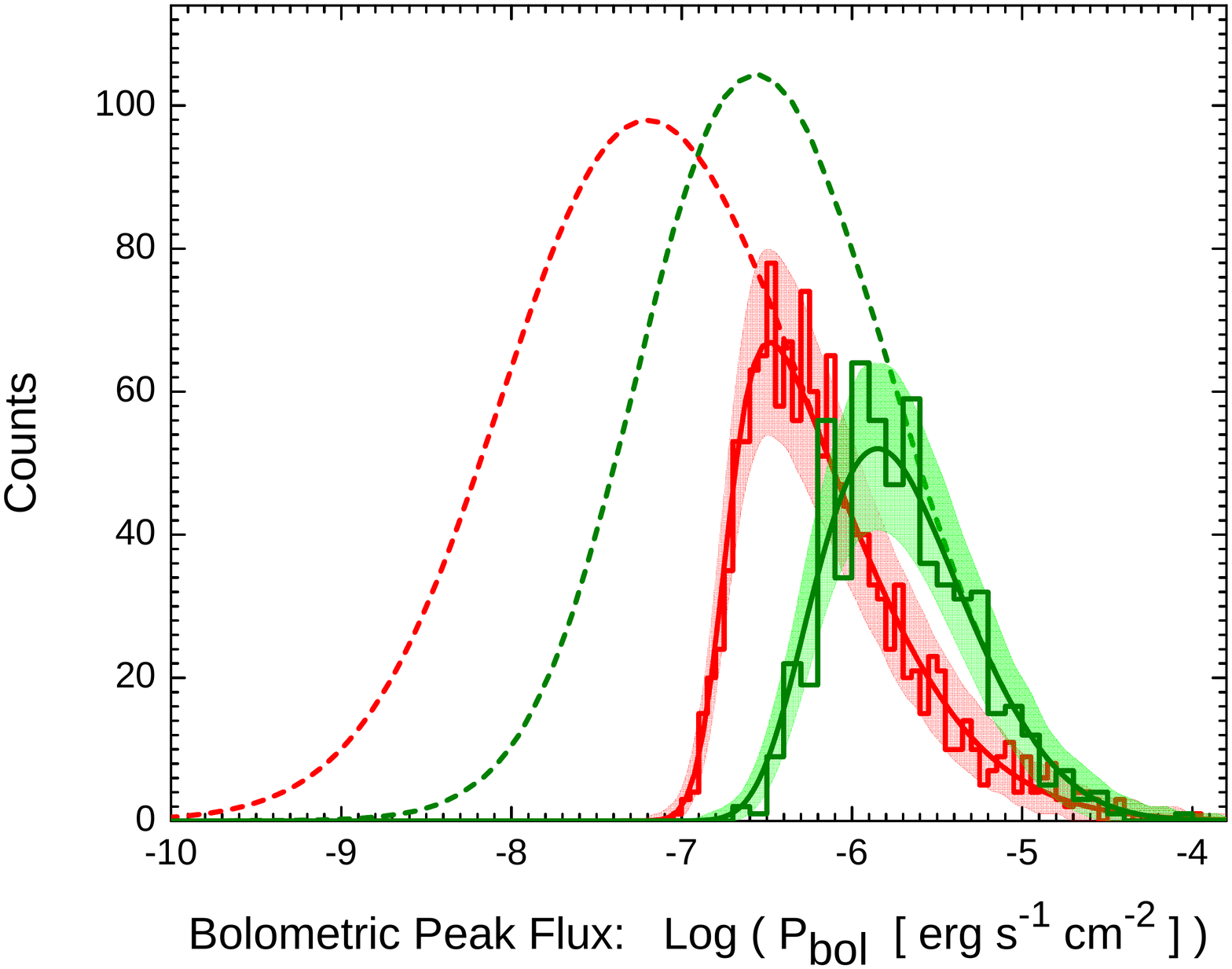} & \includegraphics[scale=0.31]{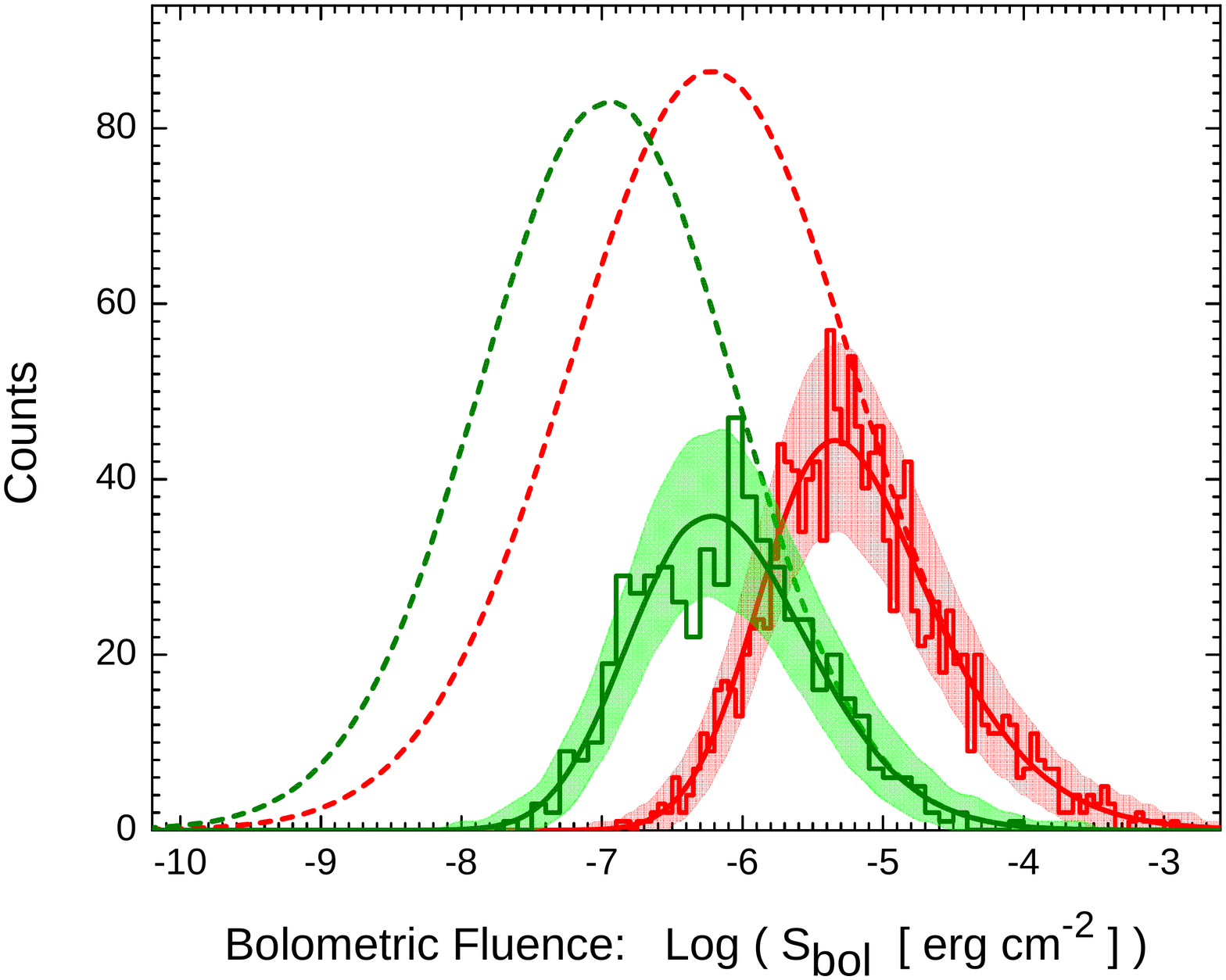} \\
                \includegraphics[scale=0.31]{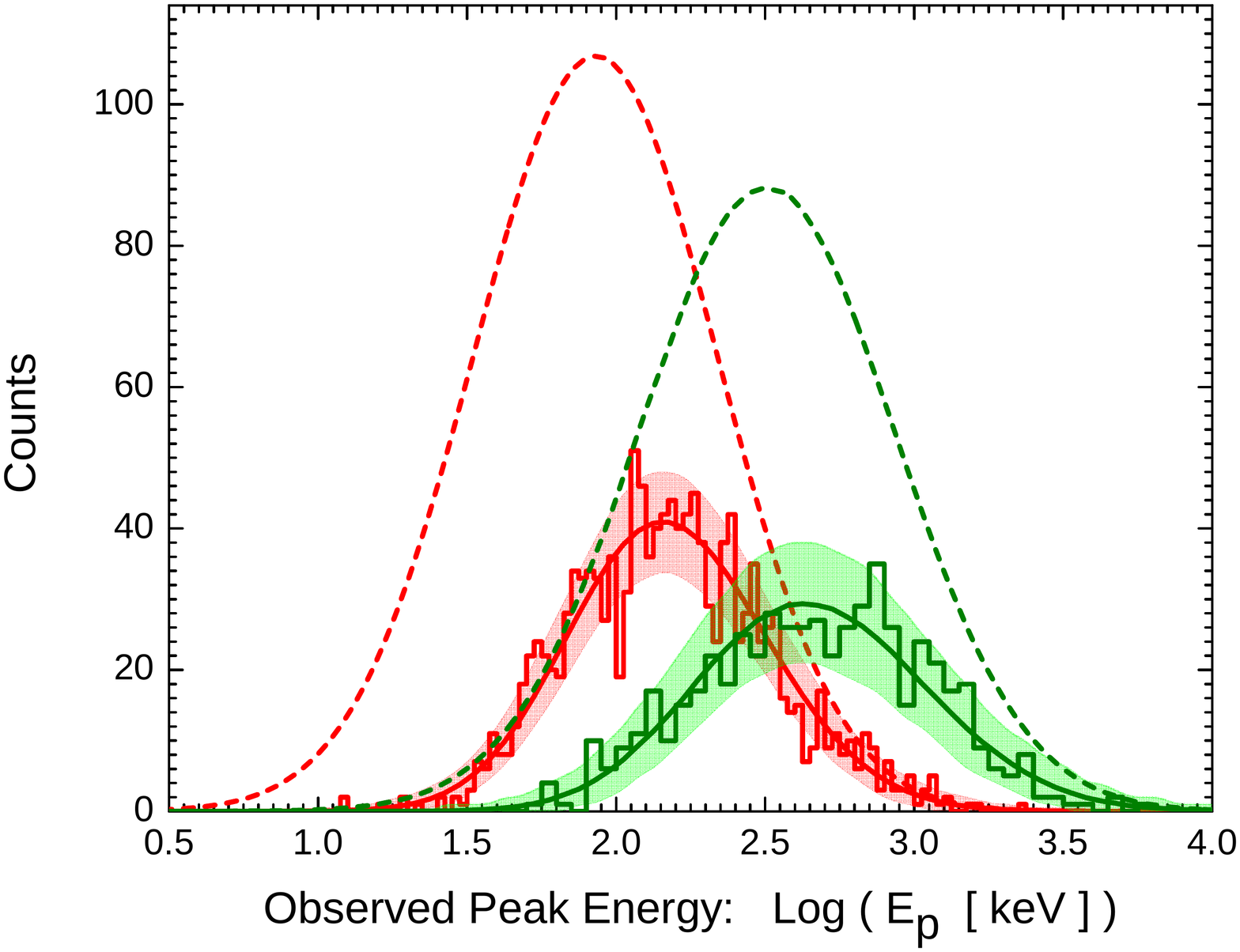} & \includegraphics[scale=0.31]{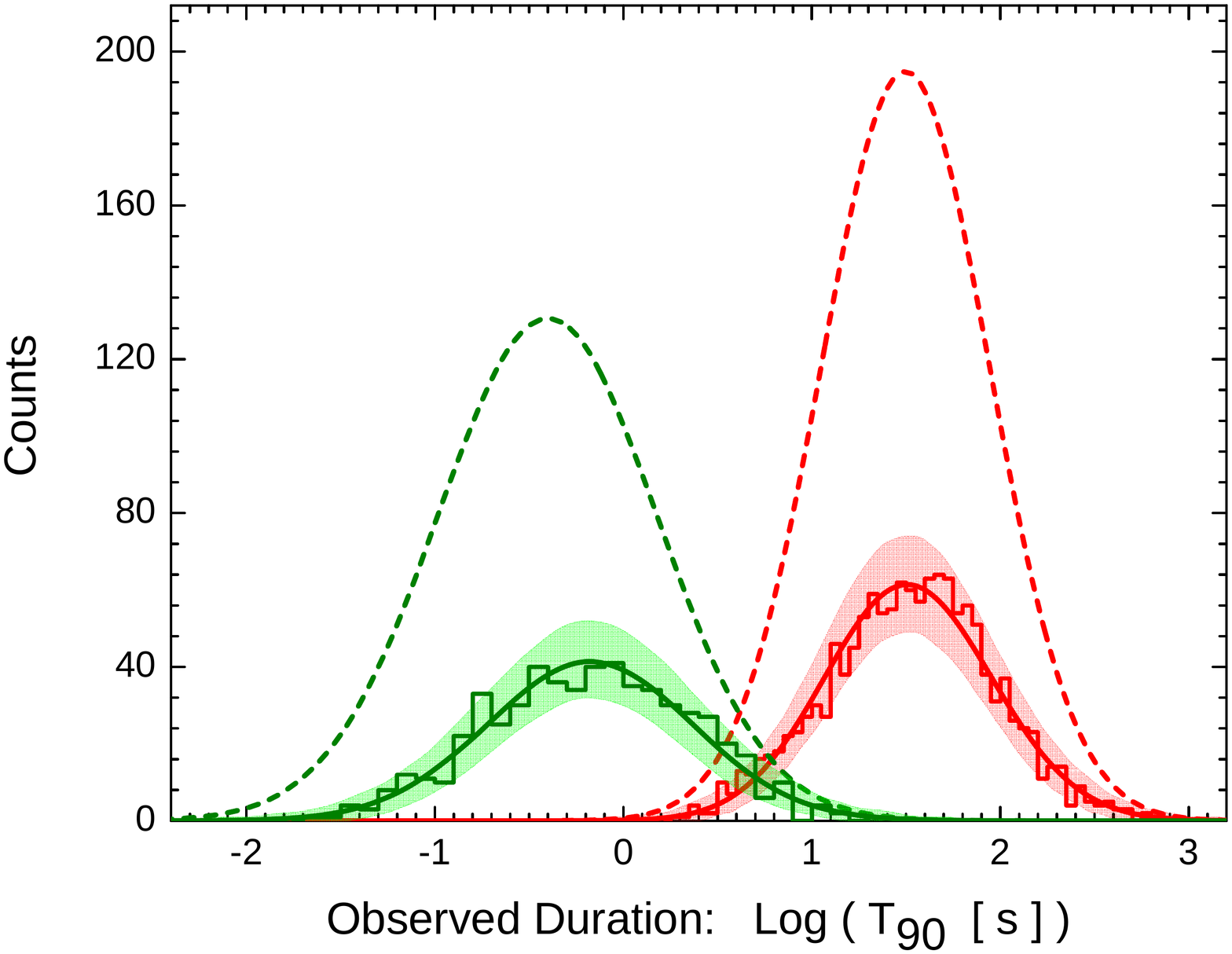} \\
            \end{tabular}
            \caption{{\bf Univariate predictions of the {\it multivariate} best-fit SGRB \& LGRB world models for BATSE catalog GRBs.} The \textcolor{red}{red} \& \textcolor{OliveGreen}{green} colors represent data and model for SGRBs \& LGRBs respectively. Each plot illustrates the projection of the multivariate GRB world model ({\it solid curves}) on the distribution of one of the four prompt gamma-ray emission variables (\textcolor{red}{red} \& \textcolor{OliveGreen}{green} {\it histograms}): peak flux $\pbol$, fluence $\sbol$, observed peak energy $\epk$ and the observed duration $\dur$, subject to BATSE detection threshold. The color-shaded areas represent the $90\%$ prediction intervals of the model for BATSE data. The {\it dashed lines} represent the predicted underlying populations of LGRBs \& SGRBs respectively. Fitting results for LGRBs are taken from \citet{shahmoradi_multivariate_2013}. For clarity, the bin size for SGRB histograms and data is twice as large as the bin size for LGRBs histograms and data in all plots. \label{fig:OFmarginals}}
        \end{figure*}

        The lack of redshift information combined with the relatively small sample size of BATSE GRBs strongly limit the variety of Goodness-of-Fit (GoF) tests that can be employed to assess the performance of the model. In addition to small sample size, marginialization of the likelihood function (Eqn. \ref{eq:likelihood}) over redshift variable, and the systematic biases in data close to detection threshold (c.f., Appendix \ref{App:SB}), have major contributions to the uncertainties of the best-fit parameters of the model (Table \ref{tab:BFP}) and increase the risk of overfitting. Following \citet{shahmoradi_multivariate_2013}, the fitting results can be first visually inspected by superposing the univariate distribution of four variables of the SGRB world model on BATSE data, as illustrated in Figure \ref{fig:OFmarginals}. For comparison, the results of LGRB world model fit to BATSE LGRB data from \citet{shahmoradi_multivariate_2013} are also shown by the red-color lines and histograms.

\begin{figure*}
            \centering
            \begin{tabular}{cc}
                \includegraphics[scale=0.31]{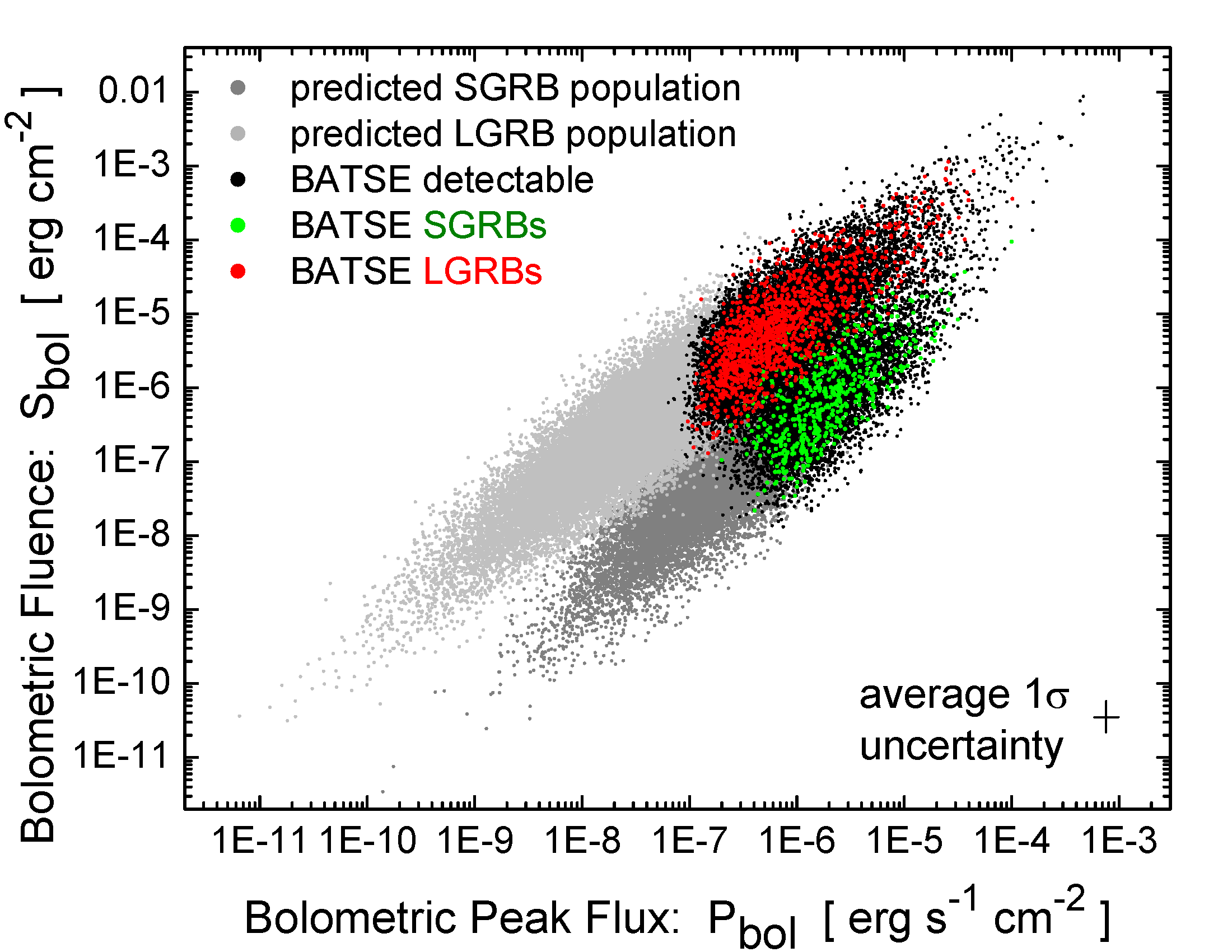} & \includegraphics[scale=0.31]{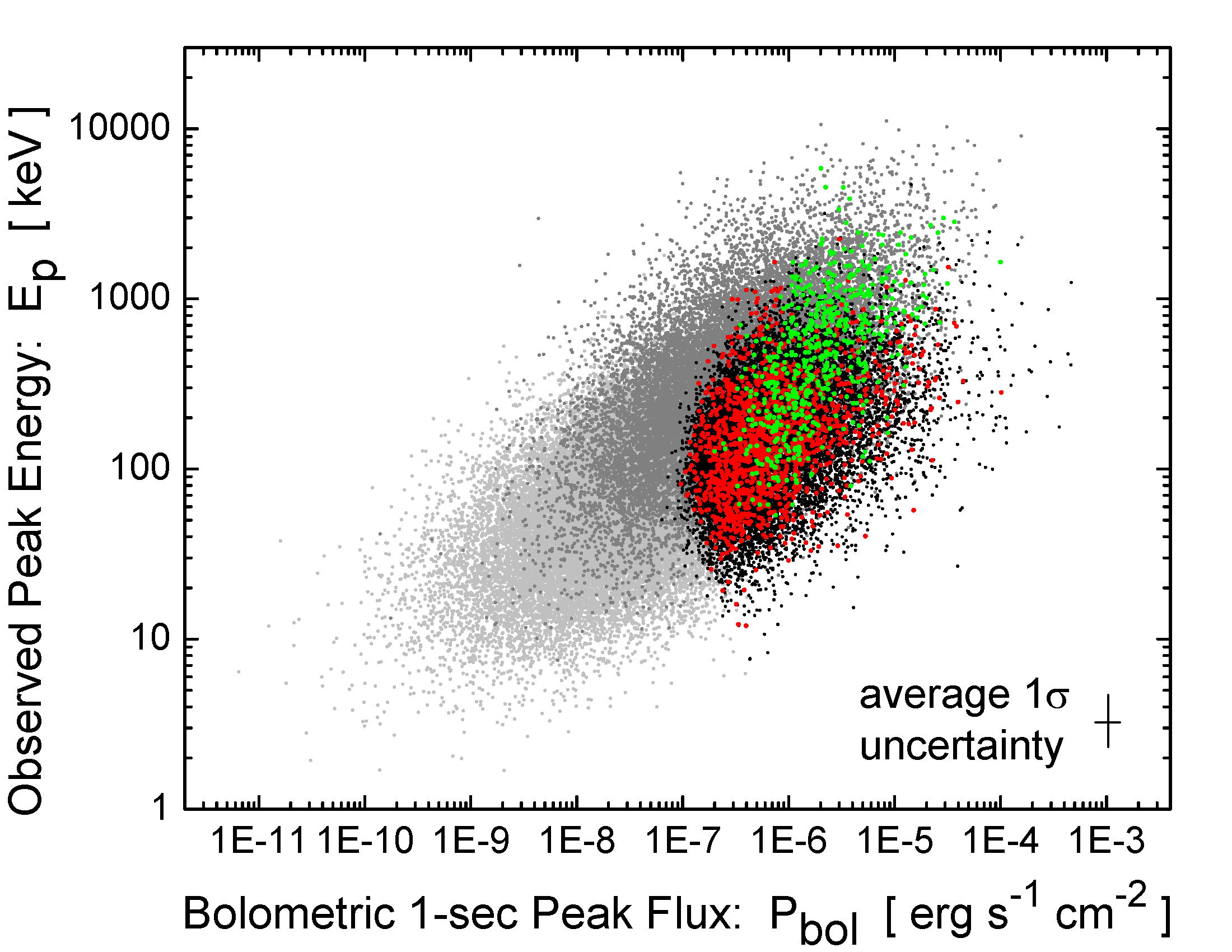} \\
                \includegraphics[scale=0.31]{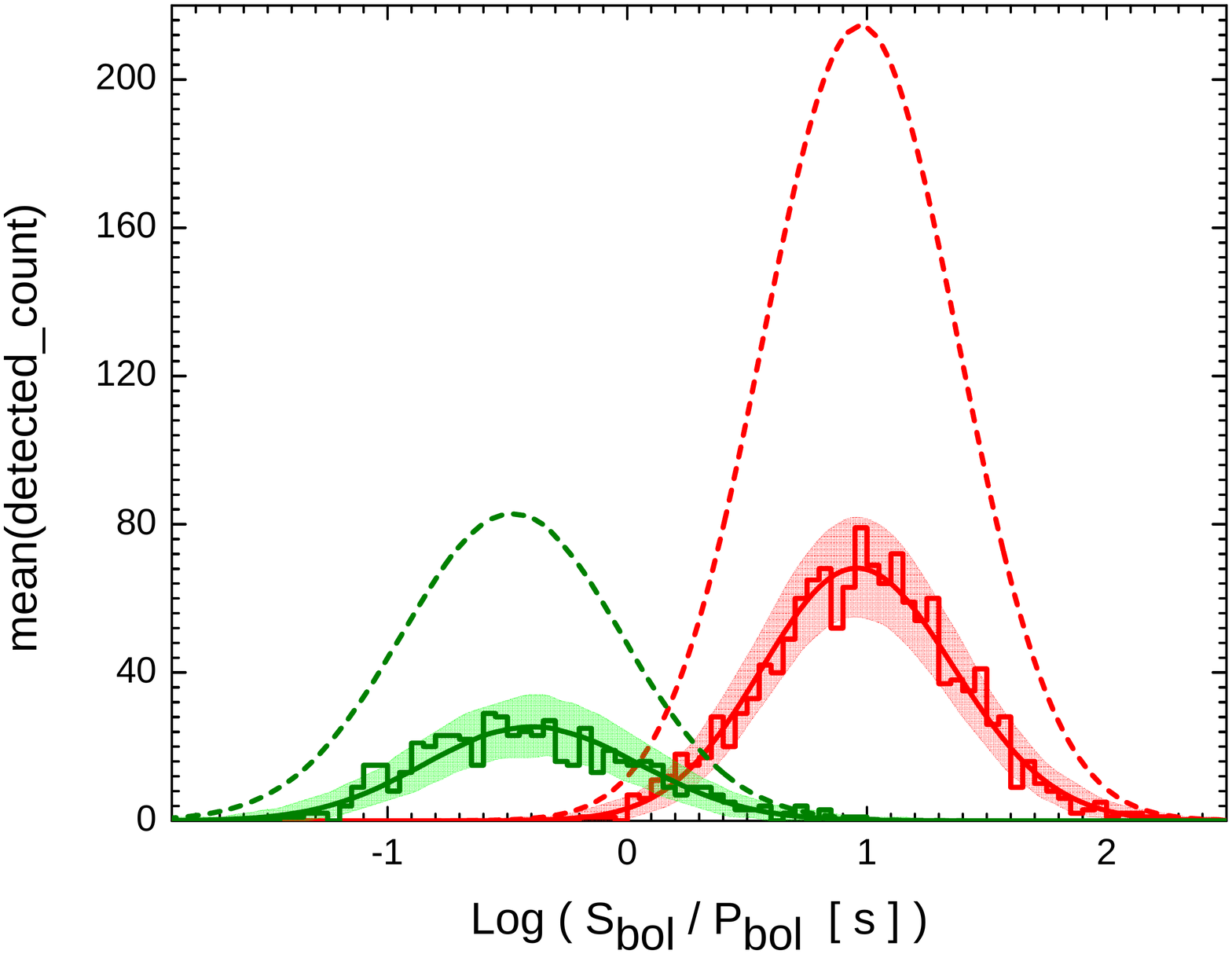}      & \includegraphics[scale=0.31]{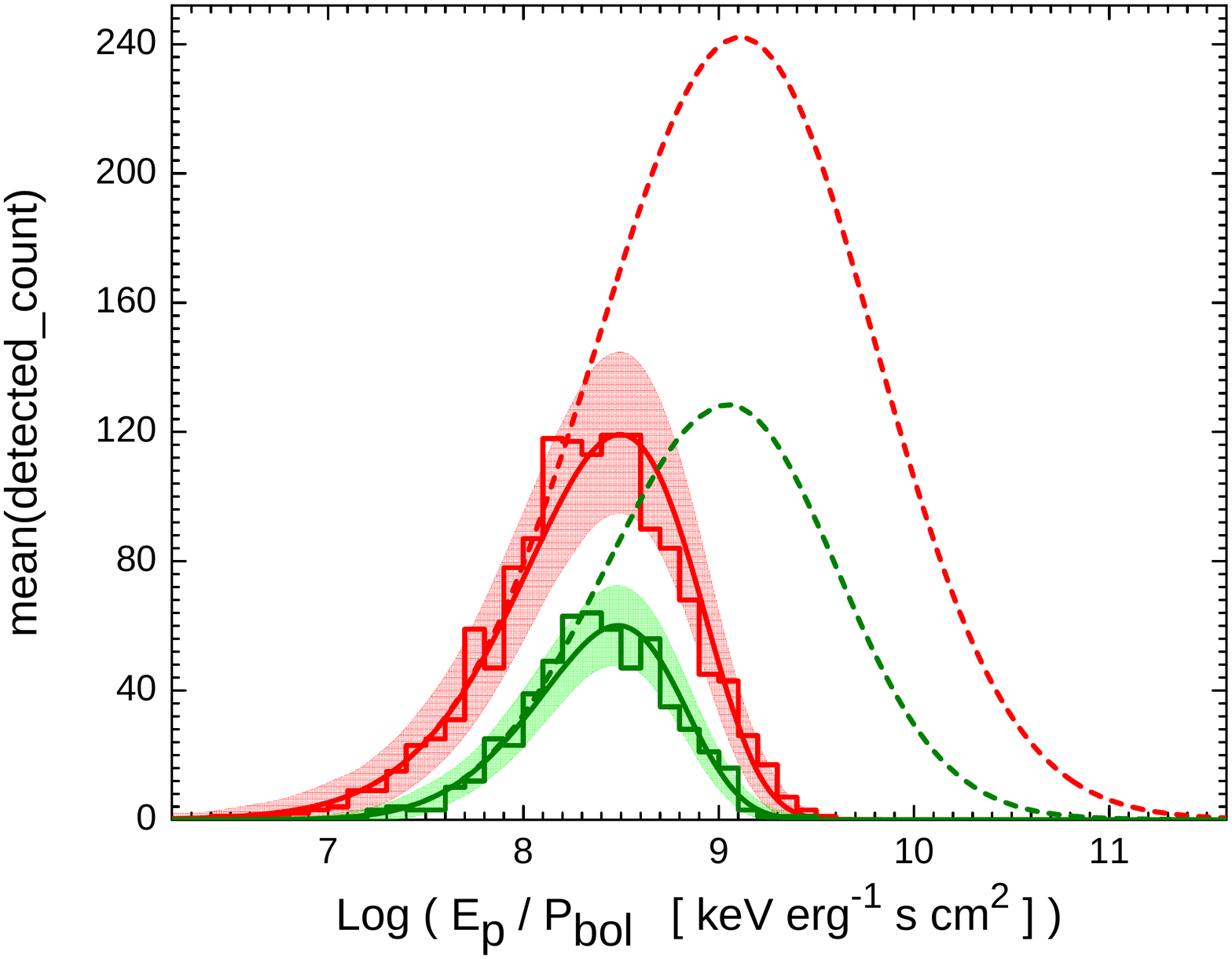}       \\
                \includegraphics[scale=0.31]{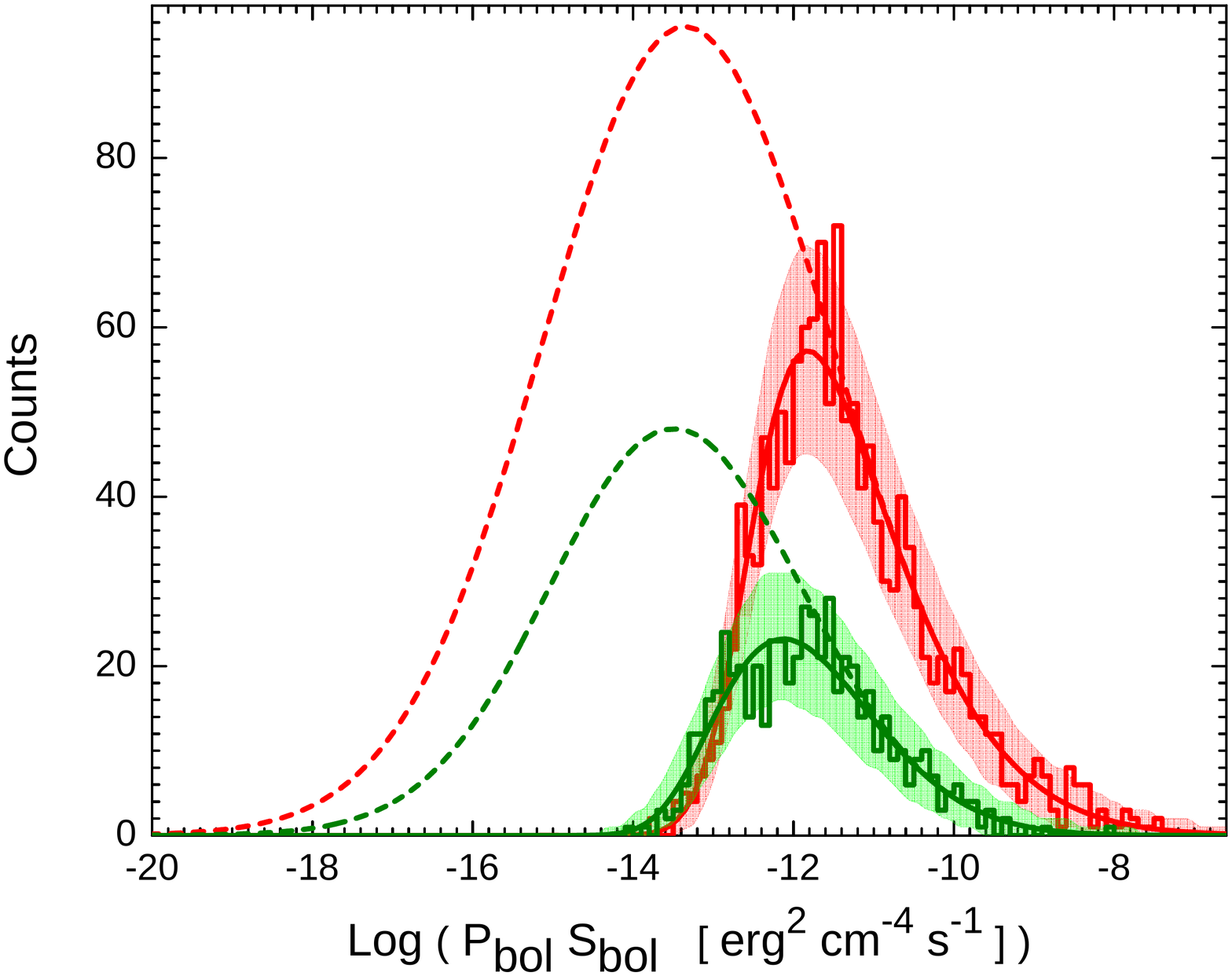}      & \includegraphics[scale=0.31]{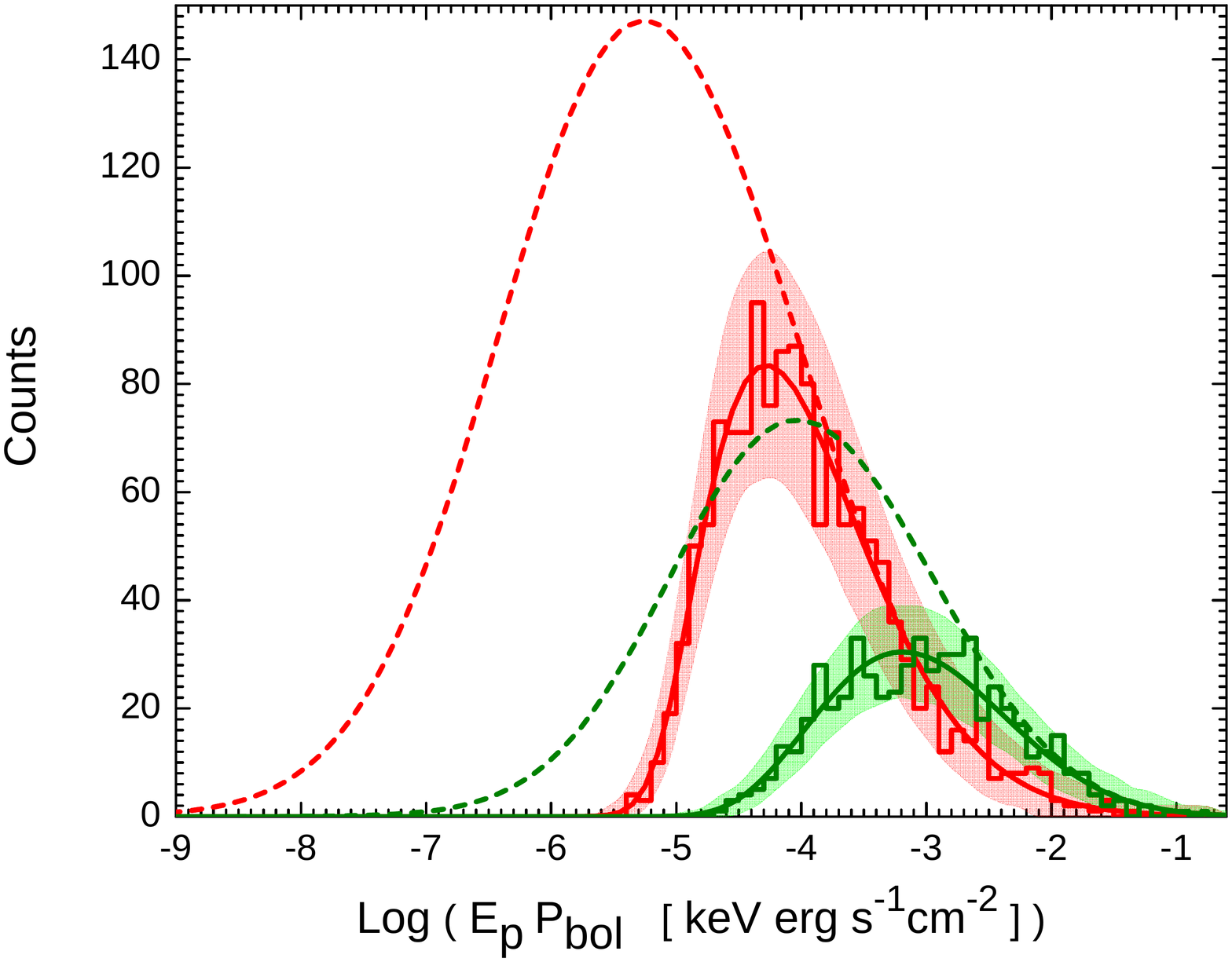}
            \end{tabular}
            \caption{{\bf Bivariate predictions of the {\it multivariate} best-fit SGRB \& LGRB world models for BATSE catalog GRBs, subject to BATSE detection threshold.} Colors, line styles and color-shaded areas bear the same meaning as in Figure \ref{fig:OFmarginals}. Evidence for potential systematic bias in peak flux measurements of BATSE GRBs close to detection threshold (Appendix \ref{App:SB}) can be seen in the center-right and bottom-left plots of the figure. Fitting results for LGRBs are taken from \citet{shahmoradi_multivariate_2013}. \label{fig:OFbivariates1}}
        \end{figure*}

\begin{figure*}
            \centering
            \begin{tabular}{cc}
                \includegraphics[scale=0.31]{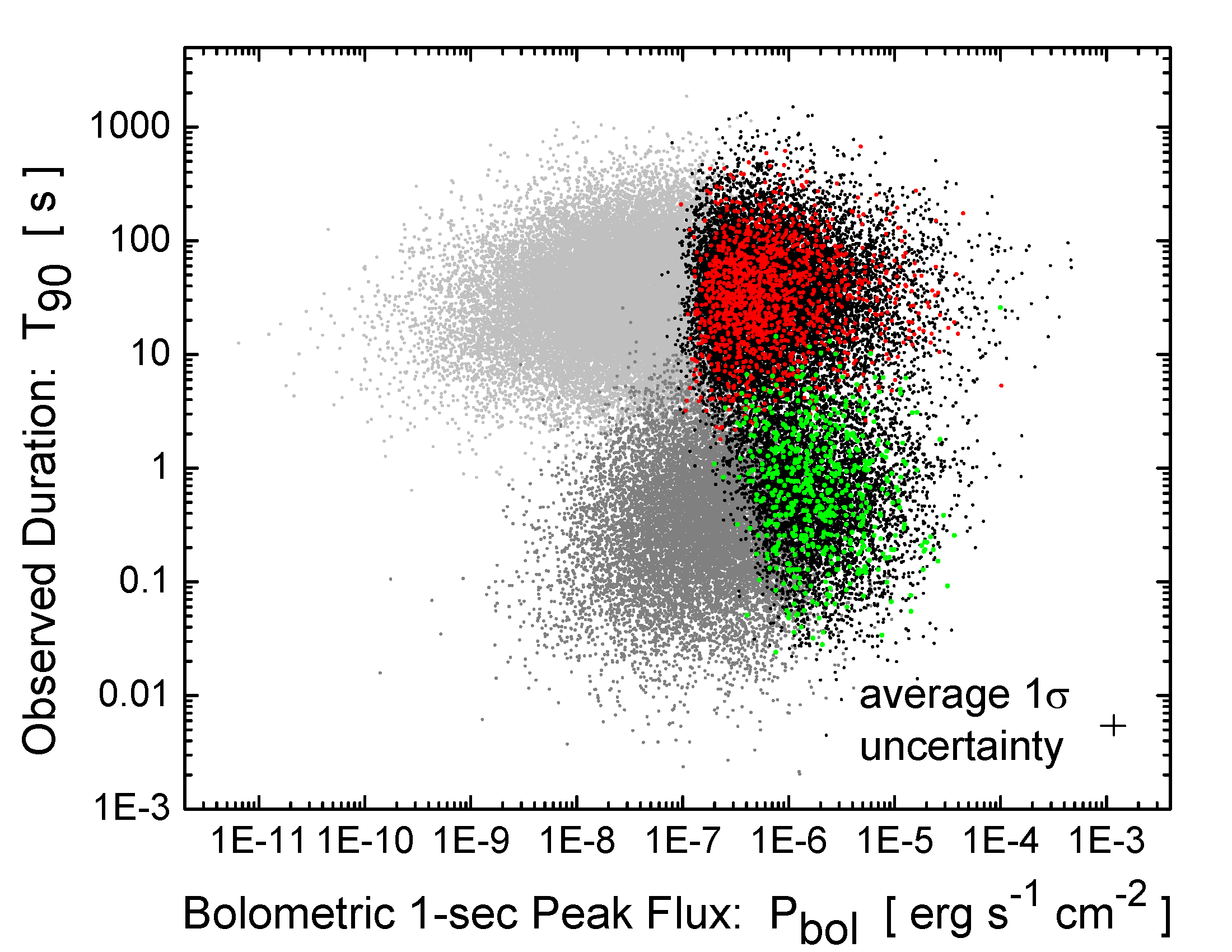} & \includegraphics[scale=0.31]{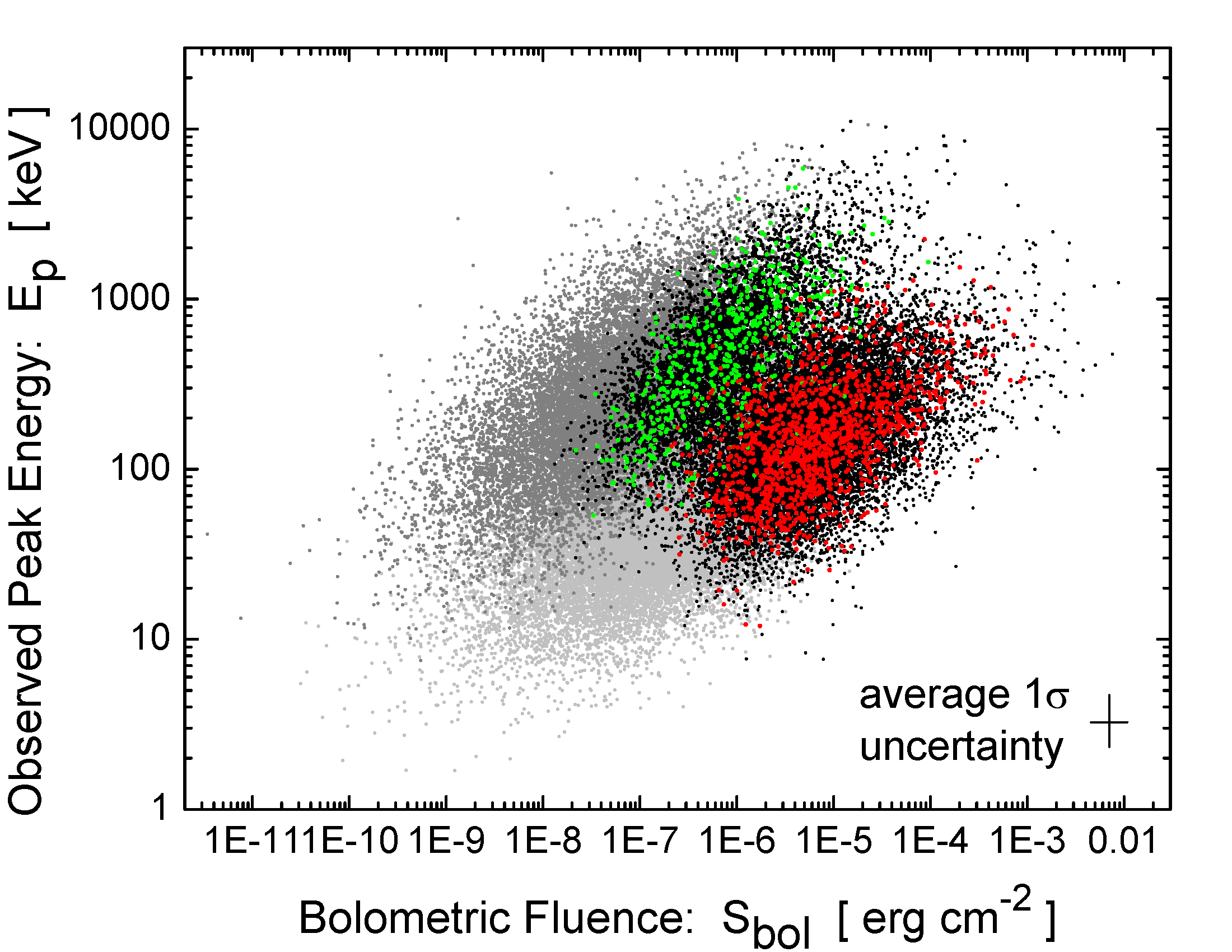} \\
                \includegraphics[scale=0.31]{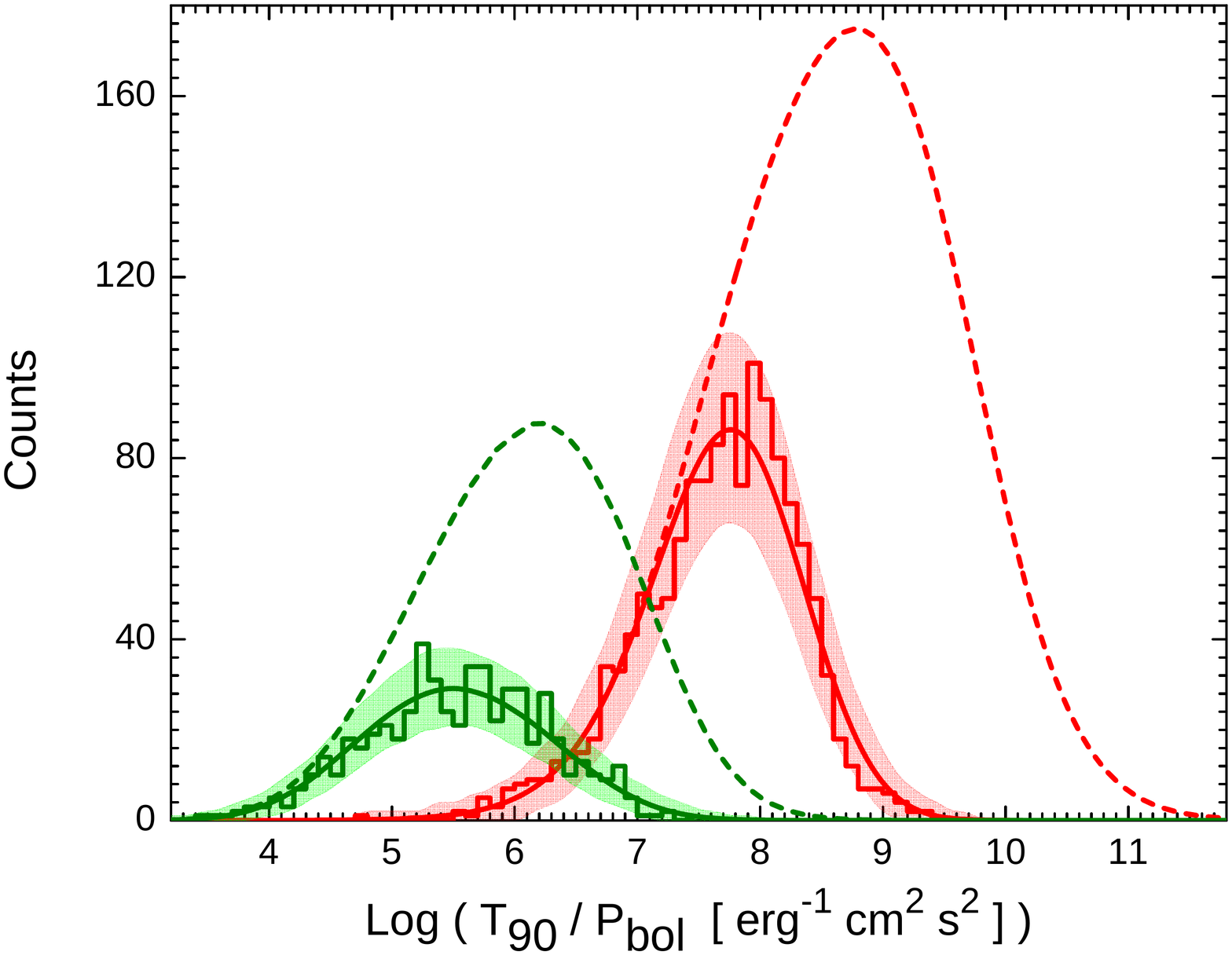}      & \includegraphics[scale=0.31]{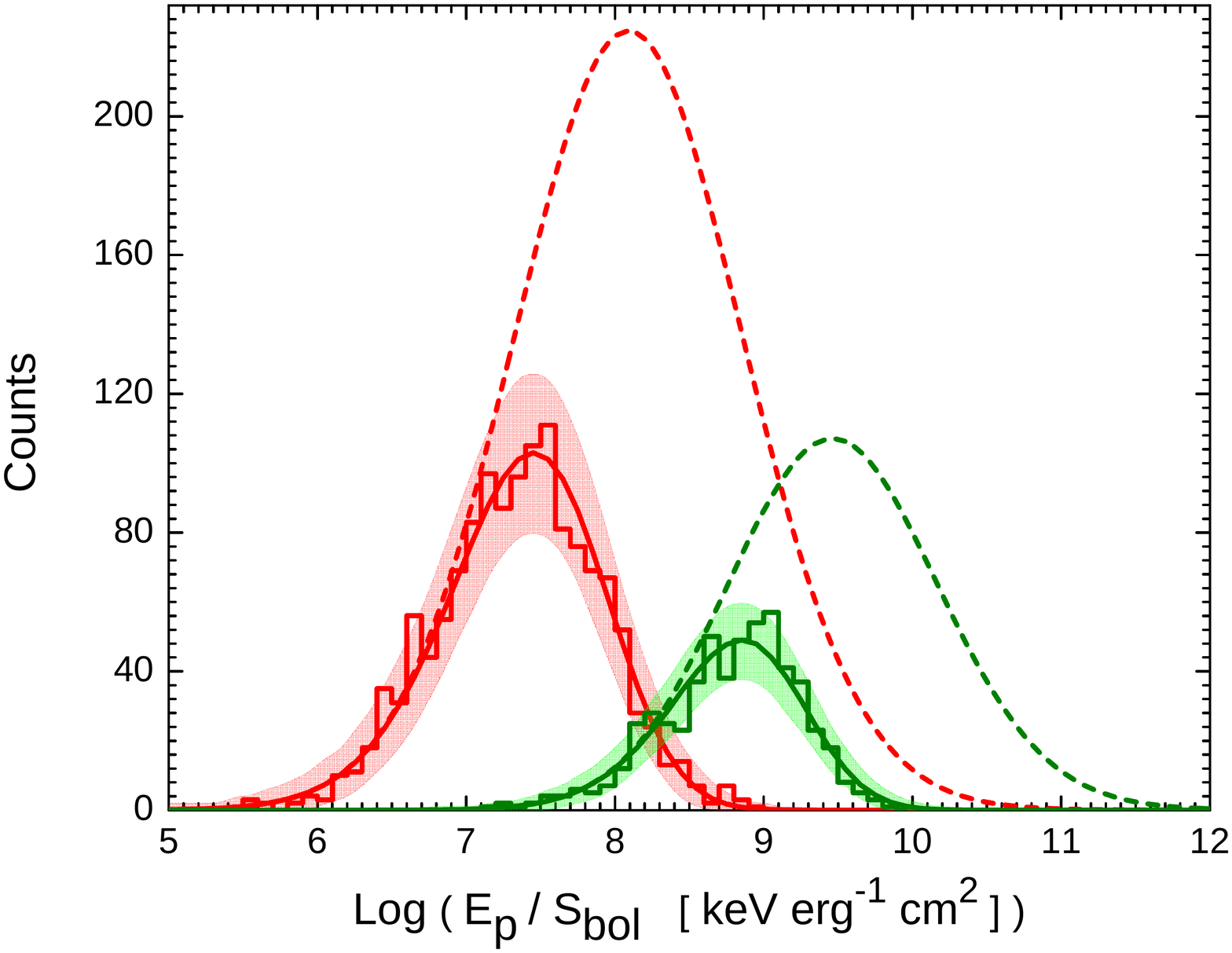}      \\
                \includegraphics[scale=0.31]{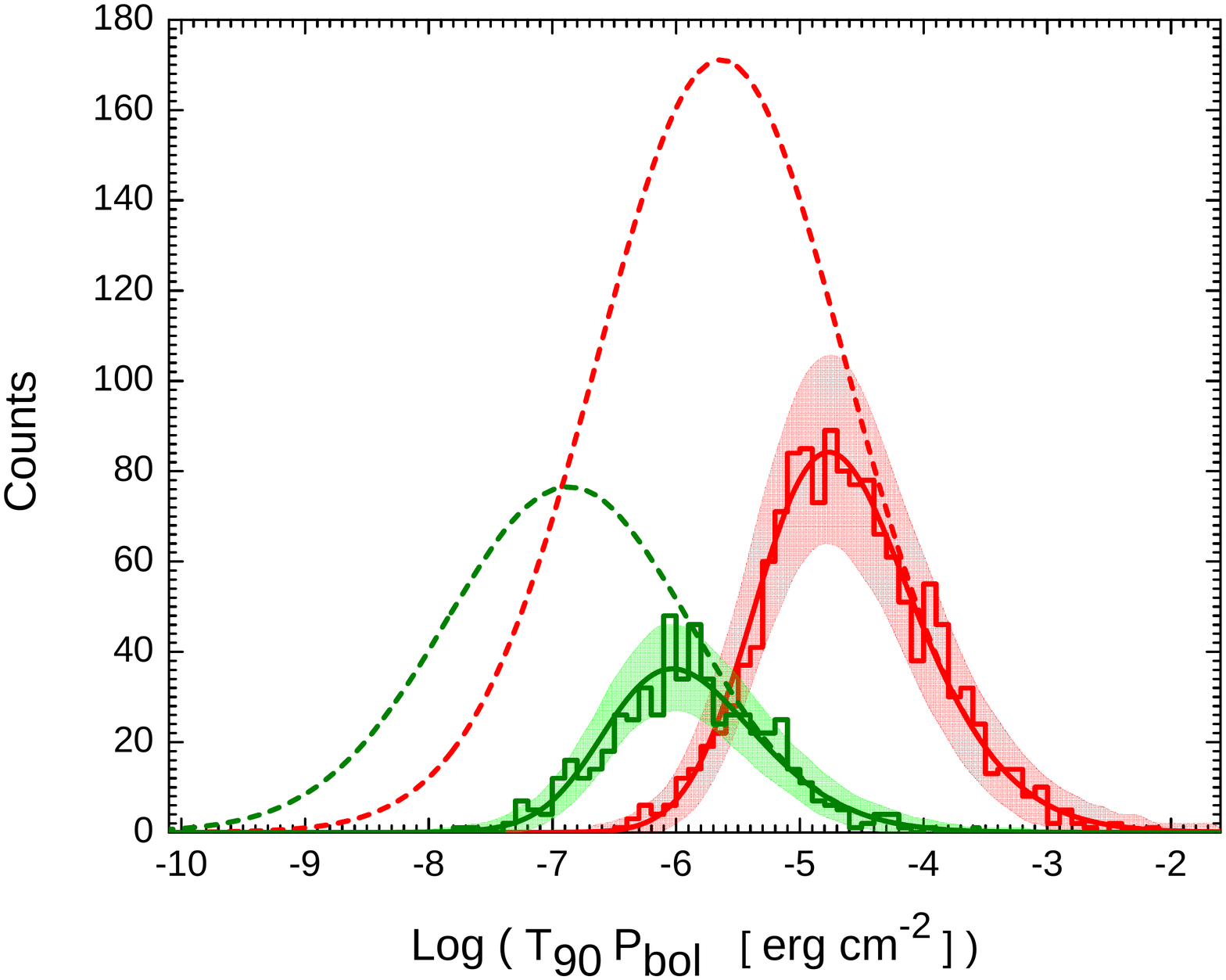}      & \includegraphics[scale=0.31]{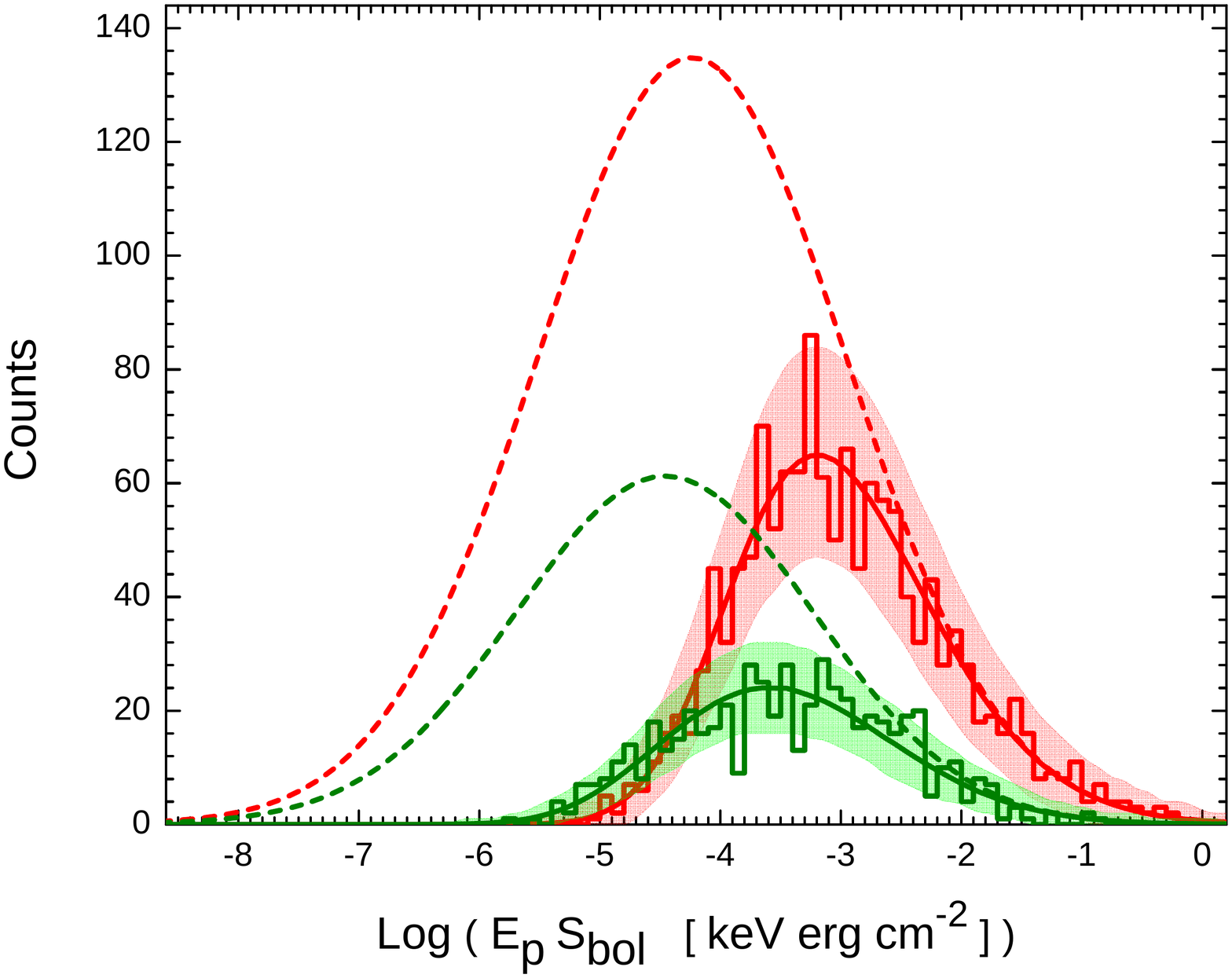}
            \end{tabular}
            \caption{{\bf Bivariate predictions of the {\it multivariate} best-fit SGRB \& LGRB world models for BATSE catalog GRBs, subject to BATSE detection threshold.} Colors, line styles and color-shaded areas bear the same meaning as in Figure \ref{fig:OFmarginals}. The apparent lack-of-fit in the top-left and center-left plots indicates the potential systematic bias in the duration ($\dur$) measurements of the longest duration BATSE GRBs close to detection threshold (c.f., Appendix \ref{App:SB}). Fitting results for LGRBs are taken from \citet{shahmoradi_multivariate_2013}. \label{fig:OFbivariates2}}
        \end{figure*}

\begin{figure*}
            \centering
            \begin{tabular}{cc}
                \includegraphics[scale=0.31]{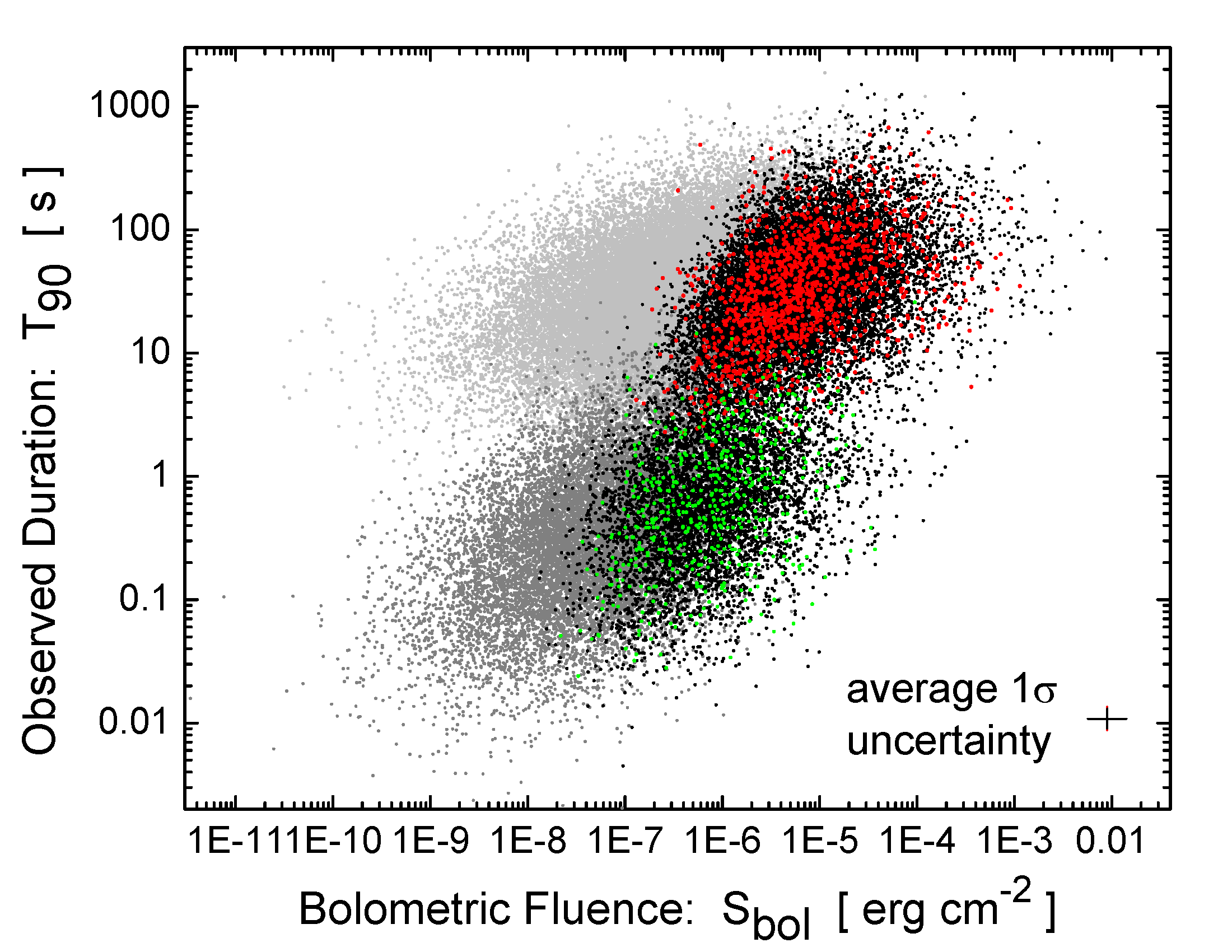} & \includegraphics[scale=0.31]{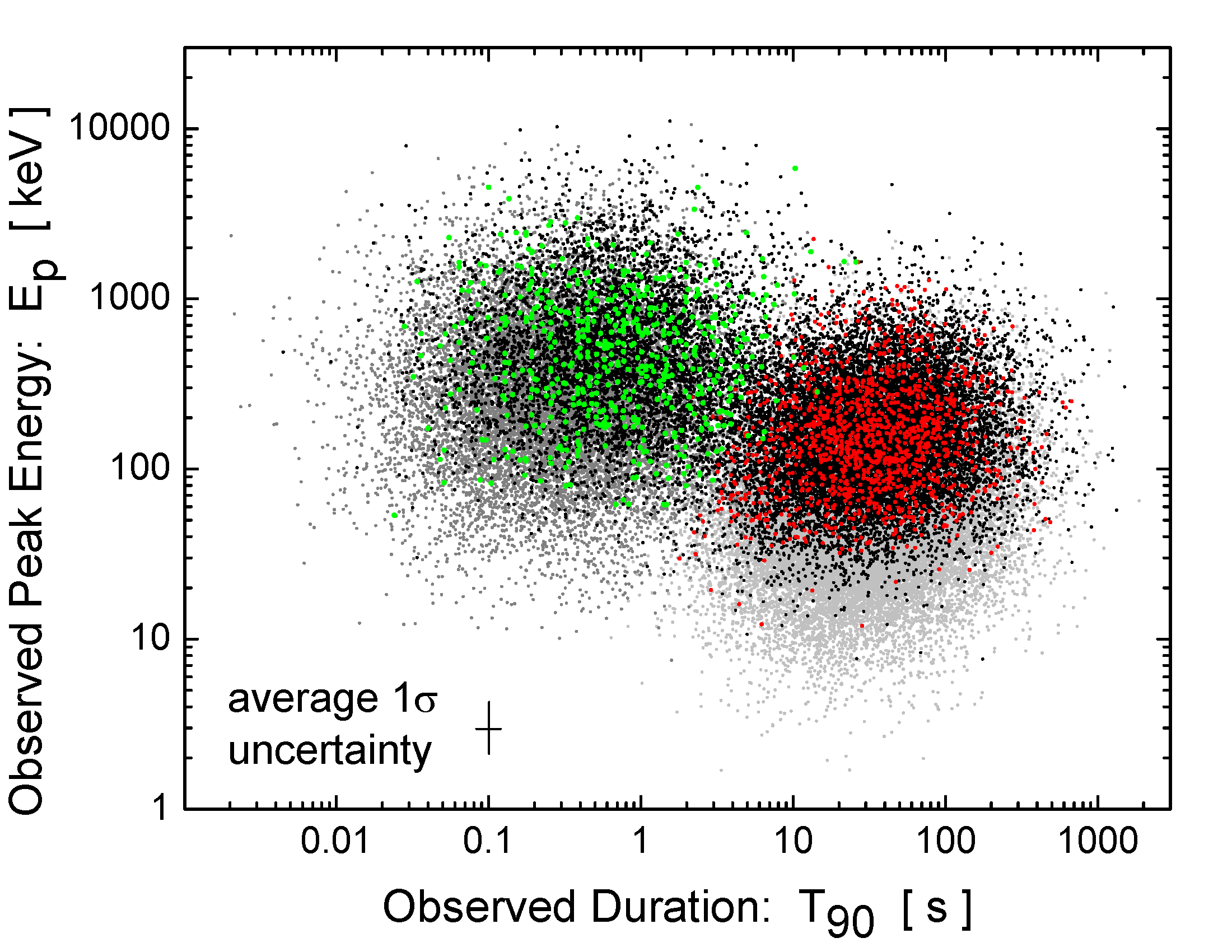} \\
                \includegraphics[scale=0.31]{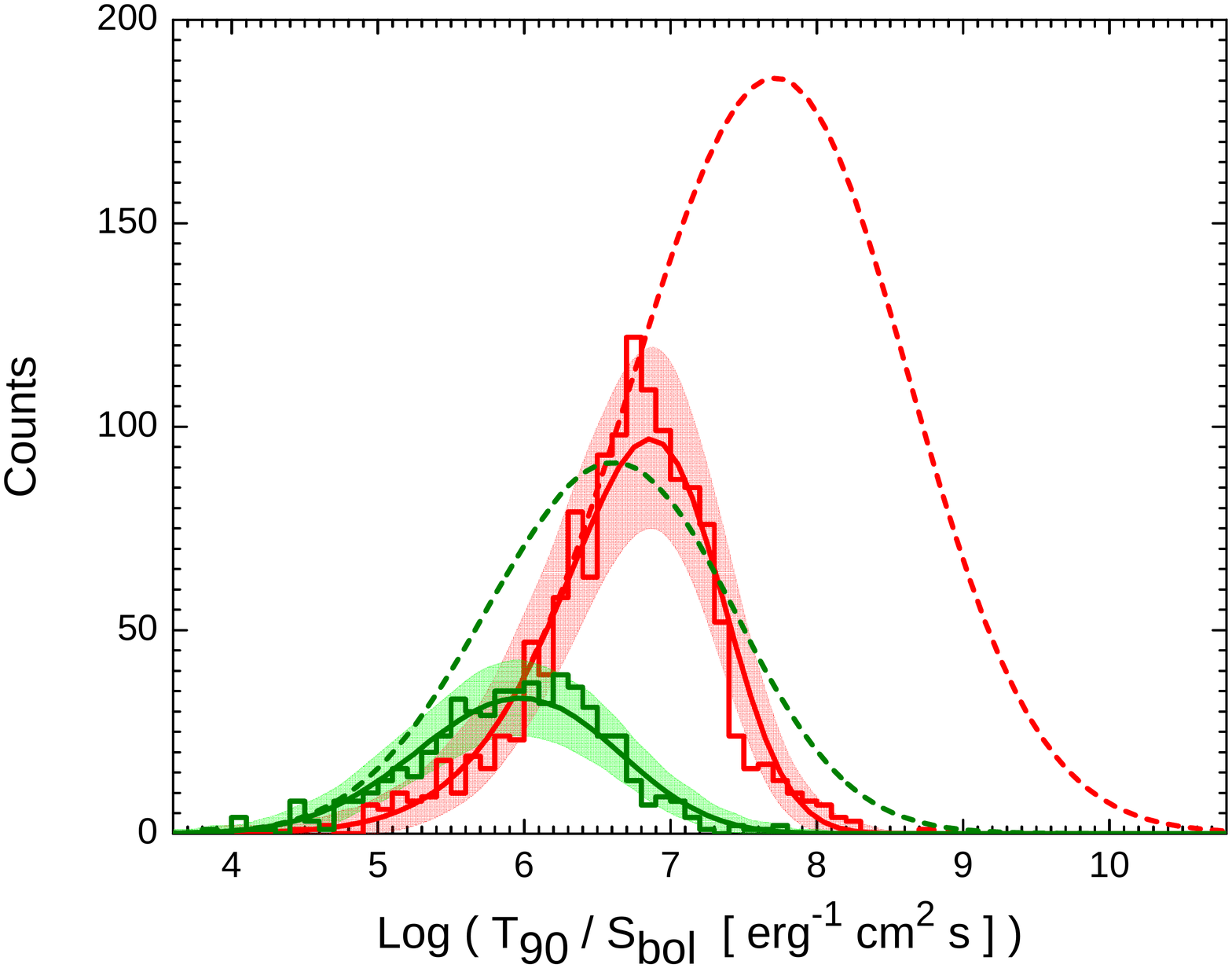}      & \includegraphics[scale=0.31]{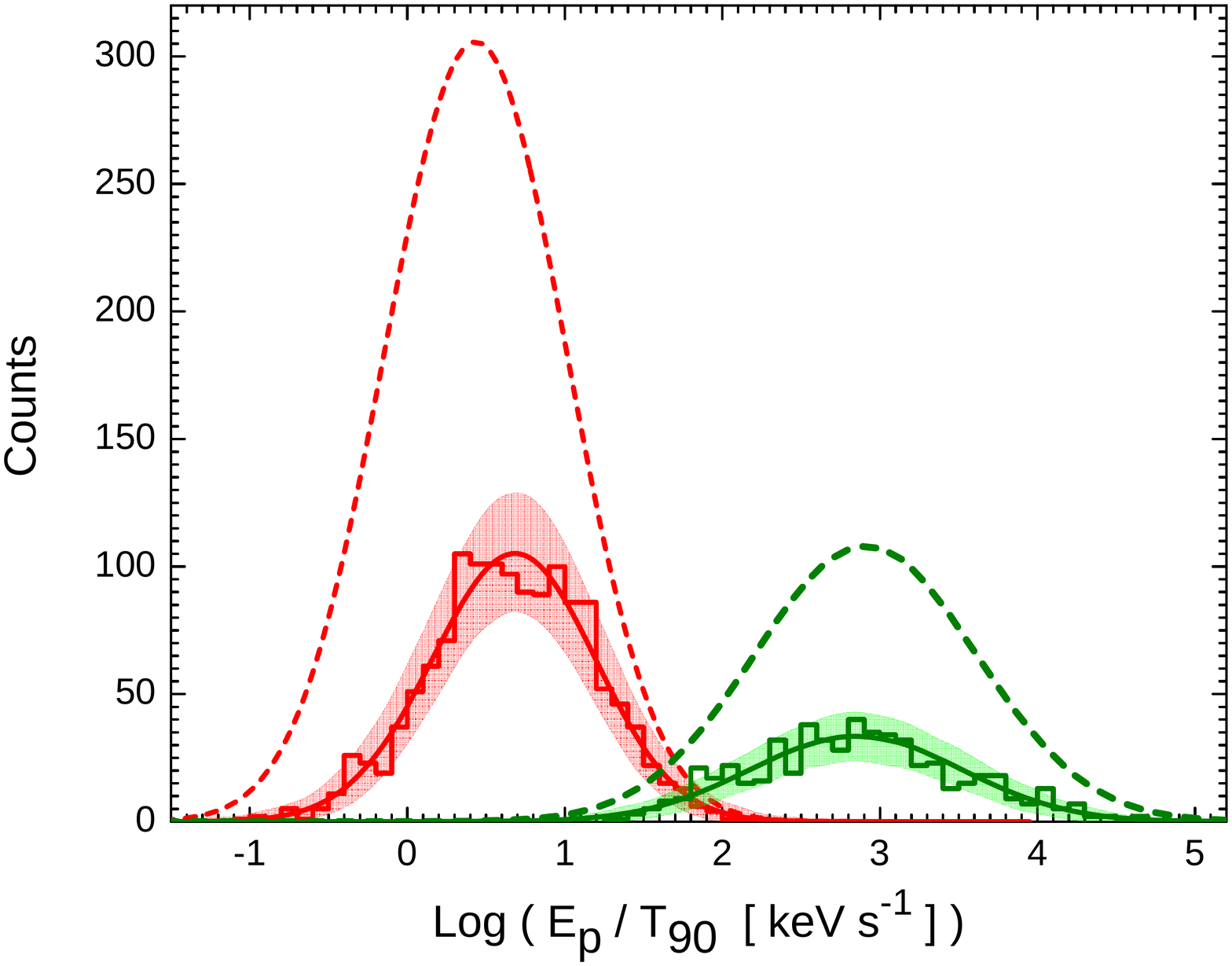}     \\
                \includegraphics[scale=0.31]{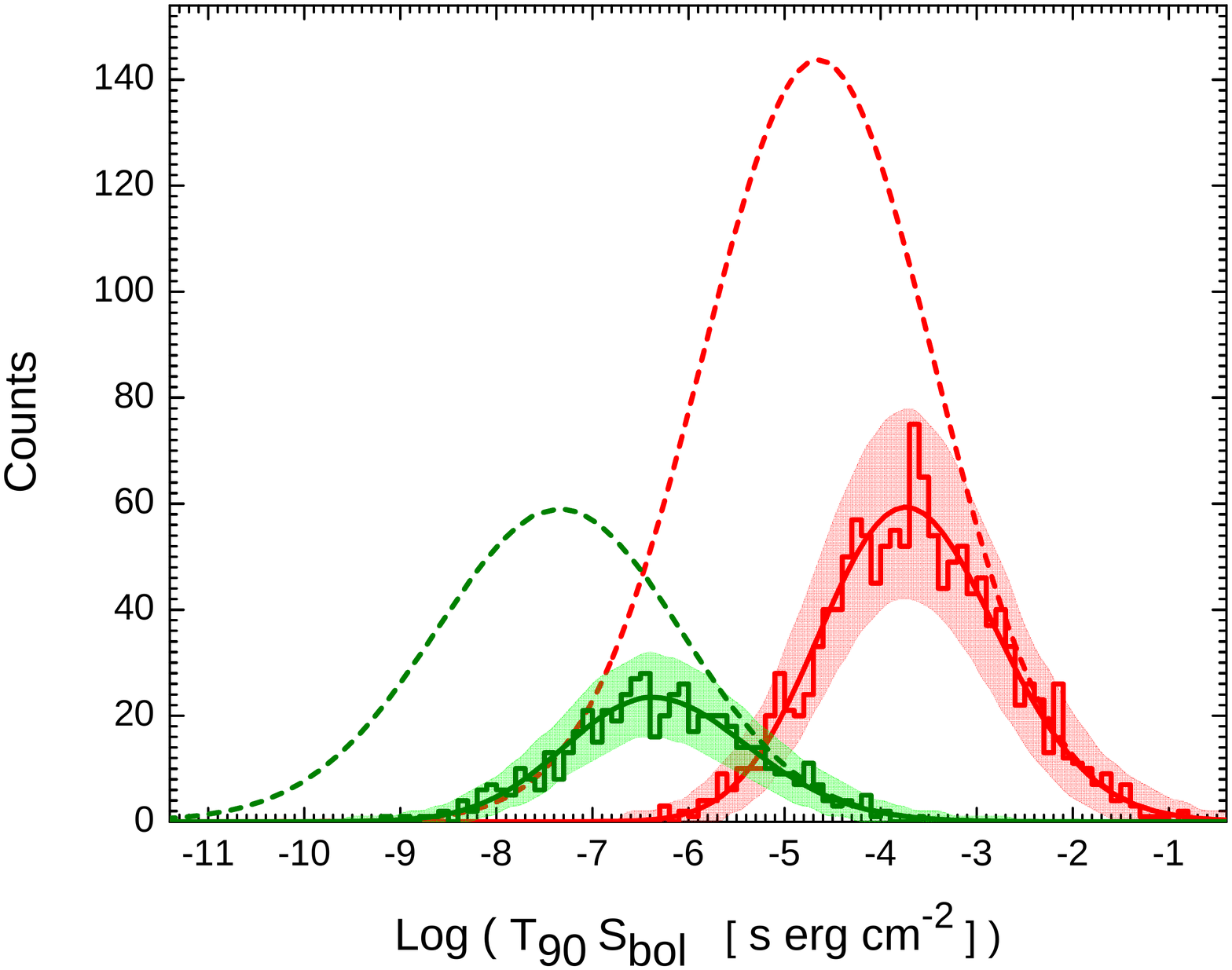}      & \includegraphics[scale=0.31]{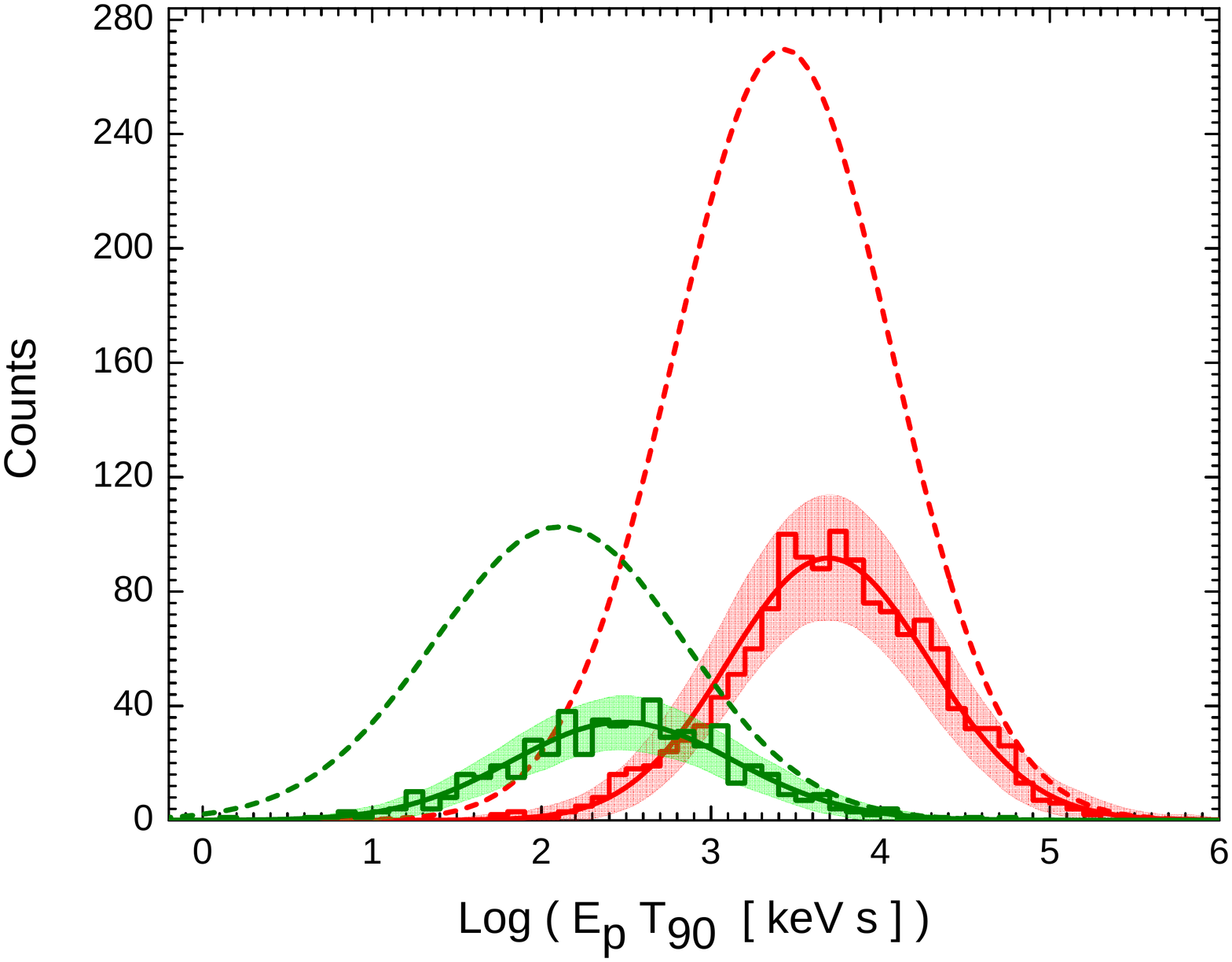}
            \end{tabular}
            \caption{{\bf Bivariate predictions of the {\it multivariate} best-fit SGRB \& LGRB world models for BATSE catalog GRBs, subject to BATSE detection threshold.} Colors, line styles and color-shaded areas bear the same meaning as in Figure \ref{fig:OFmarginals}. The apparent lack-of-fit in the top-left and center-left plots indicates the potential systematic bias in the duration ($\dur$) measurements of the longest duration BATSE GRBs close to detection threshold (c.f., Appendix \ref{App:SB}). Fitting results for LGRBs are taken from \citet{shahmoradi_multivariate_2013}. \label{fig:OFbivariates3}}
        \end{figure*}

        At the second level of GoF tests, the joint bivariate distributions of pairs of GRB variables can be compared to model predictions as illustrated in Figures \ref{fig:OFbivariates1}, \ref{fig:OFbivariates2}, \ref{fig:OFbivariates3}. The trace of BATSE trigger threshold on the populations of both SGRBs and LGRBs is visible in most bivariate distribution plots.  Following \citet{shahmoradi_multivariate_2013} and in order to ensure a good fit to the joint bivariate distributions of the four prompt emission parameters of SGRBs, we also compare the model predictions with observational data along the principal axes of the bivariate distributions, as depicted in the {\it center} \& {\it bottom} plots of Figures \ref{fig:OFbivariates1}, \ref{fig:OFbivariates2}, \ref{fig:OFbivariates3}. Although, statistically not a sufficient test for {\it multivariate} goodness-of-fit of the model to observational data, this comparison can provide strong evidence in favor of or against a good fit, at much higher confidence than solely comparing the model predictions and data via marginal distributions, as illustrated in Figure \ref{fig:OFmarginals}.

        A comparison of the bivariate model predictions with data along the principal axes immediately reveals the potential systematic biases that exist in BATSE catalog data close to detection threshold (e.g., {\it center-right} \& {\it bottom-left} plots of Figure \ref{fig:OFbivariates1}). We show in Appendix \ref{App:SB} that this bias has its origins primarily in the duration ($\dur$) and peak photon flux ($\pbol$) measurements of BATSE GRBs.

\begin{table*}
\begin{center}
\caption{Correlation Matrix of the parameters of the \textcolor{OliveGreen}{SGRB} World Model. For comparison, the correlation matrix of the \textcolor{red}{LGRB} world for a cosmic rate following SFR of \citet{li_star_2008} is reported on the lower triangle of the table \citep[c.f.,][]{shahmoradi_multivariate_2013}. \label{tab:cormat}}
\begin{scriptsize}
\begin{tabular}{|c|p{0.55cm}|p{0.55cm}|p{0.55cm}|p{0.55cm}|p{0.55cm}|p{0.55cm}|p{0.55cm}|p{0.55cm}|p{0.55cm}|p{0.55cm}|p{0.55cm}|p{0.55cm}|p{0.55cm}|p{0.55cm}|p{0.55cm}|p{0.55cm}|}
\hline
\hline
\rotatebox{55}{Parameter} & \begin{sideways}$\log(\liso)$\end{sideways} & \begin{sideways}$\log(\eiso)$\end{sideways} & \begin{sideways}$\log(\epkz)$\end{sideways} & \begin{sideways}$\log(\durz)$\end{sideways} & \begin{sideways}$\log(\sigma_{\liso})$\end{sideways} & \begin{sideways}$\log(\sigma_{\eiso})$\end{sideways} & \begin{sideways}$\log(\sigma_{\epkz})$\end{sideways} & \begin{sideways}$\log(\sigma_{\durz})$\end{sideways} & \begin{sideways}$\rho_{\liso-\eiso}$\end{sideways} & \begin{sideways}$\rho_{\liso-\epkz}$\end{sideways} & \begin{sideways}$\rho_{\liso-\durz}$\end{sideways} & \begin{sideways}$\rho_{\eiso-\epkz}$\end{sideways} & \begin{sideways}$\rho_{\eiso-\durz}$\end{sideways} & \begin{sideways}$\rho_{\epkz-\durz}$\end{sideways}  & \begin{sideways}$\mu_{thresh}$\end{sideways} & \begin{sideways}$\log(\sigma_{thresh})$\end{sideways} \\
\hline
$\log(\liso)$             &  \textcolor{black}{1.00}  &  \textcolor{OliveGreen}{0.95}  &  \textcolor{OliveGreen}{0.59}  &  \textcolor{OliveGreen}{0.77}  &  \textcolor{OliveGreen}{-0.87}  &  \textcolor{OliveGreen}{-0.78}  &  \textcolor{OliveGreen}{-0.28}  &  \textcolor{OliveGreen}{-0.42}  &  \textcolor{OliveGreen}{-0.17}  &  \textcolor{OliveGreen}{0.04}  &  \textcolor{OliveGreen}{-0.16}  &  \textcolor{OliveGreen}{-0.04}  &  \textcolor{OliveGreen}{-0.25}  &  \textcolor{OliveGreen}{-0.27}  &  \textcolor{OliveGreen}{-0.49}  &  \textcolor{OliveGreen}{-0.26}  \\
\hline
$\log(\eiso)$             &  \textcolor{red}{  0.96}  &  \textcolor{black}{1.00}  &  \textcolor{OliveGreen}{0.66}  &  \textcolor{OliveGreen}{0.83}  &  \textcolor{OliveGreen}{-0.88}  &  \textcolor{OliveGreen}{-0.86}  &  \textcolor{OliveGreen}{-0.33}  &  \textcolor{OliveGreen}{-0.48}  &  \textcolor{OliveGreen}{-0.33}  &  \textcolor{OliveGreen}{-0.05}  &  \textcolor{OliveGreen}{-0.27}  &  \textcolor{OliveGreen}{-0.11}  &  \textcolor{OliveGreen}{-0.33}  &  \textcolor{OliveGreen}{-0.35}  &  \textcolor{OliveGreen}{-0.47}  &  \textcolor{OliveGreen}{-0.25}  \\
\hline
$\log(\epkz)$             &  \textcolor{red}{  0.88}  &  \textcolor{red}{  0.90}  &  \textcolor{black}{1.00}  &  \textcolor{OliveGreen}{0.43}  &  \textcolor{OliveGreen}{-0.62}  &  \textcolor{OliveGreen}{-0.70}  &  \textcolor{OliveGreen}{-0.58}  &  \textcolor{OliveGreen}{-0.23}  &  \textcolor{OliveGreen}{-0.43}  &  \textcolor{OliveGreen}{-0.67}  &  \textcolor{OliveGreen}{-0.10}  &  \textcolor{OliveGreen}{-0.71}  &  \textcolor{OliveGreen}{-0.09}  &  \textcolor{OliveGreen}{-0.52}  &  \textcolor{OliveGreen}{-0.28}  &  \textcolor{OliveGreen}{-0.16}  \\
\hline
$\log(\durz)$             &  \textcolor{red}{  0.34}  &  \textcolor{red}{  0.43}  &  \textcolor{red}{  0.39}  &  \textcolor{black}{1.00}  &  \textcolor{OliveGreen}{-0.71}  &  \textcolor{OliveGreen}{-0.76}  &  \textcolor{OliveGreen}{-0.22}  &  \textcolor{OliveGreen}{-0.67}  &  \textcolor{OliveGreen}{-0.40}  &  \textcolor{OliveGreen}{0.06}  &  \textcolor{OliveGreen}{-0.61}  &  \textcolor{OliveGreen}{0.05}  &  \textcolor{OliveGreen}{-0.64}  &  \textcolor{OliveGreen}{-0.33}  &  \textcolor{OliveGreen}{-0.35}  &  \textcolor{OliveGreen}{-0.17}  \\
\hline
$\log(\sigma_{\liso})$    &  \textcolor{red}{  -0.90}  &  \textcolor{red}{  -0.90}  &  \textcolor{red}{  -0.81}  &  \textcolor{red}{  -0.35}  &  \textcolor{black}{1.00}  &  \textcolor{OliveGreen}{0.90}  &  \textcolor{OliveGreen}{0.36}  &  \textcolor{OliveGreen}{0.40}  &  \textcolor{OliveGreen}{0.27}  &  \textcolor{OliveGreen}{0.06}  &  \textcolor{OliveGreen}{0.19}  &  \textcolor{OliveGreen}{0.18}  &  \textcolor{OliveGreen}{0.25}  &  \textcolor{OliveGreen}{0.39}  &  \textcolor{OliveGreen}{0.32}  &  \textcolor{OliveGreen}{0.15}  \\
\hline
$\log(\sigma_{\eiso})$    &  \textcolor{red}{  -0.85}  &  \textcolor{red}{  -0.89}  &  \textcolor{red}{  -0.83}  &  \textcolor{red}{  -0.52}  &  \textcolor{red}{  0.94}  &  \textcolor{black}{1.00}  &  \textcolor{OliveGreen}{0.48}  &  \textcolor{OliveGreen}{0.51}  &  \textcolor{OliveGreen}{0.58}  &  \textcolor{OliveGreen}{0.25}  &  \textcolor{OliveGreen}{0.41}  &  \textcolor{OliveGreen}{0.33}  &  \textcolor{OliveGreen}{0.41}  &  \textcolor{OliveGreen}{0.56}  &  \textcolor{OliveGreen}{0.26}  &  \textcolor{OliveGreen}{0.12}  \\
\hline
$\log(\sigma_{\epkz})$    &  \textcolor{red}{  -0.56}  &  \textcolor{red}{  -0.60}  &  \textcolor{red}{  -0.78}  &  \textcolor{red}{  -0.32}  &  \textcolor{red}{  0.57}  &  \textcolor{red}{  0.66}  &  \textcolor{black}{1.00}  &  \textcolor{OliveGreen}{0.10}  &  \textcolor{OliveGreen}{0.36}  &  \textcolor{OliveGreen}{0.60}  &  \textcolor{OliveGreen}{0.05}  &  \textcolor{OliveGreen}{0.64}  &  \textcolor{OliveGreen}{-0.04}  &  \textcolor{OliveGreen}{0.27}  &  \textcolor{OliveGreen}{0.17}  &  \textcolor{OliveGreen}{0.11}  \\
\hline
$\log(\sigma_{\durz})$    &  \textcolor{red}{  -0.13}  &  \textcolor{red}{  -0.14}  &  \textcolor{red}{  -0.13}  &  \textcolor{red}{  -0.20}  &  \textcolor{red}{  0.13}  &  \textcolor{red}{  0.18}  &  \textcolor{red}{  0.10}  &  \textcolor{black}{1.00}  &  \textcolor{OliveGreen}{0.23}  &  \textcolor{OliveGreen}{-0.05}  &  \textcolor{OliveGreen}{0.42}  &  \textcolor{OliveGreen}{-0.09}  &  \textcolor{OliveGreen}{0.55}  &  \textcolor{OliveGreen}{0.20}  &  \textcolor{OliveGreen}{0.16}  &  \textcolor{OliveGreen}{0.08}  \\
\hline
$\rho_{\liso-\eiso}$      &  \textcolor{red}{  -0.00}  &  \textcolor{red}{  -0.10}  &  \textcolor{red}{  -0.17}  &  \textcolor{red}{  -0.52}  &  \textcolor{red}{  0.01}  &  \textcolor{red}{  0.31}  &  \textcolor{red}{  0.31}  &  \textcolor{red}{  -0.02}  &  \textcolor{black}{1.00}  &  \textcolor{OliveGreen}{0.47}  &  \textcolor{OliveGreen}{0.58}  &  \textcolor{OliveGreen}{0.37}  &  \textcolor{OliveGreen}{0.32}  &  \textcolor{OliveGreen}{0.41}  &  \textcolor{OliveGreen}{0.00}  &  \textcolor{OliveGreen}{-0.01}  \\
\hline
$\rho_{\liso-\epkz}$      &  \textcolor{red}{  -0.45}  &  \textcolor{red}{  -0.50}  &  \textcolor{red}{  -0.74}  &  \textcolor{red}{  -0.32}  &  \textcolor{red}{  0.48}  &  \textcolor{red}{  0.60}  &  \textcolor{red}{  0.83}  &  \textcolor{red}{  0.08}  &  \textcolor{red}{  0.43}  &  \textcolor{black}{1.00}  &  \textcolor{OliveGreen}{-0.01}  &  \textcolor{OliveGreen}{0.92}  &  \textcolor{OliveGreen}{-0.13}  &  \textcolor{OliveGreen}{0.36}  &  \textcolor{OliveGreen}{0.00}  &  \textcolor{OliveGreen}{0.03}  \\
\hline
$\rho_{\liso-\durz}$      &  \textcolor{red}{  0.42}  &  \textcolor{red}{  0.35}  &  \textcolor{red}{  0.31}  &  \textcolor{red}{  -0.56}  &  \textcolor{red}{  -0.45}  &  \textcolor{red}{  -0.22}  &  \textcolor{red}{  -0.14}  &  \textcolor{red}{  0.01}  &  \textcolor{red}{  0.59}  &  \textcolor{red}{  -0.05}  &  \textcolor{black}{1.00}  &  \textcolor{OliveGreen}{-0.04}  &  \textcolor{OliveGreen}{0.89}  &  \textcolor{OliveGreen}{0.46}  &  \textcolor{OliveGreen}{-0.04}  &  \textcolor{OliveGreen}{-0.06}  \\
\hline
$\rho_{\eiso-\epkz}$      &  \textcolor{red}{  -0.49}  &  \textcolor{red}{  -0.54}  &  \textcolor{red}{  -0.76}  &  \textcolor{red}{  -0.33}  &  \textcolor{red}{  0.54}  &  \textcolor{red}{  0.65}  &  \textcolor{red}{  0.84}  &  \textcolor{red}{  0.09}  &  \textcolor{red}{  0.36}  &  \textcolor{red}{  0.96}  &  \textcolor{red}{  -0.10}  &  \textcolor{black}{1.00}  &  \textcolor{OliveGreen}{-0.13}  &  \textcolor{OliveGreen}{0.54}  &  \textcolor{OliveGreen}{0.01}  &  \textcolor{OliveGreen}{0.01}  \\
\hline
$\rho_{\eiso-\durz}$      &  \textcolor{red}{  0.41}  &  \textcolor{red}{  0.35}  &  \textcolor{red}{  0.33}  &  \textcolor{red}{  -0.55}  &  \textcolor{red}{  -0.44}  &  \textcolor{red}{  -0.26}  &  \textcolor{red}{  -0.19}  &  \textcolor{red}{  0.11}  &  \textcolor{red}{  0.38}  &  \textcolor{red}{  -0.14}  &  \textcolor{red}{  0.95}  &  \textcolor{red}{  -0.16}  &  \textcolor{black}{1.00}  &  \textcolor{OliveGreen}{0.54}  &  \textcolor{OliveGreen}{0.02}  &  \textcolor{OliveGreen}{-0.02}  \\
\hline
$\rho_{\epkz-\durz}$      &  \textcolor{red}{  0.01}  &  \textcolor{red}{  -0.05}  &  \textcolor{red}{  -0.01}  &  \textcolor{red}{  -0.59}  &  \textcolor{red}{  0.05}  &  \textcolor{red}{  0.20}  &  \textcolor{red}{  0.04}  &  \textcolor{red}{  0.13}  &  \textcolor{red}{  0.35}  &  \textcolor{red}{  0.04}  &  \textcolor{red}{  0.61}  &  \textcolor{red}{  0.15}  &  \textcolor{red}{  0.66}  &  \textcolor{black}{1.00}  &  \textcolor{OliveGreen}{0.00}  &  \textcolor{OliveGreen}{-0.04}  \\
\hline
$\mu_{thresh}$            &  \textcolor{red}{  -0.66}  &  \textcolor{red}{  -0.64}  &  \textcolor{red}{  -0.59}  &  \textcolor{red}{  -0.19}  &  \textcolor{red}{  0.49}  &  \textcolor{red}{  0.44}  &  \textcolor{red}{  0.34}  &  \textcolor{red}{  0.08}  &  \textcolor{red}{  -0.06}  &  \textcolor{red}{  0.24}  &  \textcolor{red}{  -0.26}  &  \textcolor{red}{  0.26}  &  \textcolor{red}{  0.24}  &  \textcolor{red}{  -0.06}  &  \textcolor{black}{1.00}  &  \textcolor{OliveGreen}{0.74}  \\
\hline
$\log(\sigma_{thresh})$   &  \textcolor{red}{  -0.42}  &  \textcolor{red}{  -0.41}  &  \textcolor{red}{  -0.38}  &  \textcolor{red}{  -0.11}  &  \textcolor{red}{  0.29}  &  \textcolor{red}{  0.26}  &  \textcolor{red}{  0.21}  &  \textcolor{red}{  0.05}  &  \textcolor{red}{  -0.05}  &  \textcolor{red}{  0.15}  &  \textcolor{red}{  -0.16}  &  \textcolor{red}{  0.16}  &  \textcolor{red}{  -0.15}  &  \textcolor{red}{  -0.05}  &  \textcolor{red}{  0.78}  &  \textcolor{black}{1.00}  \\
\hline
\hline
\end{tabular}
\end{scriptsize}
\end{center}
\end{table*}

        This method of scanning the model and data along the principal axes of the joint bivariate distributions can be generalized to trivariate and quadruvariate joint distributions. For brevity, however, only the bivariate tests are presented here.

        \begin{figure*}
            \centering
            \begin{tabular}{cc}
                \includegraphics[scale=0.31]{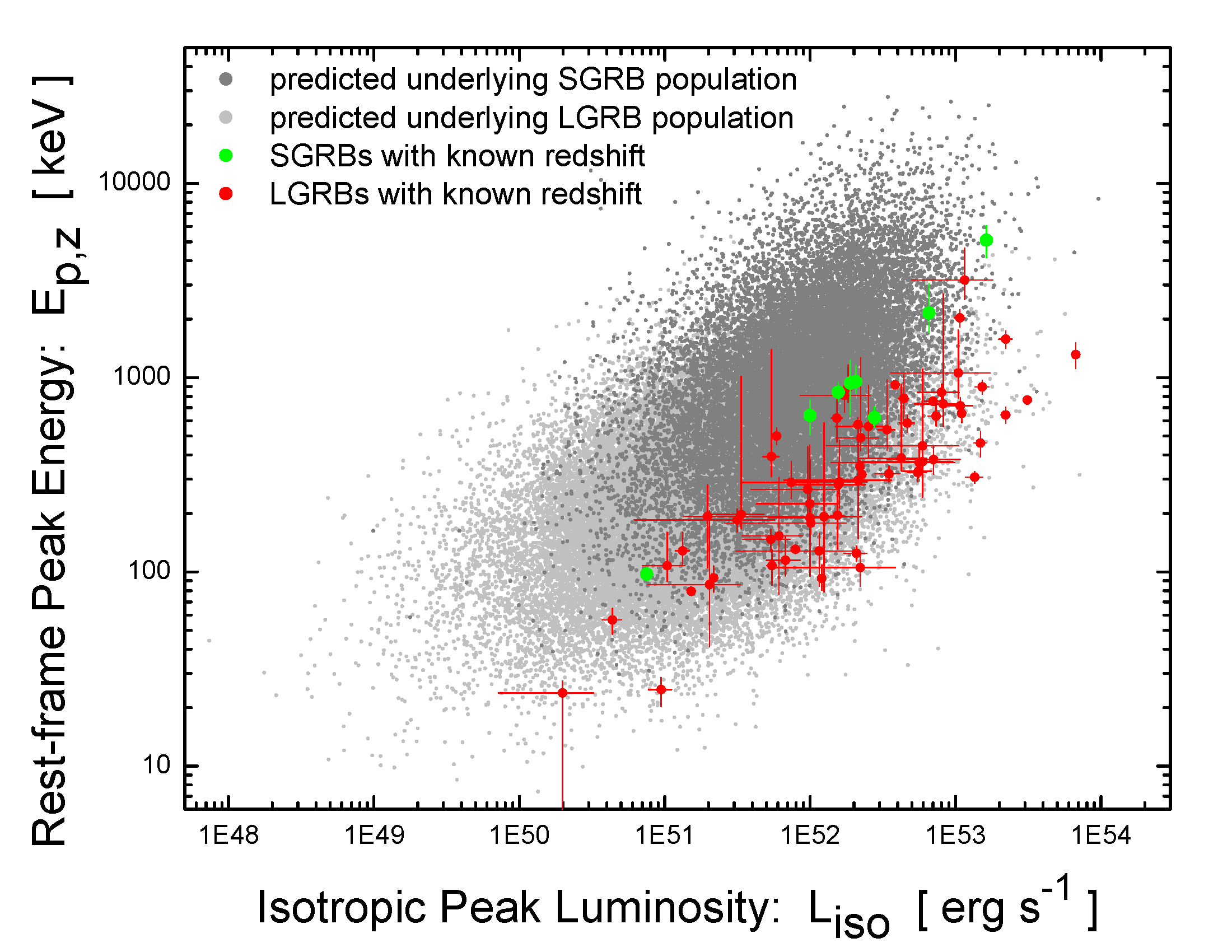} & \includegraphics[scale=0.31]{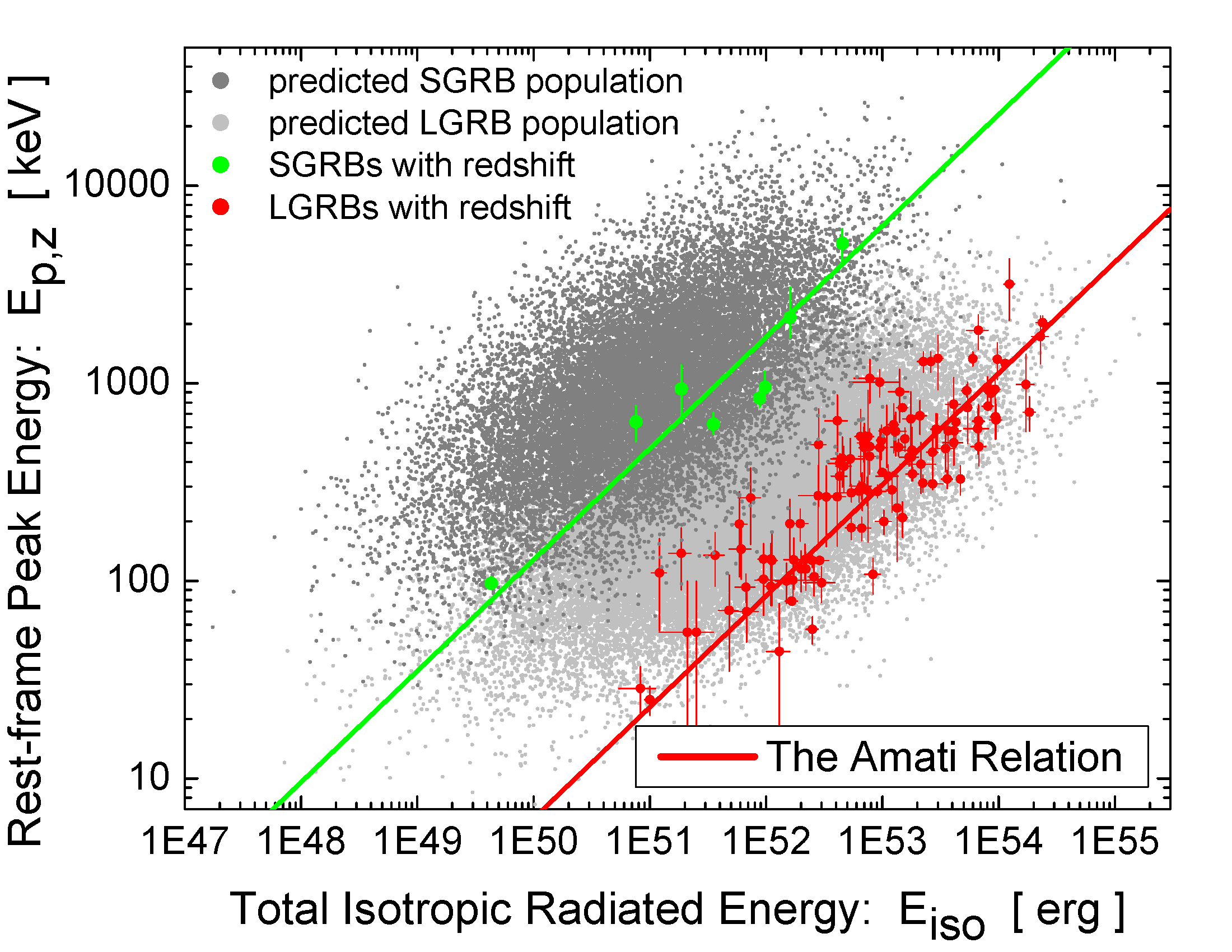}
            \end{tabular}
            \caption{{\bf Prediction of the world model for the joint bivariate distributions of SGRBs in the rest-frame planes of $\liso-\epkz$ \& $\eiso-\epkz$.} For comparison, the prediction of the world model for LGRBs population -- the Amati \& Yonetoku relations -- are also shown (c.f., \citet{shahmoradi_multivariate_2013}). The green filled circles represent SGRB data with known redshift \citep{zhang_correlation_2012, tsutsui_possible_2013} and the red filled circles represent LGRB data taken from \citet{schaefer_hubble_2007}, \citet{amati_measuring_2008},\citet{ghirlanda_e_2008}. \label{fig:rest_frame}}
        \end{figure*}

    \section{Discussion}
    \label{sec:discussion}

        Throughout previous sections, we presented and explained an elaborate analysis to constrain the energetics, luminosity function and the prompt gamma-ray correlations of short GRBs, subject to detection threshold of gamma-ray detectors, here BATSE LADs. The methodology employed in this work is similar to the LGRB population study of \citet{shahmoradi_multivariate_2013}. We have argued and shown that the intrinsic population distributions of both long and short GRBs can be well described by a multivariate log-normal model in the 4-dimensional prompt gamma-ray emission parameter space of the isotropic peak luminosity ($\liso$), total isotropic emission ($\eiso$), the spectral peak energy ($\epkz$), and the prompt duration ($\durz$). This was done by first employing a fuzzy clustering method to segregate the BATSE catalog GRBs into long and short classes (\ref{sec:samsel}), followed by a careful modelling of the effects of the BATSE detection threshold on the two GRB populations (Appendix \ref{App:BDT}).

        Ideally, if the population properties of both SGRBs and LGRBs can be well described by multivariate log-normal distribution as argued in Sec. \ref{sec:MC}, also by \citet{shahmoradi_gamma-ray_2013} \& \citet{shahmoradi_multivariate_2013}, then a multivariate log-normal {\it mixture} model ought to be used in order to simultaneously fit for the distributions of the entire BATSE catalog of LGRBs and SGRBs together. Therefore, the classification of the observed sample prior to model fitting as performed in Sec. \ref{sec:samsel} would be unnecessary and the members of the two GRB populations would be automatically determined by the best fit parameters of the mixture model. Nevertheless, our experimentation with mixture models generally led to either degenerate or very poor fitting results for BATSE GRB data. This is primarily due to the morphological differences in the lightcurves and spectra of short and long GRBs, which in turn result in challenging difficulties in modelling the BATSE triggering algorithm in a unified framework for both GRB classes in the multivariate mixture model.

        In the following sections, we discuss the key findings of the presented work and compare the results with data and findings from other gamma-ray experiments and population studies of GRBs.

        \subsection{Luminosity Function \& Isotropic Emission}
        \label{sec:LFIE}
            Despite being an ill-defined quantity, the isotropic peak luminosity ($\liso$) is among the most widely used and investigated parameters of the prompt emission of GRBs. The vague and conventional definition of $\liso$ stems from its dependence on the timescale used to define this quantity, commonly set to $t=1024ms\sim 1~$[sec] ($L_{iso,1024ms}$). Although the $L_{iso,1024ms}$ definition of the peak luminosity is more or less independent of the duration (e.g., $\durz$) of the long class of GRBs, it is strongly duration-dependent for the class of short GRBs (Figure \ref{fig:durppr}). This is primarily due to the diverse range of prompt durations of SGRBs, with many of the detected bursts lasting only a fraction of a second. Alternative definitions of the peak luminosity have been already proposed in order to provide a universal definition of the peak luminosity, independent of the duration and type of the burst, such as the {\it effective luminosity} definition of \citet{butler_cosmic_2010}. Nevertheless, such global luminosity definitions can not be employed in the presented analysis due to the specific triggering algorithm of BATSE LADs, which is defined for $3$ distinct fixed timescales: $64ms$, $256ms$ \& $1024ms$. Therefore, in order to minimize the effects of GRB duration on the definition of peak luminosity, we have used the highest resolution timescale ($64ms$) for the definition of the isotropic peak luminosity of BATSE SGRBs. By contrast, the one-second peak luminosity ($L_{iso,1024ms}$) is sufficiently accurate and almost independent of the prompt duration of almost all LGRBs, whether detected or undetected \citep[c.f.,][]{shahmoradi_multivariate_2013}.

            Given BATSE SGRB data, we find a $3\sigma$ range of $\log(\liso erg s^{-1})\in[50.0,53.0]$ for the distribution of the $64ms$ peak luminosities of short GRBs. Depending on the burst duration from long to short, the $64ms$ peak luminosity of SGRBs can be on average $1.5$ to $13$ times larger than the conventional $1024ms$ timescale definition of the peak luminosity, commonly used for LGRBs. Knowing that the most luminous bursts generally tend to be the longest, we obtain a conservative range of $\log(L_{iso,1024ms} erg s^{-1})\in[48.9,53.1]$ for short GRBs. For comparison, \citet{shahmoradi_multivariate_2013} finds a $3\sigma$ range of $\log(\liso erg s^{-1})\in[49.5,53.5]$ for the population of Long GRBs.

            It should be noted however, that the prompt-emission GRB quantities, in particular, the isotropic peak luminosity and the cosmic rates of GRBs are very sensitive to the detection threshold of the specific detector used to collect observational data, and to the systematic biases in measurements close to detection threshold (c.f., Appendix \ref{App:SB}). This has also been noted in an earlier study of the population properties of Swift LGRBs by \citet{butler_cosmic_2010}.

            As for the population distribution of the total isotropic emission of short GRBs, we obtain a $3\sigma$ range of $\log(\eiso erg)\in[48.0,53.5]$. For comparison, a $3\sigma$ range of $\log(\eiso erg)\in[49.2,54.7]$ was obtained for LGRBs by \citet{shahmoradi_multivariate_2013}. Therefore, it appears that LGRBs are on average more than one order of magnitude brighter than SGRBs. It is however, notable that the width of the distribution of $\eiso$ is approximately the same in both populations of LGRBs and SGRBs. The same also holds for the width of the distribution of $\liso$ in the two GRB classes.

        \subsection{Spectral Peak Energy \& Prompt Duration}
        \label{sec:epkdur}
            The spectral peak energies of SGRBs have long been observed to be systematically higher than the typical peak energies of the class of LGRBs \citep[e.g.,][]{kouveliotou_identification_1993-1}. The extent and significance of the difference between the two populations however, has remained a matter of speculation due to sample incompleteness and unknown selection biases in observational data of both GRB classes. The methodology presented in this work enabled us for the first time to set stringent constraints on the potential underlying {\it population} distribution of the spectral peak energies of both GRB classes in both observer and rest frames.

            For the distribution of the intrinsic spectral peak energies of short GRBs, the model predicts an approximate $3\sigma$ range of $\epkz\in[45keV,16MeV]$, with an intrinsic average peak energy of $\epkz\sim955~keV$. This corresponds to an observer-frame $3\sigma$ range of $\epkz\in[10keV,6MeV]$, with an observer-frame average peak energy of $\epkz\sim300~keV$. For comparison, \citet{shahmoradi_multivariate_2013} finds approximate $3\sigma$ ranges of $\epkz(keV)\in[20,4000]$ \& $\epk(keV)\in[5,1430]$ for the intrinsic and observer-frame spectral peak energy distributions of LGRBs, with population averages $\epkz\sim300~keV$ \& $\epk\sim85~keV$ respectively.

            In contrast, the distributions of the observed spectral peak energies of {\it BATSE catalog LGRBs \& SGRBs} as found by \citet{shahmoradi_hardness_2010} (Figure 13 therein) are $\epk\sim140(keV)$ and $\epk\sim520(keV)$ respectively, slightly larger than the inferred values for the underlying population of the two classes here in this work. Similarly, \citet{nava_spectral_2011} also find slightly larger average $\epk$ values for a sample of $438$ Fermi SGRBs and LGRBs. The discrepancy between the detected sample and the underlying population can be explained by sample incompleteness due to BATSE LAD \& Fermi GBM detection thresholds. This is line with previous findings on the role of selection effects due to gamma-ray detectors in shaping the observed properties of the two GRB classes \citep[e.g.,][]{band_testing_2005, shahmoradi_how_2009, butler_generalized_2009, shahmoradi_possible_2011, kocevski_origin_2012}.

            As for the prompt duration of short GRBs, the model predicts a $3\sigma$ range of $\durz\in[2ms,7s]$ for the intrinsic duration distribution of SGRBs with a population average of $\durz\sim180ms$. This corresponds to a $3\sigma$ dynamic range of $\dur\in[5ms,22s]$ for the duration distribution of short GRBs in the observer frame with a population average of $\dur\sim350ms$. When compared to BATSE detected sample of SGRBs with an average $\dur\sim670ms$, it is apparent that the majority of undetected SGRBs were likely among the shortest events in the population. This corroborates the early hints on the potential existence of very short-duration GRBs that could have been missed by BATSE Large Area Detectors \citep[e.g., Figure (1) \& (2) in ][]{nemiroff_gamma-ray_1998}. The minor excess predicted by the SGRB world model may also be partly attributed to the systematic biases and errors in BATSE data and the simplified model of BATSE trigger threshold for SGRBs as explained in Appendix \ref{App:BDT} \& \ref{App:SB}.

            For comparison, the LGRB world model of \citet{shahmoradi_multivariate_2013} predicts an approximate $3\sigma$ range of $\durz(s)\in[0.5,145]$ for the intrinsic duration distribution of LGRBs with a population average of $\durz\sim10s$, corresponding to an approximate $3\sigma$ dynamic range of $\dur(s)\in[1,620]$ in the observer frame with a population average of $\dur\sim30s$.

            The presented LGRB and SGRB world models also predict that intrinsically shorter duration GRBs in both classes, although likely exist, have lower chances of detection by gamma-ray instruments such as BATSE LAD and Fermi GBM. These results corroborate the recent findings of \citet{littlejohns_are_2013} \& \citet{littlejohns_investigating_2014} based on a sample of Swift LGRBs and further extend it to the population of short GRBs as illustrated in {\it right plot} of Figure \ref{fig:redshift}.

        \subsection{Temporal \& Spectral Correlations}
        \label{sec:correlations}

            There is already an extensive body of literature on the potential existence of correlations among the prompt spectral and temporal parameters of LGRBs \citep[c.f.,][ for a comprehensive review of the literature]{shahmoradi_possible_2011}. Much of the efforts so far has mainly focused on brightness-hardness types of relations, such as the Amati relation ($\eiso-\epkz$) and the Yonetoku relation ($\liso-\epkz$). Currently, the general consensus is that these relations do exist with high significance \citep[e.g.,][]{butler_cosmic_2010,shahmoradi_multivariate_2013}, but with far less strength than the original findings of \citet{amati_intrinsic_2002} and \citet{yonetoku_gamma-ray_2004}. In addition to early hints on the potential existence of duration-brightness relations \citep[e.g.,][]{horvath_gamma-ray_2005}, recently \citet{butler_cosmic_2010} also found, through an elaborate and comprehensive analysis of Swift data, some tentative signatures of a significant positive correlation between the intrinsic duration and the total isotropic emission of LGRBs. Later, \citet{shahmoradi_multivariate_2013} provided evidence in support of intrinsic duration-brightness correlations among LGRBs using independent methods and models applied to BATSE catalog data, and further showed that the brightness-duration relations (i.e., $\eiso-\durz$ \& $\liso-\durz$) are of comparable strength and significance to brightness-hardness relations (i.e., $\eiso-\epkz$ \& $\liso-\epkz$). This positive correlation is also evident in the results of the LGRB world model in Table \ref{tab:BFP}. Unlike the work of \citet{shahmoradi_multivariate_2013}, here in this work we have applied an energy band correction of the form $(1+z)^{-0.34}$ to the observed durations ($\dur$) of BATSE LGRBs \citep[e.g.,][]{gehrels_new_2006,butler_cosmic_2010}. This correction if applied, slightly relaxes the strength of the correlation between the intrinsic duration and brightness.

            Contrary to LGRBs, correlations among the prompt-emission parameters of short GRBs have been far less investigated. Recently, \citet{zhang_correlation_2012} and \citet{tsutsui_possible_2013} used a small sample of SGRBs with measured redshift to argue for the potential existence of intrinsic hardness-brightness correlations in the population of SGRBs similar to the class of LGRBs. Nevertheless, the strength and significance of these positive correlations in the underlying population of SGRBs could not be determined in their studies due to strong selection bias and sample incompleteness in observational data.

            The results of the SGRB world model, as presented in Table \ref{tab:BFP}, confirm the existence of intrinsic hardness-brightness correlations among SGRBs. Specifically, the model predicts highly significant correlation strengths of $\rho\sim0.51\pm0.10$ \& $\rho\sim0.60\pm0.06$ for $\liso-\epkz$ \& $\eiso-\epkz$ relations respectively. The two correlation strengths are very similar to the predictions of the LGRB world model for the same relations in the population of LGRBs ($\rho\sim0.45\pm0.07$ \& $\rho\sim0.58\pm0.04$, respectively). This is illustrated in the plots of Figure \ref{fig:rest_frame}.

            The similarity of the two GRB classes in the joint distributions and correlations is not limited to only hardness-brightness relations. Indeed, all four prompt-emission variables in both populations ($\liso$, $\eiso$, $\epkz$, $\durz$) appear to be similarly related to each other in both populations, as shown in Table \ref{tab:BFP}.

            A potential correlation of the form,

            \begin{equation}
                \epkz\sqrt{\durz}\propto E_{\gamma}
            \end{equation}

            has also been derived and suggested by \citet{putten_nonthermal_2008}, in which $E_{\gamma}$ stands for the beaming-angle-corrected output energy from the burst. The presented analysis is consistent with the existence of such universal relation. The strength and significance of it however, cannot be determined solely based on BATSE observational data, as it requires a knowledge of redshift and the beaming angle of individual events.  We also caution against the use of the hardness-brightness correlations, such as the Amati relation, to infer the characteristics of the inner engines of the two GRB classes. Our predictions based on BATSE catalog GRBs is that the Amati relation in its current form -- as presented by \citet{putten_origin_2014} -- is likely strongly affected by sample incompleteness. In addition, the significant overlap of the two GRB populations in hardness-brightness plots, as illustrated in Figure \ref{fig:rest_frame}, likely renders the Amati relation an ineffective tool for GRB classification.

    \section{Concluding Remarks}
    \label{sec:CR}

        The primary goal of the presented analysis was to constrain the energetics, luminosity function, and the prompt gamma-ray correlations of short-hard class of GRBs, using the wealth of information that has remained untouched in the largest catalog of GRBs available to this date, the current BATSE GRB catalog. In the following lines we summarize the steps we have taken to constrain the population properties of short-hard and similarly, long-soft GRBs (SGRBs \& LGRBs, respectively):

        \begin{enumerate}

            \item
            A sample of $565$ short-hard and $1366$ long-soft bursts were first segregated and selected from the current BATSE catalog of 2130 GRBs, for which complete data were available, including the bolometric peak flux ($\pbol$), the bolometric fluence ($\sbol$), the observed spectral peak energy ($\epk$) and the observed duration ($\dur$). The classification method is based on fuzzy clustering algorithms on the two prompt-emission variables $\epk$ \& $\dur$ which are least affected by the detection threshold of gamma-ray detectors (Sec. \ref{sec:samsel}). This methodology can be readily applied to other GRB catalogs, in particular Fermi GBM.

            \item
            We propose that the intrinsic joint distribution of the four main prompt-emission parameters of SGRBs: the isotropic peak gamma-ray luminosity ($\liso$), the total isotropic gamma-ray emission ($\eiso$), the intrinsic spectral peak energy ($\epkz$) and the intrinsic duration ($\durz$) can be well described as a multivariate (4-dimensional) log-normal distribution, once the observational data is corrected for effects of detection threshold and sample incompleteness (Sec. \ref{sec:MC}).

            \item
            The best-fit parameters of the model are then found by maximizing the likelihood function of the model given BATSE SGRB data (Eqn. \ref{eq:likelihood}) subject to the Bayesian priors of Eqn. \ref{eq:posterior}, with a SGRB rate density (Eqn. \ref{eq:ratedensity}) that is the result of the convolution of Star Formation Rate (Eqn \ref{eq:zeta}) with a log-normal binary merger delay distribution of Eqn. \ref{eq:lognormal}. The resulting best-fit parameters are summarized in Table \ref{tab:BFP}.

        \end{enumerate}

        We highlight, in the following lines, the main conclusions of the presented analysis and the important similarities and differences that we find in the prompt gamma-ray emission properties of the two classes of short-hard and long-soft GRBs.

        \begin{itemize}

            \item
            {\bf Population Distribution.}   \\ The population distributions of LGRBs and SGRBs appears to be well described by two separate multivariate log-normal distributions in the 4-dimensional parameter space of the isotropic peak gamma-ray luminosity ($\liso$), total isotropic gamma-ray emission ($\eiso$), the intrinsic spectral peak energy ($\epkz$) and the intrinsic duration ($\durz$), once corrected for the effects of detector threshold and sample incompleteness. This is in line with previous findings of \citet{shahmoradi_multivariate_2013} \& \citet{shahmoradi_gamma-ray_2013}.

            \item
            {\bf GRB Classification.}   \\ According to the predictions of our GRB world, the most accurate and the quickest method of individual GRB classification -- solely based on prompt emission properties -- appears to be the observer-frame ratio $\epk/\dur[keV~s^{-1}]$.  We find that $99\%$ of all LGRBs have $\epk/\dur\lesssim50$, and $95\%$ of all SGRBs have $\epk/\dur\gtrsim50$ (c.f., Figure \ref{fig:OFbivariates3}, {\it center right}).

            We caution against the use of other similar quantities, such as the ratio of the observed spectral peak energy to bolometric fluence $\epk/\sbol$ as proposed by \citet{goldstein_new_2010}. Although this ratio seems to be a good discriminator in the sample of {\it detected} GRBs, it is strongly affected by sample incompleteness and detector threshold effects. Figure \ref{fig:OFbivariates2} ({\it center right}) illustrates the effects of sample incompleteness on the observed distribution of this ratio.

            \item
            {\bf Energetics \& Luminosity Function.}   \\ The presented GRB world model predicts a $3\sigma$ range of $\liso(erg~s^{-1})\in[10^{50},10^{53}]$ for the $64ms$ isotropic peak luminosity of SGRBs. A translation of this range to $1024ms$ peak luminosity using Eqn. \ref{eq:ppr} approximately corresponds to $\liso(erg~s^{-1})\in[8\times10^{48},1.3\times10^{53}]$. This range is very close and similar to the predictions of the GRB world model for $1024ms$ peak luminosity distribution of LGRBs with a $3\sigma$ range of $\liso(erg~s^{-1})\in[3.2\times10^{49},3.2\times10^{53}]$.

            Also predicted by the model are the $3\sigma$ ranges of $\eiso(erg)\in[10^{48},3.2\times10^{53}]$ \& $\eiso(erg)\in[1.6\times10^{49},5.0\times10^{54}]$ for the total isotropic gamma-ray emission of SGRBs \& LGRBs respectively. The two variables $\liso$ and $\eiso$ are strongly correlated with each other in both GRB classes (c.f., Table \ref{tab:BFP}).

            \item
            {\bf Prompt Duration \& Spectral Peak Energy.}   \\ The population distribution of the {\it rest-frame} spectral peak energies ($\epkz$) of both SGRBs and LGRBs appears to be described well by log-normal distributions with population averages $\epkz\sim955~keV$ \& $\epkz\sim300~keV$, and $3\sigma$ ranges $\epkz\in[45keV,16MeV]$ \& $\epkz\in[10keV,6MeV]$ respectively. In the observer-frame, this corresponds approximately to average $\epkz\sim300~keV$ \& $\epkz\sim85~keV$ with $3\sigma$ ranges $\epk(keV)\in[20,4000]$ \& $\epk(keV)\in[5,1430]$ for the two SGRB and LGRB classes, respectively.

            While the underlying duration distribution of LGRBs ($\durz$) does not seem to be significantly affected by the detection threshold of BATSE Large Area Detectors, there is tentative evidence that very short-duration SGRBs had, in general, lower chances of detection by BATSE (Figure \ref{fig:OFmarginals}, {\it bottom right}). For the population of SGRBs, we find a $3\sigma$ range of intrinsic duration $\durz(s)\in[0.002,7]$ with a population average of $\durz\sim180ms$. In contrast, for $\durz$ distribution of LGRBs we find a $3\sigma$ range of $\durz(s)\in[0.5,145]$ with population average $\durz\sim10s$ (Sec. \ref{sec:epkdur}).

            \item
            {\bf Prompt Gamma-Ray Correlations.}   \\ All four prompt gamma-ray variables appear to be strongly and positively correlated with each other in both GRB classes, with the exception of the two variables $\epkz$ \& $\durz$ which tend to be weakly, yet positively, correlated with each other. The intrinsic {\it hardness--brightness} relations (e.g., the Amati \& the Yonetoku relations) are confirmed but with much higher dispersions than originally reported for these relations (Figure \ref{fig:rest_frame}). The presented GRB world model reveals startling similarities in the strengths of the corresponding hardness--brightness correlations in the two GRB classes. Specifically, the model predicts a Pearson's correlation strength of $\rho\sim0.6$ for $\eiso$--$\epkz$ relation and $\rho\sim0.5$ for $\liso$--$\epkz$ relation, similarly in both GRB classes.

             The presented GRB model also predicts intrinsic {\it duration--brightness} correlations that are almost identical in strength between the two GRB classes, also very similar to the correlation strengths of hardness--brightness relations (c.f., Table \ref{tab:BFP} \& Sec. \ref{sec:correlations}).

        \end{itemize}

        In summary, we have presented a mathematical model with minimal free parameters that enables us, for the first time, to constrain the main characteristics of the prompt gamma-ray emission of short--hard and long--soft GRBs, jointly and simultaneously, while paying careful attention to selection biases and sample incompleteness due to gamma-ray detector thresholds. Our model predicts a high level of similarity in the joint population distribution of the prompt-emission properties of the two GRB classes, a finding that merits further investigation of the potential similarities in the prompt emission mechanisms of both GRB classes.

    \appendix

    \section{\\ BATSE Detection Threshold} \label{App:BDT}

        \begin{figure*}
            \centering
            \begin{tabular}{cc}
                \includegraphics[scale=0.31]{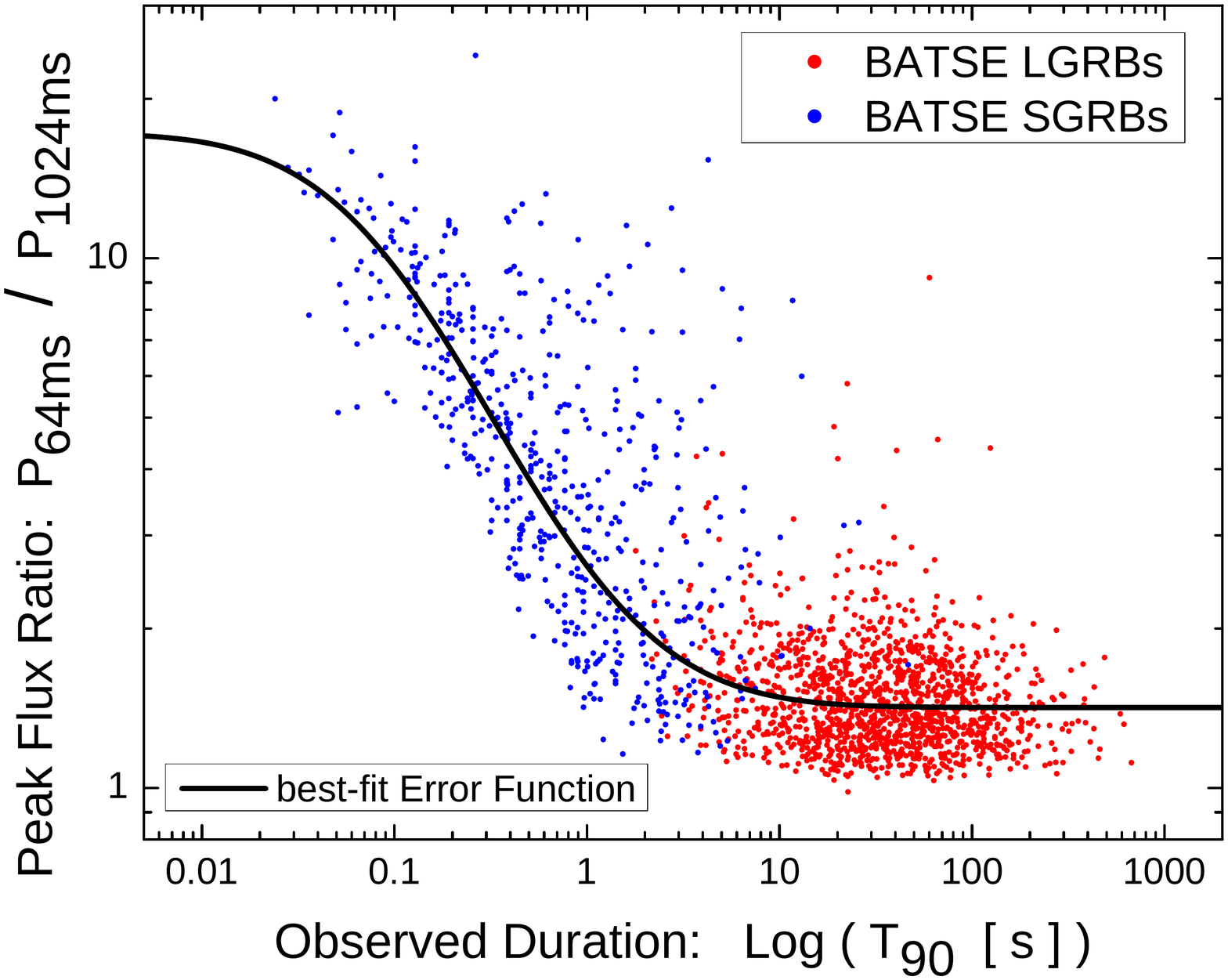} & \includegraphics[scale=0.31]{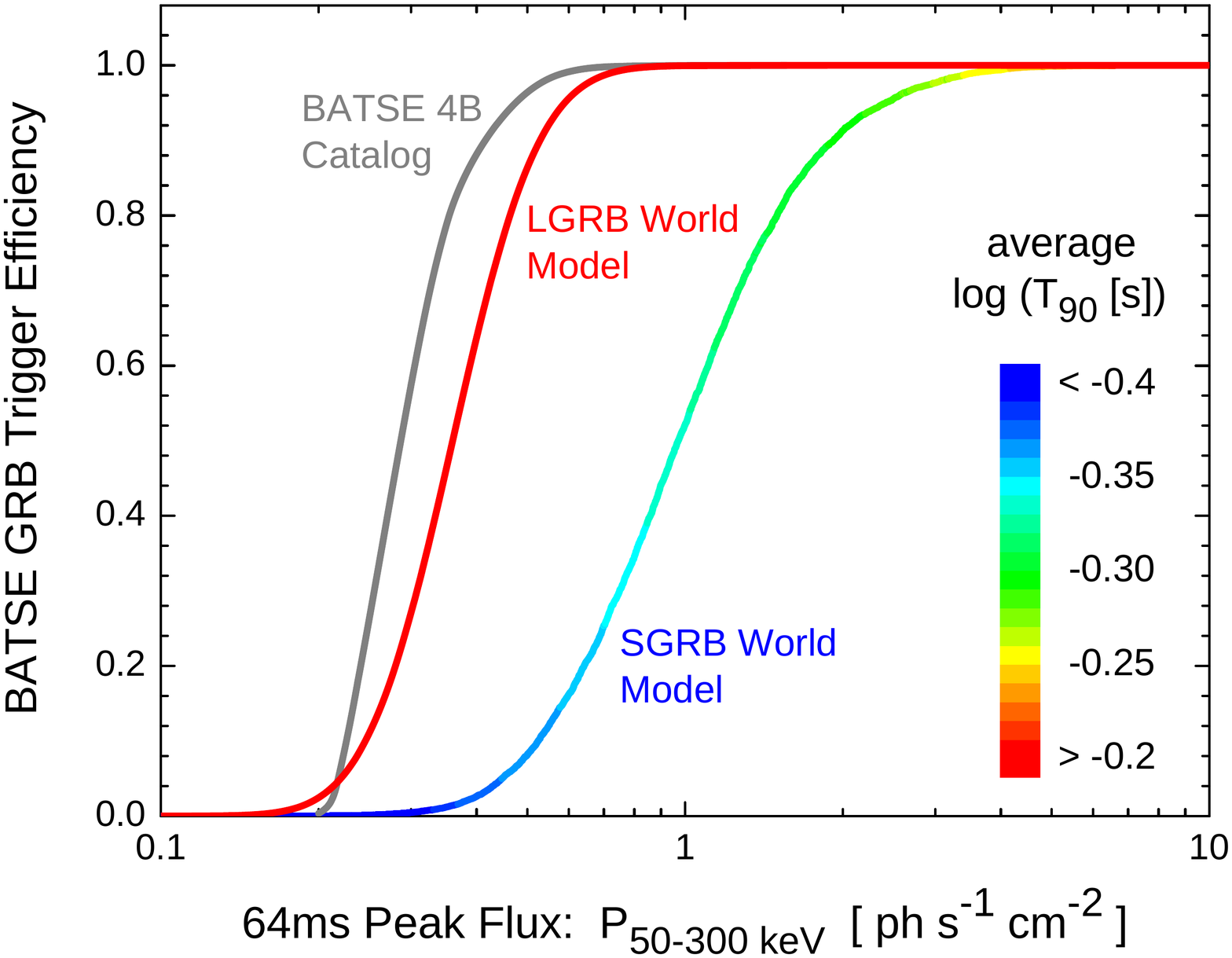}
            \end{tabular}
            \caption{{\bf Left:} An illustration of the higher detection probability of short GRBs on $64ms$ timescale peak flux towards very short durations compared to the commonly used $1024ms$ peak flux definition for LGRBs. The duration-dependence of the ratio of the two peak flux definitions highlights the inadequacy of the conventional definition of peak flux based on $1024ms$ time binning for the detection of short GRBs. {\bf Right:} The detection probabilities of BATSE GRBs according to the best-fit SGRB and LGRB world models as a function of the peak $64ms$ photon flux in BATSE detection energy range. For comparison, the nominal BATSE 4B catalog trigger efficiency \citep{paciesas_fourth_1999} for LGRBs is also shown by the grey solid line. It is evident from the efficiency curves that the very short SGRBs have on average lower detection probabilities compared to the longer duration events, while the detection probability of LGRBs is almost completely indifferent to the observed duration of the burst. This is also illustrated in the top-left plot of Figure \ref{fig:OFbivariates2}. \label{fig:durppr}}
        \end{figure*}

        An accurate modelling of the detection threshold of gamma-ray instruments is an integral part of any population study of GRBs. We have already argued in Sec. \ref{sec:MC} that modeling the trigger efficiency of gamma-ray detectors solely based on a measure of peak photon/energy flux -- as is generally done in most GRB population studies -- can potentially lead to systematic biases in the derived quantities. Although, the detection efficiency of most gamma--ray detectors depends solely on the observed peak {\it photon} flux in a limited energy window, the quantity of interest that is most often modelled and studied is the {\it bolometric} peak flux ($\pbol$). This variable depends on the observed peak photon flux and the spectral peak energy ($\epk$) for the class of LGRBs \citep[e.g., ][]{shahmoradi_multivariate_2013}, also on the observed duration (e.g., $\dur$) of the burst for the class of SGRBs. The effect of GRB duration on the peak flux measurement is very well illustrated in the {\it left} plot of Figure \ref{fig:durppr}, where we show that for GRBs with $\dur\lesssim1024ms$, the timescale used for the definition of the peak flux does indeed matter.  This is particularly important in modelling the triggering algorithm of BATSE Large Area Detectors, when a short burst could be potentially detected on any of the three different peak flux timescales used in the triggering algorithm: $64ms$, $256ms$ \& $1024ms$. Therefore, we adopt the following approach to construct a minimally-biased model of BATSE trigger efficiency for the population study of short-hard bursts.

        First, since only one definition (i.e., timescale) of the bolometric peak flux can be incorporated in the SGRB world model of Sec. \ref{sec:MC}, we use the least biased definition of peak flux for SGRBs -- the $64ms$ timescale definition -- in the GRB world model. Although, this definition is duration-independent for virtually all BATSE GRBs, it becomes an increasingly biased measure of the peak flux for very long duration GRBs ($\dur\gg1s$) close to detection threshold. We then approximate the $3$ {\it discrete} timescale trigger efficiency of BATSE LADs with a sigmoidal function that increases monotonically with increasing duration of the burst, from $64ms$ to $1024ms$. In other words, we convert the $64ms$ peak flux used in our GRB world model to an effective {\it triggering} peak flux $P_{\operatorname{eff}}~[ergs~s^{-1}]$, for which the detection efficiency of BATSE becomes duration-independent.

        To expand on this, consider an idealized GRB lightcurve containing only a single square-shaped pulse with an exact duration of $64ms$ and a signal strength that is $4$ times the required significance for its detection on a $64ms$ peak flux timescale.  In contrast, if there were only one triggering timescale $1024ms$ available on BATSE, the signal strength of this $64ms$ event would fall right on the detection threshold of BATSE LADs. Thus, a $64ms$ burst of peak flux $P_{\operatorname{64}}~[ph~s^{-1}]$ would be equivalent to an effective $1024ms$ peak flux,

        \begin{equation}
            P_{\operatorname{eff}} ~[ph~s^{-1}] = \frac{1}{4} P_{\operatorname{64}} ~[ph~s^{-1}],
        \end{equation}

        for the triggering algorithm of BATSE on a $1024ms$ timescale.

        In reality however, GRB lightcurves are far more diverse than a single square pulse. Thus in order to build a more realistic model of BATSE LAD triggering algorithm, we fit a {\it complementary} Error function of the mathematical form,

        \begin{equation}
            \label{eq:erfc}
            \operatorname{erfc}(x) = \frac{2}{\sqrt{\pi}}\int_{x}^{\infty} \operatorname{e}^{-t^2}dt,
        \end{equation}

         to the logarithm of the ratio of $64ms$ to $1024ms$ peak fluxes ($R_{\operatorname{P_{64}/P_{1024}}}$) as a function of the observed duration ($\dur$) of BATSE GRBs, as illustrated in the {\it left plot} of Figure \ref{fig:durppr}. The resulting best-fit function for $R_{\operatorname{P_{64}/P_{1024}}}$ has the form,

        \begin{eqnarray}
            \label{eq:ppr}
            \log\big(R_{\operatorname{P_{64}/P_{1024}}}\big) &\simeq& 0.15 \nonumber \\
                                                &+& 0.56 \times \operatorname{erfc}\bigg(\frac{\log\big(T_{\operatorname{90}}\big) + 0.48}{1.05}\bigg).
        \end{eqnarray}

        The effective {\it triggering} peak flux in the SGRB world model is then calculated using the following relation,

        \begin{equation}
            \label{eq:effectivePF}
            \log\big(P_{\operatorname{eff}}\big) \simeq \log\big(P_{\operatorname{64}}\big) - \frac{1}{2} \bigg( \log\big(R_{\operatorname{P_{64}/P_{1024}}}\big) - 0.15 \bigg).
        \end{equation}

        Once $P_{\operatorname{eff}}$ is obtained, we follow the approach of \citet{shahmoradi_multivariate_2013} to calculate the detection probability ($\eta$) of a given SGRB with an effective {\it triggering} peak flux $P_{\operatorname{eff}}$,

        \begin{eqnarray}
            \label{eq:eta}
            && \eta\big(\operatorname{detection}|\mu_{\operatorname{thresh}},\sigma_{\operatorname{thresh}},\liso,\epkz,\durz,z\big) \nonumber \\
            && = \frac{1}{2} + \frac{1}{2} \times \nonumber \\ && \operatorname{erf}\bigg(\frac{\log\big(P_{\operatorname{eff}}(\liso,\epkz,\durz,z)-\mu_{\operatorname{thresh}}\big)}{\sqrt{2}\sigma_{\operatorname{thresh}}}\bigg),
        \end{eqnarray}

        where $\mu_{\operatorname{thresh}}$ \& $\sigma_{\operatorname{thresh}}$ are the detection threshold parameters that are found by fitting the SGRB world model to BATSE observational data (c.f., Table \ref{tab:BFP}), and $P_{\operatorname{eff}}(\liso,\epkz,\durz,z)$ is the $1024ms$ effective triggering peak flux in BATSE energy range of detection , $50$--$300$[keV], calculated from the $64ms$ peak flux ($P_{\operatorname{64}}(\liso,\epkz,z)~[ph~s^{-1}]$) in BATSE detection energy range using Eqn. \ref{eq:effectivePF}. The connection between the rest-frame GRB parameters, $\liso \& \epkz,z$, and the $64ms$ peak flux $P_{\operatorname{64}}$ is obtained by fitting a smoothly broken power-law known as the Band model \citep{band_batse_1993} of the mathematical form,

        \begin{equation}
            \label{eq:Band}
            \Phi (E) \propto
            \begin{cases}
                E^{\alpha}~ \operatorname{e}^{\big(-\frac{(1+z)(2+\alpha)E}{\epkz}\big)} & \text{if $E\le\big(\frac{\epkz}{1+z}\big)\big(\frac{\alpha-\beta}{2+\alpha}\big)$,} \\
            	E^{\beta} & \text{if otherwise.}
            \end{cases}
        \end{equation}

        to SGRBs differential photon spectra, such that,

        \begin{equation}
            \label{eq:pf64}
            P_{\operatorname{64}}\big(\liso,\epkz,z\big)=\frac{\liso}{4\pi {D_L}^2(z)} \frac{\int_{50}^{300} \Phi ~\operatorname{d}E}{\int_{\nicefrac{0.1}{1+z}}^{\nicefrac{20000}{1+z}} E \Phi ~\operatorname{d}E},
        \end{equation}

        where $D_L(z)$ is the luminosity distance of Eqn. \ref{eq:lumdis}. In order to bring the above calculations into the realm of current computational technologies, we simplify the integration limits in the denominator of Eqn. \ref{eq:pf64} to a redshift-independent energy range $[0.1\text{keV},20\text{MeV}]$ and fix the low-- \& high-- energy photon indices of the Band model (Eqn. \ref{eq:Band}) to their corresponding population averages $\alpha=-1.1$ \& $\beta=-2.3$. \citet{butler_cosmic_2010} show that these simplifications result in an uncertainty of $<0.05$dex in the estimated peak flux, which is negligible compared to the existing systematic biases in BATSE data (c.f., Appendix \ref{App:SB} and uncertainties in the spectral peak energy estimates of \citet{shahmoradi_hardness_2010} used in this work).  The resulting best-fit model of BATSE detection efficiency as a function of $P_{\operatorname{64}}$ for the class of short-hard bursts is illustrated and compared to the detection efficiency of long-soft bursts in the {\it right} plot of Figure \ref{fig:durppr}.

    \section{\\ Systematic Biases in BATSE GRB Data} \label{App:SB}

        \begin{figure*}
            \centering
            \begin{tabular}{cc}
                \includegraphics[scale=0.31]{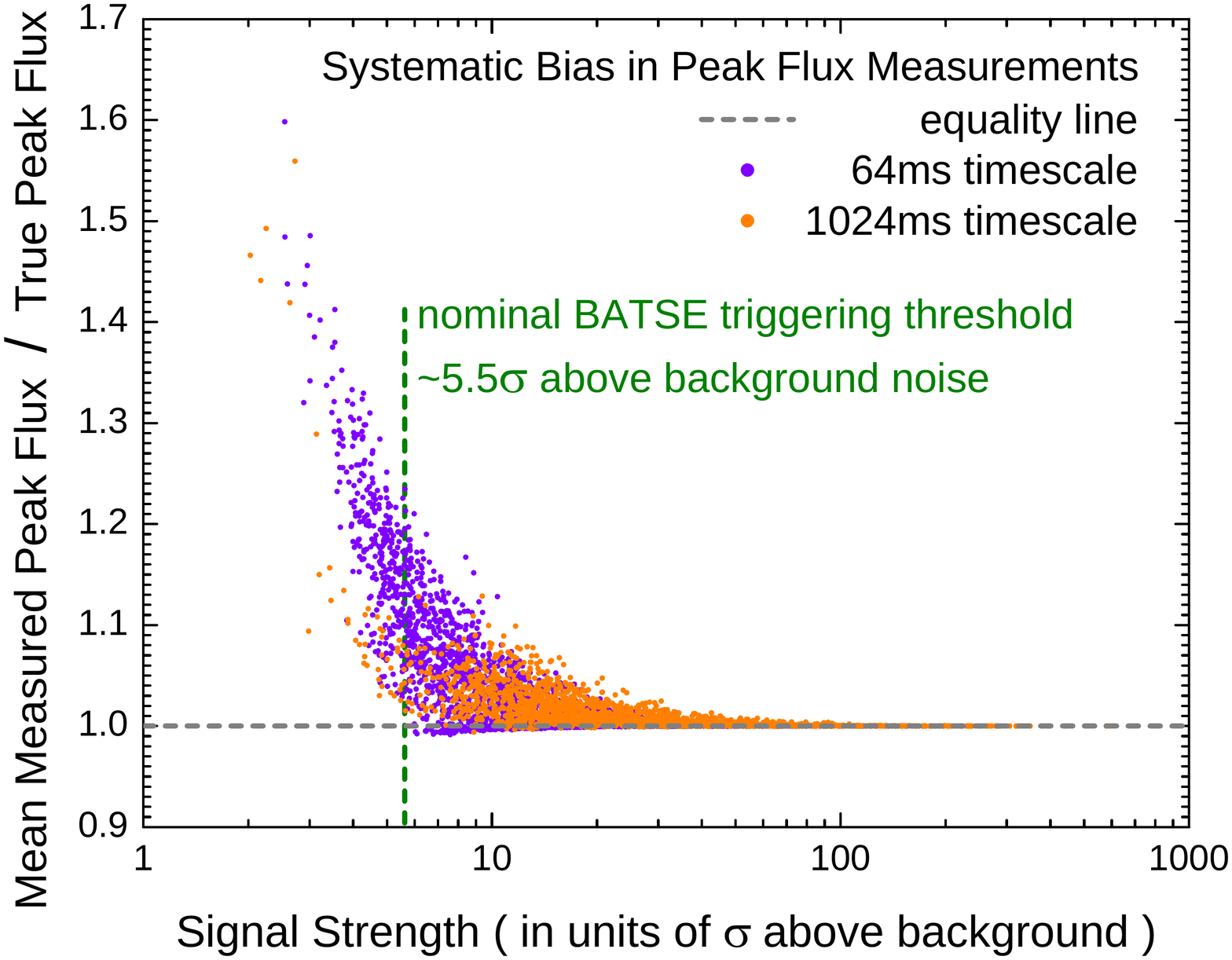} & \includegraphics[scale=0.31]{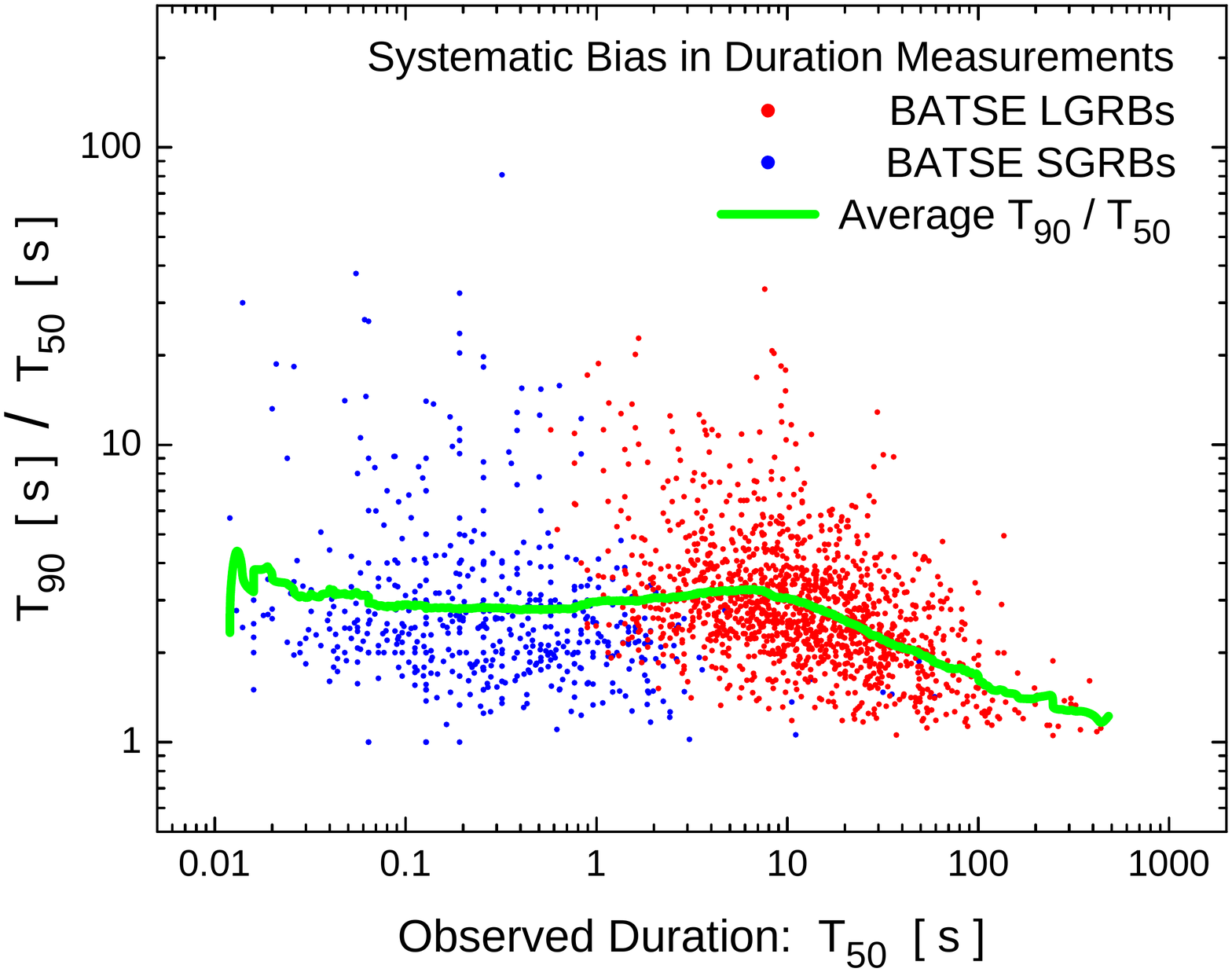}
            \end{tabular}
            \caption{{\bf Left:} An illustration of the existing systematic bias in the peak flux measurements of BATSE GRBs and possibly other GRB catalogs, such as Swift BAT \& Fermi GBM. The horizontal axis represents the {\it peak-flux} signal strength of a given BATSE GRB lightcurve, in units of the standard deviation ($\sigma$) of the background photon counts fluctuations. For comparison, the nominal BATSE triggering threshold is shown by the green vertical dashed line. {\bf Right:} An illustration of the existing systematic bias in the duration measurements of BATSE GRBs, in particular $\dur$ definition at very long durations. At durations longer than $\dur \sim 30 [s]$ corresponding to $T_{50} \sim 10 [s]$ close to detection threshold, the prompt durations of most GRBs tend to be systematically underestimated in BATSE catalog \citep[c.f., ][]{koshut_systematic_1996, hakkila_fluence_2000}. This systematic bias is also evident in the top-left \& center-left plots of both Figures \ref{fig:OFbivariates2} \& \ref{fig:OFbivariates3}. \label{fig:systematicbiases}}
        \end{figure*}

        As argued by \citet{hogg_maximum_1998}, astronomical catalogs and surveys are prone to systematic biases in measurements, in particular, close to the detection threshold of the observational instruments. The BATSE catalog of GRBs is no exception to such biases, also noted by BATSE team and others \citep[e.g.,][]{nemiroff_gross_1994, paciesas_fourth_1999, stern_off-line_2001, stern_evidence_2002}. Although, throughout this work we relied on BATSE catalog data in their original form, here we present the results of our search for the potential signatures of systematic biases in BATSE data, which will pave the way for more accurate and rigorous population studies of GRBs in future.

        In order to identify the extent of systematic bias in BATSE GRB data close to detection threshold, we first calculate the average background photon counts for each individual BATSE GRB lightcurve, in each of the four main energy channels of BATSE Large Area Detectors. we then subtract the calculated average background gamma-ray photon counts from the corresponding BATSE GRB lightcurves. The background-subtracted lightcurves are then used to calculate the peak photon fluxes of all GRBs in the sample, in three $64ms$, $256ms$, \& $1024ms$ timescale definitions. Although the calculated peak fluxes are already contaminated and biased by background noise, we assume they represent the `true peak fluxes' of BATSE GRBs and use them as our reference to simulate and investigate the effects of background noise in the calculation of peak flux at very low signal-to-noise ratios. To do so, we add synthetic background noise to the entire time-bins of each of the background-subtracted lightcurves. The noise count ($n$) for each time-bin in a given GRB lightcurve is drawn from the Poisson distribution,

        \begin{equation}
            \label{eq:poisson}
            P(n|\lambda) = \frac{\lambda^n}{n!}\operatorname{e}^{-\lambda}.
        \end{equation}

        The mean of the noise count ($\lambda$) for each BATSE GRB lightcurve is set to the original mean background photon counts found in each of the original lightcurves. We then subtract the average background counts ($\lambda$) from the newly obtained lightcurves and measure the new peak fluxes in three timescales $64ms$, $256ms$ \& $1024ms$. This procedure of adding and subtracting synthetic background noise is then repeated $10000$ time for each GRB lightcurve to obtain a sample of peak flux measurements for each of BATSE catalog GRB. The mean of this sample is then compared to the original `true peak flux' measured after the initial background subtraction. The results are illustrated in the {\it left} plot of Figure \ref{fig:systematicbiases}. It is evident that peak the flux estimates become increasingly biased with decreasing Signal-to-Noise Ratio (SNR) in BATSE data. This result corroborates the findings of \citet{hogg_maximum_1998}, who argue that flux measurements in astronomical surveys tend to be overestimated at very low SNR.

        We have also tested the BATSE fluence and spectral peak energy data for the potential existence of systematic biases at low SNR, for which we find no significant evidence. We identify, however, a systematic bias in the duration measurements of BATSE catalog GRBs, specifically, in very long $\dur$ measurements. This is illustrated in the {\it right} plot of Figure \ref{fig:systematicbiases}, where we show that the $\dur$ measures of BATSE LGRBs are likely systematically underestimated at very long durations, corresponding to durations $T_{50}\gtrsim10[s]$. As argued by \citet{koshut_systematic_1996, hakkila_fluence_2000, kocevski_lack_2012}, the $\dur$ definition of GRB duration seems to be prone to systematic underestimations for very long duration low-SNR GRBs, since it is more likely that the late-time weak signals in the lightcurves would vanish in the background noise. The effects of this duration bias are also evident in the predictions of the LGRB world model shown in Figures \ref{fig:OFbivariates2} \& \ref{fig:OFbivariates3} ({\it top--left \& center--left} plots in both figures).

\bibliographystyle{mn2e}
\bibliography{manuscript_MNstyle}

\label{lastpage}

\end{document}